\pdfoutput=1
\documentclass[iop, apj, twocolappendix, numberedappendix]{emulateapj}
\usepackage{xspace}
\usepackage{amsmath}
\usepackage{framed} 
\usepackage{txfonts}
\usepackage{epstopdf}
\usepackage{color}
\usepackage{rotating}
\usepackage{natbib}
\usepackage{ulem}
\usepackage{xspace}
\usepackage[colorlinks=true,urlcolor=blue,linkcolor=blue,citecolor=blue]{hyperref}

\newdimen\hssize
\newdimen\hdsize
\def\gcm3{\mathrm{g} / \mathrm{cm}^3}
\def\m200m{M_{\rm 200m}}

\def\gtsima{$\; \buildrel > \over \sim \;$}
\def\ltsima{$\; \buildrel < \over \sim \;$}
\def\prosima{$\; \buildrel \propto \over \sim \;$}
\def\gsim{\lower.7ex\hbox{\gtsima}}
\def\lsim{\lower.7ex\hbox{\ltsima}}
\def\simgt{\lower.7ex\hbox{\gtsima}}
\def\simlt{\lower.7ex\hbox{\ltsima}}
\def\simpr{\lower.7ex\hbox{\prosima}}

\def\btheta{\boldsymbol{\theta}}

\usepackage{etoolbox}
\makeatletter
\patchcmd{\NAT@citex}
  {\@citea\NAT@hyper@{\NAT@nmfmt{\NAT@nm}\NAT@date}}
  {\@citea\NAT@nmfmt{\NAT@nm}\NAT@hyper@{\NAT@date}}
  {}
  {}
\patchcmd{\NAT@citex}
  {\@citea\NAT@hyper@{%
     \NAT@nmfmt{\NAT@nm}%
     \hyper@natlinkbreak{\NAT@aysep\NAT@spacechar}{\@citeb\@extra@b@citeb}%
     \NAT@date}}
  {\@citea\NAT@nmfmt{\NAT@nm}%
   \NAT@aysep\NAT@spacechar%
   \NAT@hyper@{\NAT@date}}
  {}
  {}
\patchcmd{\NAT@citex}
  {\@citea\NAT@hyper@{%
     \NAT@nmfmt{\NAT@nm}%
     \hyper@natlinkbreak{\NAT@spacechar\NAT@@open\if*#1*\else#1\NAT@spacechar\fi}%
       {\@citeb\@extra@b@citeb}%
     \NAT@date}}
  {\@citea\NAT@nmfmt{\NAT@nm}%
   \NAT@spacechar\NAT@@open\if*#1*\else#1\NAT@spacechar\fi%
   \NAT@hyper@{\NAT@date}}
  {}
  {}
\makeatother

\def\avrg#1{\left\langle #1 \right\rangle}

\newcommand{\revise}[1]{{\textcolor{black}{#1}}}
\newcommand{\reviseplus}[1]{{\textcolor{black}{#1}}}

\shorttitle{
Mass-richness relation of SDSS clusters
}
\shortauthors{Murata, Nishimichi, Takada et~al.}
\slugcomment{To be submitted to the Astrophysical Journal}

\begin{document}
\def\figdir{.}
\def\figext{pdf}
\title{
Constraints on the mass-richness relation 
from the abundance \\ and weak lensing of SDSS clusters
}
\author{
Ryoma Murata\altaffilmark{1, 2}\email{ryoma.murata@ipmu.jp}, 
Takahiro Nishimichi\altaffilmark{1, 3}, 
Masahiro Takada\altaffilmark{1}, 
Hironao Miyatake\altaffilmark{4, 1},
Masato Shirasaki\altaffilmark{5},
Surhud More\altaffilmark{1},\\
Ryuichi Takahashi\altaffilmark{6},
and Ken Osato\altaffilmark{2}
}

\affil{
$^1$ Kavli Institute for the Physics and Mathematics of the Universe (WPI),
The University of Tokyo Institutes for Advanced Study (UTIAS),
\\ The University of Tokyo, 5-1-5 Kashiwanoha, Kashiwa-shi, Chiba, 277-8583, Japan; \texttt{ryoma.murata@ipmu.jp}\\
$^{2}$ Department of Physics, University of Tokyo, 7-3-1 Hongo, Bunkyo-ku, Tokyo 113-0033 Japan\\
$^{3}$ CREST, JST, 4-1-8 Honcho, Kawaguchi, Saitama, 332-0012, Japan\\
$^{4}$Jet Propulsion Laboratory, California Institute of Technology, Pasadena, CA 91109, USA\\
$^{5}$Division of Theoretical Astronomy, National Astronomical Observatory of Japan, 2-21-1 Osawa, Mitaka, Tokyo 181-8588, Japan\\
$^{6}$Faculty of Science and Technology, Hirosaki University, 
3 Bunkyo-cho, Hirosaki, Aomori, 036-8561, Japan
}

\begin{abstract}
We constrain the scaling relation between optical richness ($\lambda$) and halo
mass ($M$) for a sample of SDSS redMaPPer galaxy clusters within the context of
the {\it Planck} cosmological model. We use a forward modeling approach where we
model the probability distribution of optical richness for a given mass, $P(\ln
\lambda| M)$. To model the abundance and the stacked lensing profiles, we use
an emulator specifically built to interpolate the halo mass function and the
stacked lensing profile for an arbitrary set of halo mass and redshift, 
which is calibrated based on a suite of high-resolution $N$-body simulations.
We apply our method to 8,312 SDSS redMaPPer clusters with $20\le \lambda \le
100$ and $0.10\le z_{\lambda}\le0.33$, and show that the log-normal
distribution model for $P(\lambda|M)$, with four free parameters, well
reproduces the measured abundances and lensing profiles simultaneously. The
constraints are characterized by the mean relation, $\left\langle
\ln{\lambda}\right\rangle(M)=A+B\ln(M/M_{\rm pivot})$, with
$A=3.207^{+0.044}_{-0.046}$ and $B=0.993^{+0.041}_{-0.055}$
(68\%~CL), where the pivot mass scale $M_{\rm pivot}=3\times 10^{14}
h^{-1}M_\odot$, and the scatter
$\sigma_{\mathrm{\ln\lambda}|M}=\sigma_0+q\ln(M/M_{\rm pivot})$ with
$\sigma_0=0.456^{+0.047}_{-0.039}$ and
$q=-0.169^{+0.035}_{-0.026}$.  We find that a large scatter in halo
masses is required at the lowest richness bins ($20\le \lambda \lesssim 30$) in
order to reproduce the measurements. Without such a large scatter, the model
prediction for the lensing profiles tends to overestimate the measured
amplitudes. This might imply a possible contamination of intrinsically
low-richness clusters due to the projection effects. Such a low-mass halo
contribution is significantly reduced when applying our method to the sample of
$30\le \lambda \le 100$. 
\end{abstract}

\section{INTRODUCTION}
\label{sec:intro}

The abundance of galaxy clusters \revise{is} a powerful tool \revise{to constrain} the primordial fluctuations as well as cosmological parameters: e.g., see \citet{Whieetal:93} for the \revise{pioneering} theoretical work, \citet{Ekeetal:96}, \citet{KitayamaSuto:97} and 
\citet{Vikhlininetal:09} for the cosmological 
constraints using the sample of X-ray clusters,  
\citet{PlanckSZ:16} for the sample of clusters selected via the Sunyaev-Zel'dovich (SZ) effect, 
and \citet{Rozoetal:10} for the sample of optically-selected clusters. In particular, the time evolution of the cluster abundance, if measured precisely, can be used to constrain properties of dark energy that govern the cosmic accelerating expansion \citep{Haiman:2001,Lima:2005,TakadaBridle:07,OguriTakada:2011} \citep[also see][for a thorough review]{Weinberg:2013}. Ongoing and upcoming wide-area galaxy surveys promise to further improve cluster-based cosmology, if systematic errors are under control. Such surveys include the Subaru Hyper Suprime-Cam (HSC) Survey\footnote{\url{http://hsc.mtk.nao.ac.jp/ssp/}} \citep{HSCOverview:17},
the Dark Energy Survey\footnote{\url{https://www.darkenergysurvey.org}}, and ultimately the Large Synoptic Survey Telescope\footnote{\url{https://www.lsst.org}}, the Euclid satellite mission\footnote{\url{http://sci.esa.int/euclid/}}, and the WFIRST satellite mission \citep{Spergel:2013}.

\revise{The theoretical predictions for cluster observables are primarily given by the
the halo mass for a given cosmological model.  Therefore, to attain the full
potential of cluster cosmology with upcoming wide-area surveys,} we need to
infer the ``masses'' of individual clusters or the average mass \revise{of} a sample of
clusters. In many cases we \revise{have to} resort to mass-observable relations, which
are calibrated by intensive observations \revise{of a subsample of clusters} under
consideration. Well-calibrated, unbiased mass-observable \revise{relations can allow
us} to infer \revise{the} cluster masses on individual or statistical basis, thereby
bridging \revise{the gap} between theory and measurements \revise{and} to obtain cosmological
constraints from the comparison.

\revise{The} weak gravitational lensing \citep[][for a review]{Bartelmann:2001} \revise{has
emerged as} a powerful \revise{observable} to constrain cluster masses. \revise{Gravitational
lensing causes a} coherent distortion pattern in the shapes of galaxies \revise{that lie
in the background of galaxy clusters. Measurements of such distortions allow a
direct measure of the projected mass density profile for the clusters, which in
turn can be used to calibrate the cluster mass-observable relations}
\citep{Johnston:2007,Okabe:2010,Okabeetal:13,Hoekstraetal:15,vanUitertetal.16,Miyatake:2016,Battagliaetal:16,Simet:2016,Melchior:2016}.
However, \revise{the} weak lensing signal is too noisy \revise{to measure for individual
clusters and is measurable only in a statistical sense,} e.g., stacking shapes
of many background galaxies for a sample of clusters. To obtain an unbiased
estimation of the mass for the sampled clusters, we need to take into account
the underlying distribution of clusters masses for the sample. In addition to
this, the sample is usually affected by the measurement errors as well as
intrinsic scatters in the mass-observable relation. These effects need to be
carefully taken into account. 

The main purpose of this paper is to \revise{calibrate} the mass-observable relation
from a joint measurement of the abundance (number counts) and the stacked
cluster \revise{weak} lensing profiles. \revise{We develop and} apply our method to the SDSS
redMaPPer cluster catalog that is constructed by identifying overdensities of
red-sequence galaxies with similar colors as galaxy clusters from the SDSS 
$ugriz$ photometries
\citep{Rykoff:2014,Rozo:2014:redMaPPer2,Rozo:2015:redMaPPer3,Rozo:2015}
\citep[most recently][for the details of the method]{Rykoff:2016}. Since the
cluster finder gives an estimation of \revise{the} optical richness, $\lambda$, for each
cluster, we will constrain the scaling relation between the optical richness
and mass for the clusters. 
In this paper we \revise{develop a} \textit{forward} modeling approach, where 
we constrain the probability distribution of richness for a given halo mass, $P(\ln \lambda|M)$. 
This is \revise{in contrast with previous studies} \citep{Baxter:2016,Simet:2016,Melchior:2016,Jimenoetal:17}, where the backward modeling approach is employed to constrain the probability of mass for a given richness, 
$P(\ln M|\lambda)$. 
The forward modeling approach has \revise{several} \revise{advantages}.
\revise{Firstly, we can use the abundance measurements more easily to constrain the mass-observable relation,
as \cite{Saroetal2015} constrained $P(\ln \lambda|M)$ from the abundance measurements of SZ selected clusters after matching to redMaPPer clusters.
Secondly, $P(\ln M|\lambda)$ can be inferred from $P(\ln \lambda|M)$ once the halo mass function $P(\ln M)$ is given, based on the Bayes theorem,
while the opposite transformation, i.e, inferences of $P(\ln \lambda|M)$ from $P(\ln M|\lambda)$, is not straightforward, because
this requires knowledge of the richness function $P(\ln \lambda)$ over the whole range of $\lambda$,
which is not generally available, or is at least very noisy (and possibly
affected by contamination), for richness below a threshold richness in cluster
catalogs.  Thirdly, the forward modeling is convenient to generate mock
catalogs of clusters by populating halos in $N$-body simulations with galaxies,
e.g. to test systematics in a cluster-finding algorithm.} 

In order to accurately constrain halo masses for a sample of the SDSS redMaPPer
clusters, we need to properly model the mass density profile around the
clusters for an assumed cosmological model \revise{(the {\it Planck} cosmological model
\citep{Planck:2015} in our case)}.  For this \revise{purpose} we use a halo
\textit{emulator}, which is an interpolator for the halo mass
function and the halo-matter cross-correlation function for an input set of
parameters (halo mass, redshift, \revise{and} correlation separation length or wavenumber).
\revise{The emulator} is built using a suite of high-resolution $N$-body simulations and their halo
catalogs (Nishimichi et~al. in prep.). The $N$-body simulation-calibrated emulator properly models 
both the 1-halo term, the 2-halo term and the transition regime of the mass profile, each of which \revise{is sensitive}
to \revise{the} halo mass. In addition, the relation of halo concentration with \revise{mass}, more generally variations in the halo profiles in the 1-halo
regime, is automatically included in the emulator prediction. On the other hand, if an analytical model of the halo profile
such as the Navarro-White-Frenk model \citep{Navarro:1996} is \revise{adopted}, one needs
to model the halo mass-concentration relation in order to compute the lensing profile, and such an analytical 
halo model ceases to be accurate at scales in the transition regime between the 1- and 2-halo terms, unless a sufficient 
number of free parameters are introduced \citep{Diemer:2014}. \revise{In} order to model statistical errors in the abundance and 
the stacked lensing measured from the SDSS data over the survey footprint (about 10,000 ${\rm deg}^2$), we use 108 mock catalogs 
of the redMaPPer clusters and source galaxies for the entire SDSS footprint, which are generated based on the full-sky ray-tracing 
simulations and \revise{the resultant} halo catalogs \citep{Shirasaki:2016,Takahashietal:17}. Thus we 
extensively use $N$-body simulation calibrated models and mocks in preparation for high-precision cosmology with cluster
observables that \revise{will become available with} upcoming surveys. Although we apply this method to the SDSS redMaPPer clusters, it is applicable
to any cluster sample including the high-redshift cluster catalog that is built based on the deep Subaru HSC data \citep{Ogurietal:17}.

This paper is organized as follows. 
In Section~\ref{sec:m_observable}, we describe details of cluster observables we measure from the SDSS data: the cluster abundance and the stacked 
lensing profile. 
In Section~\ref{sec:m_modeling}, we introduce a model to describe the mass-richness relation, $P(\ln \lambda|M)$, and then
formulate the method to model the cluster observables in a forward modeling approach way, where all the observables in a given richness bin are
modeled from $P(\ln \lambda|M)$ for an assumed cosmology. In Section~\ref{sec:m_halo_emulator}, we describe the halo emulator to model the lensing profile and the halo mass function for the {\it Planck} cosmology, and then, in Section~\ref{sec:m_mock_covariance}, describe details of the mock catalogs of the SDSS 
data which are used to estimate the error covariance matrix for the cluster observables. 
In Section~\ref{sec:m_results}, we show the results obtained by applying the method to the SDSS data, which include constraints on the mass-richness
relation, and compare our results with previous work in \citet{Simet:2016}. 
In Section~\ref{sec:m_discussion}, we 
discuss the impact of possible residual systematic errors on our results. 
Section~\ref{sec:conclusion} is devoted to conclusion and summary.
Throughout this paper we use natural units in which the speed of light is set equal to one, $c=1$.
\revise{We adopt $M\equiv M_{\rm 200m}= 4\pi (R_{200{\rm m}})^3\bar{\rho}_{\rm m0}\times 200/3$ for the halo mass definition, where $R_{\rm 200m}$ is the spherical halo boundary radius within which the mean mass density is $200$ times the present-day mean mass density.
As for the fiducial cosmological model, we adopt the \textit{Planck} cosmology \citep{Planck:2015}:
$\Omega_{\rm b0}h^2=0.02225$ and $\Omega_{\rm c0}h^2=0.1198$ for the density parameters of baryon and cold dark matter, respectively,
$\Omega_{\Lambda}=0.6844$ for the cosmological constant, and $\ln (10^{10}A_s)=3.094$ and $n_s=0.9645$ for the primordial power spectrum.}

\section{DATA AND CLUSTER OBSERVABLES}
\label{sec:m_observable}

In this section we describe details of the data and cluster observables
used in this paper: the abundance and stacked cluster lensing.  

\subsection{The SDSS redMaPPer cluster catalog}
\label{sec:dataredmapper}

\begin{figure}
	\centering \includegraphics[width=0.495
    \textwidth]{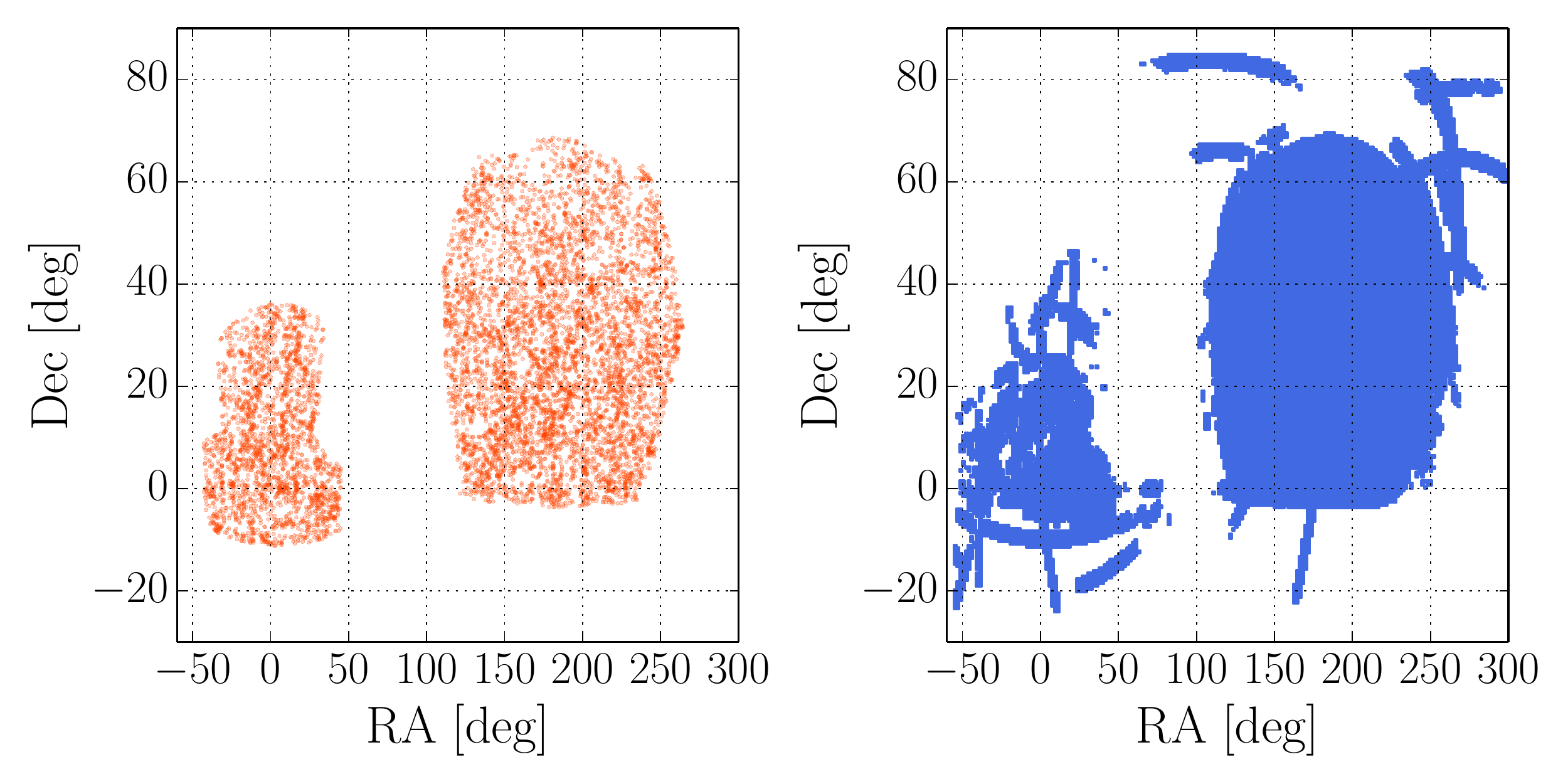} 
	\caption{{\it Left}: The distribution of SDSS redMaPPer clusters in the celestial coordinates, where each dot corresponds to a cluster \citep{Rykoff:2016}. 
We use $8,312$ clusters with richness and redshift range of $20 \leq \lambda \leq 100$ and 
$0.10 \leq z_{\lambda} \leq 0.33$, which makes
the sample nearly volume-limited, over 
the total area of  $\Omega_{\rm tot}=10,401~\mathrm{deg^2}$.  
{\it Right}: The distribution of galaxies in the SDSS shape catalog,
used for the stacked lensing analysis, taken from  \citet{Mandelbaum:2013}.
There are approximately 39 million source galaxies used for the lensing analysis in this paper, which
have a slightly different distribution from that of the redMaPPer clusters
based on the different cuts of the data (image quality and depth). 
For illustrative purpose, we show the distribution after assigning the source galaxies to rectangular pixels in the RA and Dec coordinates.
}
\label{fig:footprint}
\end{figure}

We use the catalog of galaxy clusters identified from
the SDSS DR8 photometric galaxy catalog \citep{Aihara:2011} by the 
\textit{red}-sequence \textit{Ma}tched-filter
\textit{P}robabilistic \textit{Per}colation (redMaPPer) cluster finder algorithm of v6.3 
 \citep{Rykoff:2014, Rozo:2015,Rykoff:2016}. 
We also refer to the website\footnote{The catalog can be found at
http://risa.stanford.edu/redmapper/.
} for further details.
The cluster finder uses the \revise{SDSS} $ugriz$ magnitudes and their errors to
identify overdensities of red-sequence galaxies with similar colors.
For each cluster, the catalog contains an optical richness estimate
$\lambda$, a photometric redshift $z_\lambda$, the angular position and
centering probabilities of five candidate central galaxies.  Throughout
this paper we use the position of the most probable central galaxy
in each cluster region as a proxy of the cluster center.
We denote the centering probability for the chosen central galaxy in each cluster region 
by $p_{\rm cen}$.
A separate member galaxy catalog provides a list of
members for each cluster, with assigned membership
probability, $p_{\mathrm{mem} }$.
\revise{The} typical  photometric redshift error, 
$\sigma(z_{\lambda})$, \revise{has been shown to be of order}
0.01 \citep{Rykoff:2016}.  In addition, \revise{the} typical richness estimate error, $\sigma(\lambda)$, is 
about $3$. Following \citet{Miyatake:2016}, in this
paper we adopt the parent cluster catalog consisting of 8,312 redMaPPer
clusters with $20\leq \lambda \leq 100$ and $0.10\leq z_\lambda\leq 0.33$,
which is an approximately volume-limited sample \citep[also
see][]{More:2016}. The average and median redshift of the clusters are
$0.24$ and $0.25$, respectively. 
We assume $\Omega_{\mathrm{tot}} = 10,401\ \mathrm{deg}^2$ for the total
survey area which takes into account the cut of contiguous high-quality
data based on the masks defined in the BOSS analysis
\citep{Dawson:2013}, compared to the original DR8 imaging catalog
covering approximately $14,000~$deg$^2$
\citep[see Section~2.2 in][for details]{Rykoff:2016}.
Figure~\ref{fig:footprint} shows the SDSS survey footprint, where the
redMaPPer clusters are defined.

The optical richness $\lambda$ for each cluster is estimated as the sum
of membership probability $p_{ \mathrm{mem} }$ for all the potential member
galaxies, as described in \cite{Rozo:2009} and \cite{Rykoff:2012}:
\begin{equation}
	\frac{\lambda}{S} = \sum_{\mathrm{gals}} p_{\mathrm{mem}},
\end{equation}
where $S$ is a scale factor to correct for the effects of masks and
depth variation on optical richness estimation.

Since a cluster is a finite size object and the projected size depends
on richness and redshift of a cluster, a detection efficiency of the
cluster depends on the richness and redshift as well as
observational effects such as the survey boundary and masked regions.
The redMaPPer finding algorithm adopts \revise{a} richness-dependent, circular aperture to
estimate \revise{the} richness for each cluster \citep{Rykoff:2012, Rozo:2015}:
\revise{
\begin{equation}
R_{\rm aperture}(\lambda, z)=1.0\times \left( \frac{\lambda}{100} \right)^{0.2}(1+z)~~[h^{-1}{\rm Mpc} ],
\label{eq:aperture}
\end{equation}
in comoving coordinates.}
In order to estimate the most probable richness for each cluster, 
the richness is iteratively determined \revise{by} varying the aperture
size.  

To measure the stacked cluster lensing profiles, we use
120 times larger number of random points (therefore one million points)
than the number of clusters in our samples, which was defined in
\citet{Rykoff:2016}.
These random points incorporate the survey geometry, depth variations
(hence the detection rate of a cluster), and distributions of clusters
in redshift and richness; accordingly each random point has the assigned richness and redshift 
from the clusters in the real catalog.
Furthermore, \citet{Rykoff:2016} used the random catalog to estimate a
detection efficiency 
for each cluster with $\lambda$ and $z_{\lambda}$:
\begin{equation}
 w_{\mathrm{rand}}(\lambda,z_{\lambda})=\frac{n_{\rm samp}(\lambda, z_{\lambda})}{n_{\rm keep}(\lambda, z_{\lambda})}.
\label{eq:w_r}
\end{equation}
Here $n_{\rm samp}(\lambda,z_{\lambda}) \sim 1000$ is the total number
of random points, 
which are injected into the survey footprint, for each of the clusters (here 
the cluster with $(\lambda,z_\lambda)$),
and $n_{\rm keep}(\lambda, z_{\lambda})$ is the number of the random
points that pass the mask and richness threshold cuts, $
f_{\rm mask}\le0.2$ and $\lambda/S\ge20$, where $f_{\rm mask}$ is the weighted area
fraction of masks
within the aperture around the chosen position. 
The detection efficiency is given by $1/w_{\rm rand}$ and
has a lower value
for lower richness clusters, because such clusters are more affected by
the survey boundary, masks and depth variations.
We can use this to estimate an effective area for clusters with $\lambda$ and $z_\lambda$:
\begin{equation}
\Omega_{\rm eff}(\lambda,z_\lambda)=\frac{\Omega_{\rm tot}}{w_{\rm
 rand}(\lambda,z_\lambda)},
 \label{eq:Omega_eff}
\end{equation}
where $\Omega_{\rm eff}(\lambda,z_\lambda)\le\Omega_{\rm tot}$.

\textit{}

\subsection{Cluster abundance}
\label{sec:abundance}

\begin{deluxetable*}{ccccccccccc}
\tablewidth{0pt}
\tablecaption{Binning scheme for the redMaPPer clusters and characteristics
in each bin
  \label{tab:binning}}
 \tablehead{\colhead{
bin index (abundance)
}
 & \colhead{
bin index (lensing)
}
 & \colhead{$\lambda_{\mathrm{min} }$}
 & \colhead{$ \lambda_{\mathrm{max} }$}
 & \colhead{$\left< \lambda \right>$}
 & \colhead{$z_{\mathrm{min} }$}
 & \colhead{$z_{ \mathrm{max} }$}
 & \colhead{$\left<z_{\lambda} \right>$}
 & \colhead{$\left<p_{ \mathrm{cen} } \right>$}
 & \colhead{$N^{\rm raw}_{\lambda_\alpha }$} 
 & \colhead{$N^{\rm corr}_{\lambda_\alpha}$}}
 \startdata
 1& 1 &   20.0 & 25.0 & 22.3 & 0.10 & 0.33 & 0.25 & 0.87 & 3133 & 3488.4 (11.3\%) \\ 
 2& 1 & 25.0 & 30.0 & 27.2 & 0.10 & 0.33 & 0.24 & 0.86 & 1762 & 1790.8 (1.6\%) \\ 
 3& 2 &  30.0 & 35.0 & 32.3 & 0.10 & 0.33 & 0.24 & 0.86 & 1146 & 1164.1 (1.6\%)\\ 
 4& 2 &  35.0 & 40.0 & 37.4 & 0.10 & 0.33 & 0.24 & 0.86 & 734 & 745.7 (1.6\%)\\     
 5& 3 &   40.0 & 47.5 & 43.5 & 0.10 & 0.33 & 0.24 & 0.87 & 596 & 605.2 (1.5\%)\\ 
 6& 3 &  47.5 & 55.0 & 51.0 & 0.10 & 0.33 & 0.24 & 0.88 & 381 & 386.8 (1.5\%)\\ 
 7& 4 &  55.0 & 77.5 & 63.6 & 0.10 & 0.33 & 0.24 & 0.87 & 434 & 440.4 (1.5\%)\\ 
 8& 4 &  77.5 & 100 & 85.8 & 0.10 & 0.33 & 0.23 & 0.89 & 126 & 127.8 (1.4\%) 
  \enddata  
\tablecomments{
In this paper we use the abundance and lensing
measurements in eight and four richness bins, respectively, 
where 
each richness bin is defined by $\lambda_{\rm min}$ and $\lambda_{\rm max}$ 
as denoted in the third and fourth columns. For the redshift range, 
we use the same cut, $0.10\le z_{\lambda}\le 0.33$ for all the bins.
The fifth and eighth columns give the mean richness 
and the mean redshift of the clusters in each richness bin. 
The ninth column
gives the mean of the centering probability 
of the chosen central galaxies in each sample. 
The
tenth column shows the number of clusters, 
and the eleventh shows the number after correcting for the effective survey area due to 
the effects of survey boundary 
and masks according to equation~\eqref{eq:N_alpha}.
The number in parenthesis in the eleventh column shows
the ratio of the abundances before and after the correction.
}
 \end{deluxetable*}
As the first cluster observable, we consider the abundance of 
redMaPPer clusters, the total number of clusters observed across the
entire SDSS footprint in a given richness bin. Throughout this paper, we
\revise{use} a single redshift bin, i.e. $0.10 \leq z_\lambda\leq 0.33$,
and do not consider \revise{smaller bins in redshift}. Instead we divide the clusters
into eight richness bins, as given in Table~\ref{tab:binning}. In addition, we use the point estimate of richness and redshift for the clusters to calculate the abundance in this section and lensing profile in Section~\ref{sec:lensing_measurement}.

We use the abundance of the clusters in each
richness bin correcting for the detection efficiency: 
\begin{equation}
 \widehat{N}_{\lambda_\alpha }= \sum_{l; {\lambda_l\in \lambda_{\alpha} }}\frac  {\Omega_{\rm tot}}{\Omega_{\rm
  eff}(z_l,\lambda_l)},
  \label{eq:N_alpha}
\end{equation}
where $\lambda_\alpha$ denotes the $\alpha$-th richness bin
in Table~\ref{tab:binning}, and the summation runs over all the clusters
residing in the richness bin. The factor $\Omega_{\rm tot}/\Omega_{\rm
eff}$ corrects for the detection efficiency we
discussed. Equation~\eqref{eq:N_alpha} gives an estimator of the abundance of
the clusters we could observe under the perfect conditions for the
survey area of $\Omega_{\rm tot}$, i.e. the abundance without masking
and depth-variation effects.  With this correction, we need not consider
the effects of survey boundary and masks on the model prediction.
The number of clusters in the $\alpha$-th
richness bin before and after the detection efficiency correction are given by
$N_{\lambda_\alpha}^{\rm raw}$ and $N_{\lambda_\alpha}^{\rm corr}$, respectively, in Table~\ref{tab:binning}.
The
correction is larger for a lower richness bin; the correction is about
11\% for the lowest richness bin ($20\le \lambda\le 25$).
In order to study the impact of the lowest richness bin on our
results, \revise{in what follows, we will consider} two samples defined by $20\le
\lambda\le 100$ and $30\le \lambda\le 100$, respectively, and compare the results.
\subsection{Stacked cluster lensing}
\label{sec:lensing_measurement}

The cross-correlation of positions of the redMaPPer clusters with shapes of
background galaxies 
allows us to measure
the average excess mass density profile around the clusters
-- hereafter the stacked cluster lensing profile. 
We use the catalog of source galaxy shapes in
\citet{Mandelbaum:2013} \citep[also see][]{Reyes:2011}, constructed after the
carefully-tuned selection based on the imaging quality, data reduction
quality, galactic extinction, apparent magnitude, photometric redshift and galaxy size.  We use
approximately the 39 million source galaxies. The area coverage is approximately $9,000~\mathrm{deg^2}$ and the footprint is overlapped with that of the
redMaPPer catalog on the sky as shown in Figure~\ref{fig:footprint}.

The galaxy shapes in this catalog are measured \revise{using the} re-Gaussianization
technique \citep{Hirata:2003}, and the systematic uncertainties on shape
measurements \revise{have} been investigated in detail \citep{Mandelbaum:2005}.
Following \cite{Mandelbaum:2013}, we use the photometric redshift
derived from a template-fitting method; the Zurich Extragalactic
Bayesian Redshift Analyzer, or $\mathtt{ZEBRA}$
\citep{Feldmann:2006}. For each galaxy, we use the best-fitting
photometric redshift estimate after marginalizing over the SED
templates. 
We correct for a
bias in the lensing signal due to the photometric redshift errors 
by using the method in 
\citet{Nakajima:2012}.
For the redshift binning ($0.10
\leq z_{\lambda} \leq 0.33$), \revise{the} debias factor is 
found to be about $1.08$ for all the richness bins, and we applied these
correction factors for \revise{all} the lensing signals. 

To estimate the excess surface mass density profile for a sample
of redMaPPer clusters in a given richness bin, we use the estimator defined in
\citet{Mandelbaum:2013}:
\begin{eqnarray}
	\widehat{ \Delta\Sigma_{\lambda_\beta}}(R)&=&
	\frac{1}{2 \mathcal{R}}\left[
	\frac{1}{N_{ls, \beta}(R)}\left.
	\sum_{l,s; \lambda_l\in \lambda_\beta}
    w_{ls} \Sigma_{\rm cr}(z_l,z_s)\epsilon_{+}({\btheta}_s)\right|_{R=\chi_l|\btheta_l-\btheta_s|}\right.
	\nonumber\\
 	&&\left.
    -\frac{1}{N_{rs, \beta}(R)} \left.
	\sum_{r,s;\lambda_r\in \lambda_\beta}
	w_{rs} \Sigma_{\rm cr}(z_r,z_s)
   	\epsilon_{+}({\btheta}_s)\right|_{R=\chi_r|\btheta_r-\btheta_s|}
   	\right],\nonumber\\
	\label{eq:DSigma_est}
\end{eqnarray}
where the index $\beta$ denotes the $\beta$-th richness bin as given in
Table~\ref{tab:binning}, the subscripts $s, l$ or $r$ stand for
\textit{source}, \textit{lens} (cluster) or \textit{random}, and $\epsilon_{+}$ is the
tangential component of source galaxy ellipticity 
with respect to the cluster center or the random
point. 
The shear {\it responsivity}
\citep{Bernstein:2002} is given as $\mathcal{R}\simeq 0.87$
\citep{Mandelbaum:2013}, which represents the \revise{statistically-averaged} response
of galaxy ellipticities to the lensing shear in the weak lensing regime.
The summation runs over all pairs of source galaxies and clusters 
(or randoms) 
in a given projected separation $R=\chi_l|\btheta_l-\btheta_s|$ (or
$R=\chi_r|\btheta_r-\btheta_s|$) to within the bin width, where
$\chi_l\equiv \chi(z_l)$ (or $\chi_r\equiv \chi(z_r)$) is the comoving
angular diameter distance to each cluster (or random).  As given in
Table~\ref{tab:binning}, we divide the redMaPPer clusters into four
richness bins, where we adopt \revise{a binning scheme similar} to that in
\citet{Simet:2016} \revise{to enable comparison to their work}.
\revise{We use} 19 radial bins that are
equally spaced logarithmically from $0.2\
h^{-1}\mathrm{Mpc}$ to $50\ h^{-1}\mathrm{Mpc}$. The lensing
efficiency function is defined for a system of lens and source at $z_l$
and $z_s$ 
for a flat universe as
\begin{equation}
 \Sigma_{\rm cr}^{\revise{-1}}(z_l,z_s)=4\pi G (1+z_l)\chi_l \left( 1-\frac{\chi_l }{\chi_s}\right). 
\label{eq:sigma_cr}
\end{equation}
The second term in equation~\eqref{eq:DSigma_est} denotes the
lensing signals around the random points. As stressed in
\citet{Sheldonetal:04} and \citet{Mandelbaum:2005} \citep[also see][for
a recent detailed study]{Singhetal:16}, the subtraction of the
{\it random} signal from the signals around the clusters (the first term)
allows us to measure the {\it excess} mass density profile around the
clusters with respect to the background density. The random
subtraction can also correct for an additive bias in the estimated shear 
e.g. due to PSF ellipticity errors \citep{Mandelbaum:2005}.
For a given subsample of the clusters, we use \revise{those} random points, which
\revise{fall} in the same richness and redshift bins, to estimate the random
signal.
The lens-source pair weight $w_{ls}$ is given as
\begin{equation}
 w_{ls} = \frac{\Omega_{\rm tot}}{\Omega_{\rm  eff}(z_l,\lambda_l)}
  \Sigma^{\revise{-2}}_{\mathrm{cr}}(z_l, z_s) w_s,
 \label{eq:lsweight}
\end{equation}
where $w_s$ is the source weight as
\begin{equation}
 w_s=\frac{1}{\sigma_e^2+\sigma_{\rm SN}^2},
  \label{eq:shape_weight}
\end{equation}
where $\sigma_e$ is the 
measurement error of galaxy
ellipticity and $\sigma_{\rm SN}$ is the intrinsic ellipticity amplitude, 
for
which we used \revise{a} fixed value, $\sigma_{\rm SN}=0.365$ following
\citet{Reyes:2011}. The factor of $\Omega_{\rm
eff}(\lambda_l,z_l)/\Omega_{\rm tot}$ in equation~\eqref{eq:lsweight} corrects
for the effective area as we did for the abundance estimator in
equation~\eqref{eq:N_alpha}.
Similarly, the weight $w_{rs}$ for the random
points is given as
\begin{equation}
 w_{rs} = \left(\frac{\Omega_{\rm tot}}{\Omega_{\rm  eff}(z_r,\lambda_r)}\right)^2
  \Sigma^{\revise{-2}}_{\mathrm{cr}}(z_r, z_s) w_s.
 \label{eq:rsweight}
\end{equation}
\revise{The extra factor} $\Omega_{\rm tot}/\Omega_{\rm eff}$ in this weight is from the fact that the number of
random points in each richness bin accounts for the detection efficiency
in the construction as described around equation~\eqref{eq:w_r}.
\revise{Note that, although the richness and redshift distributions of the {\it original} random catalog is
different from that of actual clusters, the distributions of the random catalog after
accounting for the correction factor $\Omega_{\rm tot}/\Omega_{\rm eff}$ become the same as that of actual clusters.}
The denominator of the first term on the right hand side of
equation~\eqref{eq:DSigma_est} is the weighted number of cluster-source pairs in each separation bin,
defined as
\begin{equation}
 N_{ls, \beta}(R)\equiv \left.\sum_{l,s;
 \lambda_l \in \lambda_\beta}w_{ls}\right|_{R=\chi_l|\btheta_l-\btheta_s|},
\end{equation}
and $N_{rs, \beta}(R)$ in the second term is the weighted number of random-source pairs, similarly
to the above equation.

Following \citet{Mandelbaum:2005} \citep[also
see][]{Mandelbaum:2013, Miyatake:2015},
we use the redMaPPer random catalog to correct for a possible dilution
effect, which might arise from a contamination of unlensed, member galaxies
\revise{in} the source galaxy catalog
due to imperfect photometric redshifts.
To correct for this, we multiply the measured lensing signals 
by the \textit{boost factor}, estimated \revise{as}
\begin{equation}
	C_{\beta}(R) = \frac{N_{r, \beta}}{N_{l, \beta}} \frac{ N_{ls, \beta}(R)    }{ N_{rs, \beta}(R) },
\label{eq:boost}
\end{equation}
where $N_{r, \beta}$ and $N_{l, \beta}$ are the number of random points and clusters after
correcting for an effective area: $N_{r, \beta}\equiv\sum_{r; \lambda_r\in
\lambda_\beta}(\Omega_{\rm tot}/\Omega_{\rm eff})^2$ and $N_{l, \beta}\equiv
\sum_{l;\lambda_l\in \lambda_\beta}(\Omega_{\rm tot}/\Omega_{\rm eff})$,
respectively. 
We show the boost factor for the clusters in each richness bin
in Appendix~\ref{app:boost}.

As described above,
we include the $\Omega_{\rm eff}(\lambda_l, z_l)$ correction for the lensing weight 
in equations~\eqref{eq:lsweight} and \eqref{eq:rsweight} to properly account for the detection efficiency.
This correction is however not important for the lensing measurement 
since the stacked lensing is the average of lensing profile from the clusters 
(the correction factor 
cancels out to some extent in the numerator and denominator in each term of the estimator, equation~\ref{eq:DSigma_est}).
The correction changes the lensing profile measured for the lowest richness bin ($20 \le \lambda \le 30$),   
only by a level of 3\% at most in the amplitude from that without the weight correction. 
In addition, the correction directions are in both positive and negative sides 
at different radii. 
This effect
is negligible for the higher richness bins ($30 \le \lambda \le 100$). 
We thus confirmed
that the $\Omega_{\rm eff}$ correction is important only for 
the abundance measurement as shown in Table~\ref{tab:binning},
given the error bars we will show below.

\section{Forward Modeling of Cluster Observables}
\label{sec:m_modeling}

In this paper, we \revise{adopt} a \textit{forward} modeling approach to model cluster
observables for the fiducial \textit{Planck} cosmology \citep{Planck:2015}. 
In this method we model
the probability distribution function of optical richness for halos with
a given mass, $P(\ln \lambda|M)$.
An alternative approach is a backward approach, where $P(\ln M|\lambda)$, the probability
distribution of halo mass for a given richness, is considered as
studied in the previous works
\citep[e.g.,][]{Baxter:2016,Simet:2016, Melchior:2016}.

\subsection{Mass-richness relation}
\label{sec:scaling}
Following \citet{Lima:2005} \citep[also see][]{OguriTakada:2011},
we assume that the probability distribution of the {\it observed} richness
$\lambda$ for halos with a fixed mass
$M$ is given by a log-normal distribution:
\begin{equation}
 P(\ln \lambda|M)\mathrm{d}\ln \lambda \equiv \frac{1}{\sqrt{2\pi}\sigma_{\ln \lambda|M}}
  \exp\left( -\frac{x^\revise{2}(\lambda,M)}{2\sigma_{\ln\lambda|M}^2}\right)
  \mathrm{d}\ln \lambda,
\label{eq:p_lambda}
\end{equation}
where  $x(\lambda,M)$ models the mean relation 
with two model parameters $A$ and $B$ as
\begin{equation}
 x(\lambda,M) \equiv \ln\lambda-\left[
   A + B \ln\left(\frac{M}{M_{\mathrm{pivot} } } \right)\right].
  \label{eq:lambda_M}
\end{equation}
Hence $x(\lambda,M)=0$, where the probability peaks at a fixed halo
mass, gives the mean relation. Or equivalently the mean relation is
defined by the following average:
\begin{eqnarray}
\langle \ln \lambda\rangle (M) &\equiv & \int_{-\infty}^{+\infty} {\rm d}\ln \lambda~ P(\ln \lambda|M)\ln \lambda \nonumber\\
&=&  A +  B\ln
\left(\frac{M}{M_{\mathrm{pivot}}}\right).
\label{eq:mean_relation}
\end{eqnarray}
Throughout this paper we \revise{adopt}
$M_{\mathrm{pivot}}=3\times10^{14}\ h^{-1}M_{\odot}$ for the pivot mass scale,
roughly corresponding to \revise{the} typical mass scale of the redMaPPer
clusters.  
We assumed that the mean relation simply follows a power-law
relation between $\lambda$ and $M$ (a linear relation between $\ln \lambda$ and $\ln M$): $A$
determines the normalization and $B$ specifies a power-law index for
the halo mass dependence.
In addition, we assume that the scatter in the richness around
the mean relation at a fixed halo mass is 
modeled by two parameters, the normalization
$\sigma_{0}$ and the mass dependence $q$, as
\begin{equation}
\sigma_{\ln\lambda|M} = \sigma_0 + q \ln\left(\frac{M}{M_{\mathrm{pivot} } }\right),
    \label{eq:scatter_M}
\end{equation}
 where $q=0$ means that the scatter is independent of halo mass.  In our
treatment, $\sigma_{\ln \lambda|M}$ should be considered as a {\it total} scatter, including contributions of the richness measurement errors and the intrinsic scatters, as we will discuss later.

Hence we model the mass-richness relation in
equation~\eqref{eq:p_lambda} by four model parameters: $\{A,B,\sigma_0,
q\}$. We will explore the best-fitting model parameters that can
reproduce both the abundance and lensing profile measurements
simultaneously. In this fiducial model, we ignore possible redshift
dependence of the mass-richness relation for simplicity.

\subsection{Abundance in richness bin}
\label{sec:model_abundance}
Once the mass-richness relation $P(\ln \lambda|M)$ is given, we can
compute a model prediction for the abundance of the redMaPPer clusters
for the \textit{Planck} cosmology.  For the $\alpha$-th richness bin
$(\lambda_{\alpha, {\rm min}}\le \lambda \le \lambda_{\alpha,{\rm max} })$ and a redshift range $(z_{\rm min} \le z \le z_{\rm max})$, 
the abundance of the clusters for the total survey area is given as
\begin{eqnarray}
&N_{\lambda_\alpha}&
    \equiv  \Omega_{\rm tot}
        \int_{z_{\rm min}}^{z_{\rm max}}\!\mathrm{d}z~ \frac{ \mathrm{d^2} V}{ \mathrm{d}z \mathrm{d}\Omega}
    \int_{M_{\rm min}}^{M_{\rm max}}\!\!\mathrm{d}M~
    \frac{\mathrm{d}n}{\mathrm{d} M}
\int_{ \revise{\lambda_{\alpha,{\rm min}}} }^{ \revise{\lambda_{\alpha,{\rm max}}}}\!\!\ \revise{\frac{ {\rm d}\lambda }{\lambda}}~
        P(\ln \lambda|M)\nonumber\\
    &&\hspace{-1.5em}\revise{=}\Omega_{\rm tot}  \int_{z_{\rm min}}^{z_{\rm max}}\!\!\
    \mathrm{d}z~ \frac{\mathrm{\chi}^{\revise{2}}(z)}{H(z)}
    \int_{M_{\rm min }}^{M_{\rm max }}\!\!\mathrm{d}M~
    \frac{\mathrm{d}n}{\mathrm{d} M}\,
    S(M|\lambda_{\alpha,{\rm min}},\lambda_{\alpha,{\rm max}}),
    \label{eq:richnessfunc_model}
\end{eqnarray}
where $\chi^{\revise{2}}(z)/H(z)$ is the comoving volume per unit redshift interval
and per unit steradian, $\mathrm{d}n/\mathrm{d}M$ is the halo mass
function in the mass range $[M, M+\mathrm{d}M]$ at redshift $z$.
The selection function of halo mass in the richness bin, $S(M|\lambda_{\alpha, {\rm min} },\lambda_{\alpha,{\rm max}})$,  
is obtained by 
integrating the log-normal distribution 
$P(\ln \lambda | M)$ over the richness range as
\begin{eqnarray}
 S(M|\lambda_{\alpha,{\rm min}},\lambda_{\alpha,{\rm max}})
  &\equiv &\int_{\ln \lambda_{\alpha,{\rm min}}}^{\ln \lambda_{\alpha,\rm{max}}}
  \!\! \mathrm{d}\ln \lambda~
  P(\ln \lambda|M)
 \nonumber \\
  &&\hspace{-6em}=
 \frac{1}{2}  \left[
	   {\rm erf}\left(\frac{x(\lambda_{\alpha,{\rm
	max}}, M)}{\sqrt{2}\sigma_{\ln
	\lambda|M}}\right)
    -{\rm erf}\left(\frac{x(\lambda_{\alpha,{\rm
	       min}}, M)}{\sqrt{2}\sigma_{\ln\lambda|M}}\right)\right],
 \label{eq:selection_func}
\end{eqnarray}
where ${\rm erf}(x)$ is the error function.  

\subsection{Stacked cluster lensing in richness bin}
\label{sec:model_lensing}

In this subsection we describe details of how we model the stacked lensing
profile for the redMaPPer clusters based on the forward modeling approach.
The stacked lensing profile for halos with mass $M$ and at redshift
$z_l$ probes the {\it average} radial profile of matter distribution
around the halos, $\rho_{\rm hm}(r; M, z_l)$. Due to statistical
isotropy, the average distribution is one-dimensional, given as a
function of separation from the halo center.
\revise{Here $r$ is in the comoving coordinates.}
For convenience of the
following discussion we express the average mass density profile in
terms of the cross-correlation function between the halo distribution and the matter
density fluctuation field, $\xi_{\rm hm}(r;M,z_l)$:
\begin{equation}
 \rho_{\rm hm}(r;M,z_l)=\bar{\rho}_{\rm m0}
  \left[1+\xi_{\rm hm}(r;M,z_l)\right].
  \label{eq:rho_hm}
\end{equation}
Note that we used the present-day mean mass density, $\bar{\rho}_{\rm
m0}$, in the above equation \revise{since 
we use} the comoving
coordinates rather than the physical coordinates. 
The cross-correlation is related to the cross-power spectrum $P_{\rm
hm}(k; M, z_l) $ via the Fourier transform as
\begin{equation}
\xi_{\rm hm}(r;M,z_l)=\int_{0}^{\infty}\!\!\frac{k^2\mathrm{d}k}{2\pi^2}~ P_{\rm
 hm}(k;M,z_l)j_0(kr), 
\end{equation}
where 
$j_0(x)$ is the zeroth-order spherical Bessel function.
The lensing fields are obtained from a  projection of the
three-dimensional profile along the line-of-sight direction. Once $\xi_{\rm
hm}(r)$ or $P_{\rm hm}(k)$ is given, the {\it average} surface mass
density profile is given as
\begin{eqnarray}
 \Sigma(R;M,z_l)&=&\bar{\rho}_{\rm
  m0}\int_{-\infty}^{\infty}\!\mathrm{d}
  \chi~\xi_{\rm hm}\left(r=\sqrt{R^2+\chi^2}; M, z_l\right)\nonumber\\
 &=&\bar{\rho}_{\rm m0}\int_{0}^{\infty}\!\frac{k\mathrm{d}k}{2\pi}~P_{\rm
  hm}(k;M,z_l)J_0(kR), 
  \label{eq:Sigma_M}
\end{eqnarray}
where $J_0(x)$ is the zeroth-order Bessel function
\citep[e.g.][]{OguriTakada:2011, Hikage:2012, Hikage:2013}, and
$R$ is the projected separation from the halo center \revise{in the comoving coordinates}.

Similarly, the excess surface mass density profile around halos, which
is a direct observable from the weak lensing measurement, is given as
\begin{eqnarray}
 \Delta\Sigma(R;M,z_l)&=&
\avrg{\Sigma(R;M,z_l)}_{<R}-\Sigma(R;M,z_l)  \nonumber\\
  &=&
  \bar{\rho}_{\rm m0} \int_{0}^{\infty}\!\frac{k\mathrm{d}k}{2\pi}~
  P_{\rm hm}(k;M,z_l)J_2(kR),
  \label{eq:dSigma_M}
\end{eqnarray}
where $\avrg{\Sigma(R; M, z_l)}_{<R}$ denotes the average of $\Sigma(R; M, z_l)$ within a
circular aperture of radius $R$, and $J_2(x)$ is the second-order Bessel function. 

Taking into account the distribution of halo masses and redshifts for
the clusters in the $\beta$-th richness bin $(\lambda_{\beta,{\rm
min}} \le \lambda \le {\lambda_{\beta,{\rm max}}})$, 
similarly to equation~\eqref{eq:richnessfunc_model}, 
we can compute a model
prediction for the stacked lensing profile as
\revise{
\begin{eqnarray}
\Delta\Sigma_{\lambda_{\beta}}(R) 
 &\equiv &
\frac{1}{N_{\Delta\Sigma}(R; \lambda_{\beta,{\rm min}},\lambda_{\beta,{\rm
max}})}
	\int_{z_{\rm min}}^{z_{\rm max}}\!\!\mathrm{d}z
\!\!	\int_{M_{\rm min}}^{M_{\rm max} }
    \!\!\mathrm{d}M~  
	\frac{\mathrm{\chi}^{\revise{2}}(z)}{H(z)}
\nonumber \\ 
&& \times
w_{l}(z; R) 
\frac{\mathrm{d}n}{\mathrm{d}M}S(M|\lambda_{\beta,{\rm min}},\lambda_{\beta,{\rm max}}) 
   \Delta\Sigma(R; M, z)
\nonumber \\
&& \times \left[1 +
	 \left\langle\frac{1}{\Sigma_{\mathrm{cr}}}\right\rangle_{\beta}\hspace{-0.5em}(R)~
	 \Sigma(R; M, z) \right]. 
	 \label{eq:lensing_model}
\end{eqnarray}
}
The term in the square bracket on the r.h.s. accounts for nonlinear
contribution of the reduced shear that might not be negligible at very
small radii \citep[e.g.][]{Johnston:2007}, where $\left<
1/\Sigma_{\rm cr}\right>_{\beta}(R)$ is measured from the pairs of redMaPPer
clusters and source galaxies in each radial bin for the $\beta$-th richness bin.
The weight \revise{$w_l(z_l;R)$} is
introduced to take into account the dependence of lens redshift \revise{at}
each radial bin \revise{on} the lensing profile measurement in 
equation~\eqref{eq:DSigma_est}, and is defined as
\begin{equation}
 w_{l}(z_{l}; R) =
  {\avrg{ w_s \Sigma_{\mathrm{cr}}^{-2}(z_l, z_s)\Bigr|_{R=\chi_l|\btheta_l-\btheta_s|}
}_{z_s} },
\end{equation}
with equations~\eqref{eq:sigma_cr} and \eqref{eq:shape_weight}.  We
compute $w_l$ as follows. First we divide the lens-source pairs into
nine lens redshift bins, which \revise{are} linearly spaced in $z_{l} \in [0.10,
0.33]$. Secondly we estimate the weight $w_l$ from the average over all
the sources in each redshift and radial bin. Then we interpolate them
linearly as a function of $z_{{l}}$ for each radial
bin. Even if we ignore the weight in equation~\eqref{eq:lensing_model},
it does not largely change the model prediction (only by about $1\%$ in the lensing profile amplitude).
The normalization
factor in the denominator of equation~\eqref{eq:lensing_model} is
similar to the abundance prediction in
equation~\eqref{eq:richnessfunc_model}, but is defined by taking into
account the weight $w_l$ as
\begin{eqnarray}
N_{\Delta\Sigma}(R; \lambda_{\beta,{\rm min}},\lambda_{\beta,{\rm max}}) 
  &\equiv &\int_{z_{\rm min}}^{z_{\rm max}}\!\!\mathrm{d}z
 \int_{M_{\rm min}}^{M_{\rm max}}\!\!\mathrm{d}M~
 \frac{\mathrm{\chi}^{\revise{2}}(z)}{H(z)}w_{l}(z; R)
\nonumber \\ 
 &&\times
\frac{\mathrm{d}n}{\mathrm{d}M}
S(M|\lambda_{\beta,{\rm min}},\lambda_{\beta,{\rm max}}).
\label{eq:N_dsigma}
\end{eqnarray}
Furthermore, although we employ the most probable central galaxy in each
cluster region as a proxy of the cluster center, the central galaxy
might be off-centered from the true center. Following \citet{OguriTakada:2011} and
\citet{Hikage:2012,Hikage:2013} \citep[also see][]{Moreetal:2015}, we \revise{will marginalize over} the effect
of off-centered clusters on the lensing profiles $\Sigma(R; M, z_l)$ and $\Delta
\Sigma(R; M, z_l)$, by modifying the halo-matter cross-power spectrum as
\begin{equation}
 P_{\rm hm}(k; M, z_l)\rightarrow\left[
f_{{\rm cen}, \beta}+(1-f_{{\rm cen}, \beta})\tilde{p}_{\rm off}(k;R_{\rm off, \beta})
\right]P_{\rm hm}(k; M, z_l)
\label{eq:pk_off}
\end{equation}
when computing equations~\eqref{eq:Sigma_M} and \eqref{eq:dSigma_M} for
the clusters in the $\beta$-th richness bin.
Here $f_{{\rm cen}, \beta}$ is a parameter to model a fraction of the
centered clusters in the $\beta$-th richness bin, while $(1-f_{{\rm cen}, \beta})$
is a fraction of the off-centered clusters. The function of ${p}_{\rm
off}(r; R_{\rm off, \beta})$ is the normalized radial profile of
off-centered ``central'' galaxies with respect to the true center for
which we assume a Gaussian distribution given as $p_{\rm off}(r;
R_{\rm off, \beta}) =\exp(-r^2/2R_{\rm off, \beta}^2)/[(2\pi)^{3/2}R_{\rm off, \beta}^3]$, where $R_{\rm off, \beta}$ is a parameter to model the typical off-centering radius.
The Fourier transform of $p(r; R_{\rm off, \beta})$, $\tilde{p}_{\rm off}(k ;
R_{\rm off, \beta})$, is given as 
$\tilde{p}_{\rm off}(k; R_{\rm off, \beta})=\exp(-k^2R_{\rm off, \beta}^2/2)$.
We parametrize the off-centering radius relative to the
richness-dependent aperture radius 
in equation~\eqref{eq:aperture}
by a dimension-less parameter $\alpha_{\rm off}$ as
$R_{\rm off, \beta}=\alpha_{\rm off} R_{\lambda_\beta}$, where $R_{\lambda_\beta}$
is the weighted average of the richness-aperture radii, estimated as
\revise{
\begin{equation}
\widehat{R}_{\lambda_\beta} = \frac{\sum_{l; \lambda_l\in \lambda_\beta} 
[\Omega_{\rm tot}  / \Omega_{\rm eff}(\lambda_l, z_l)  ]  \times R_{\rm aperture}(\lambda_l, z_l) }{\sum_{l; \lambda_l\in \lambda_\beta} [\Omega_{\rm tot} / 
\Omega_{\rm eff}(\lambda_l, z_l) ]}~ [h^{-1} \mathrm{Mpc}]\revise{.}
\label{eq:R_lambda_beta}
\end{equation}
}
As \revise{is evident} from equations~\eqref{eq:Sigma_M}, \eqref{eq:dSigma_M}, 
\eqref{eq:lensing_model} and \eqref{eq:pk_off}, once the halo mass
function $\mathrm{d}n(M, z)/\mathrm{d}M$ and the three-dimensional
halo-matter cross-correlation $\xi_{\rm hm}(r;M,z)$ or $P_{\rm
hm}(k;M,z)$ are provided, we can compute the model prediction for the
stacked cluster lensing profile in each richness bin. The model is
specified by nine parameters in total for the cluster sample
of $20\le \lambda\le 100$: the four parameters $\{A, B, \sigma_0, q\}$
for the mass-richness relation $P(\ln \lambda|M)$ and the five 
parameters $\{f_{{\rm cen}, \beta}, \alpha_{\rm off}\}$ for the
off-centering effect for four richness bins ($\beta=1, 2, 3 $ or $4$ as
given in Table~\ref{tab:binning}). For the sample of $30\le \lambda\le
100$ based on the same binning scheme, we will use eight parameters (three $f_{{\rm cen}, \beta}$-parameters instead of four).
For the Fourier transform
we use the FFTLog algorithm \citep{Hamiltonetal:00}, which allows a quick, but sufficiently accurate computation of the relevant quantities.

\section{$N$-body Simulation Based Emulator and Covariance}
\label{sec:emurator}
To have accurate model predictions for the redMaPPer cluster
observables, we use an {\it emulator} for the halo mass function and the
halo-matter cross-correlation function for the \textit{Planck} cosmology, 
which is built based on a set of high-resolution, 
cosmological $N$-body simulations (Nishimichi et~al. in prep.).
For the error covariance matrix that models statistical uncertainties in
the cluster observables, we use mock catalogs of the SDSS data\revise{, which we describe in detail in this section}.

\subsection{Halo emulator}
\label{sec:m_halo_emulator}
In order to estimate model parameters in an unbiased way from the
measurements of cluster observables (abundances and lensing profiles in this paper),
the model predictions have to be as accurate as \revise{the} precision of the
measurements. 
Nishimichi et~al. (in prep.) developed a scheme for
predicting statistical quantities of halos, which include the mass
function, the halo-matter cross power spectrum, and the halo
auto-spectrum, as a function of halo mass, redshift, wavenumber (or separation length), 
and cosmological
models, based on a large set of high-resolution, cosmological $N$-body
simulations. The emulator is named \texttt{Dark Emulator}. We briefly
summarize details of the emulator below.

\revise{Each of the $N$-body simulations used in to construct the emulator were run with}
the parallel Tree-Particle Mesh code $\mathtt{Gadget2}$
\citep{Springel:2005}. In this paper, we use an emulator constructed
based on 24 realizations of the simulations for one particular
cosmology, the \textit{Planck} cosmology \citep{Planck:2015}: \revise{$\Omega_{\rm b0}h^2=0.02225$} and $\Omega_{\rm c0}h^2=0.1198$
for the density parameters of baryon and 
CDM, respectively, 
$\Omega_{\Lambda}=0.6844$ for the cosmological constant, 
and $\ln (10^{10}A_s)=3.094$ and
$n_s=0.9645$ for the primordial power spectrum. 
We assume a flat
geometry, and include $\Omega_{\nu0}h^2=0.00064$ corresponding to
$m_{\nu,{\rm tot}}=0.06~${\rm eV} for the sum of three-flavor neutrino
masses.
\revise{The baseline model of \cite{Planck:2015} 
employed
this value 
from the lower bound for the normal mass hierarchy inferred in 
the neutrino oscillation experiments \citep[e.g., see][for review]{PDG2014}.}
For this model, $\Omega_{\rm m0}=0.3156$ 
for the present-day non-relativistic matter
density, and $\sigma_8=0.831$ for the present-day rms mass density
fluctuations within a top-hat sphere of $8~h^{-1}$Mpc radius. Note that
our simulations did not include the effect of massive neutrinos in $N$-body simulations, 
and we include $\Omega_{\nu 0}$ to compute the linear matter power spectrum to
set the initial conditions for $N$-body simulations. \revise{All the matter component
corresponding to a density of ${\Omega_{\rm m0}}$ was assumed to be
collisionless and modelled as a single species.}

\revise{The simulations track the evolution of} $2048^3$ $N$-body particles
\revise{in} a box size of $1~h^{-1}\mathrm{Gpc}$ on a side, where the mass
resolution is about $1\times10^{10}~ h^{-1}M_{\odot}$.  \revise{The initial
conditions were generated using} second-order Lagrangian perturbation
theory to compute the initial displacements of $N$-body particles at the
initial redshift $z_i=59$ \citep{Nishimichi:2009}, using the linear
matter power spectrum computed with $\mathtt{CAMB}$ \citep{Lewis:2000}
for the 
above fiducial $\Lambda$CDM 
cosmology. 
We stored the outputs of each
$N$-body realizations in 21 redshift bins in the range of 
$0\leq z\le 1.47$,
equally stepped by the linear growth rate for the fiducial model.
In this paper, we use 24 realizations of the $N$-body simulations with
the fiducial cosmology, effectively corresponding to the simulation
volume of $24~(h^{-1}\mathrm{ Gpc})^3$. This is sufficiently large
compared to the SDSS volume, which is $\sim 2~(h^{-1}{\rm Gpc})^3$.  Our
simulations are sufficient to accurately estimate the halo mass function
and the halo-matter cross-correlation function in each halo mass bin.

To identify dark matter halos in each simulation output, we used
$\mathtt{Rockstar}$ \citep{Behroozi:2013} that identifies dark matter
halos and subhalos based on a clustering of $N$-body particles in phase
space.  
Throughout this paper  we adopt $M\equiv M_{\rm 200m}= 4\pi (R_{200{\rm m}})^3\bar{\rho}_{\rm
m0}\times 200/3$ for the halo mass definition, where $R_{\rm 200m}$ is
the spherical halo boundary radius \revise{in comoving units} within which the 
mean mass density is
$200$ times $\bar{\rho}_{\rm m0}$. 
We use the potential minimum as the halo center proxy for each halo.
Our definition of halo mass includes all the $N$-body particles within the
boundary $R_{\rm 200m}$ around the halo center (i.e. including
particles even if those are not gravitationally bound by the halo).
Every member particle in a halo is counted once; if a separation between
different halos (their centers) is smaller than the sum of their $R_{\rm
200m}$ radii, we assign member particles in the overlapping spherical
regions to the halo of larger mass. That is, once dark matter particle
is counted as a member particle for one halo, we exclude the particle
from the list, and then define other halos from the remaining particles.
The minimum halo mass in the halo catalog is about $10^{12}h^{-1}M_\odot$,
while
the maximum halo mass with enough statistics is around
$2 \times 10^{15} h^{-1}M_{\odot}$.
Throughout this paper we set $M_{\rm min}=10^{12}~h^{-1}M_\odot$ and 
$M_{\rm max}=2\times 10^{15}~h^{-1}M_\odot$ for the minimum and maximum halo masses to evaluate the halo mass integration 
in the model predictions for the cluster observables (e.g., equation~\ref{eq:richnessfunc_model}).

Using the catalogs of halos and $N$-body particles in each simulation
realization, we compute the halo mass function
($\mathrm{d}n/\mathrm{d}M$) and the halo-matter cross-correlation
($\xi_{\rm hm}$) in different bins of halo mass, redshift and separation length. 
For the
halo mass binning, we employ 20 logarithmically-spacing bins in one
decade of halo mass, i.e.  $\Delta\log_{10}M=0.05$, for each
redshift output.  We computed $\xi_{\rm hm}(r; M, z)$ using a direct
summation method at small separation, $r<5~h^{-1}{\rm Mpc}$ and using
the FFT method at larger separations (more exactly, we estimated
$P_{\rm hm}(k; M, z)$ for the latter method, and then Fourier-transformed back
it to $\xi_{\rm hm}(r; M, z)$). 
We use $\xi_{\rm hm}(r; M, z)$ to compute the
projected mass density profiles, $\Sigma(R; M, z)$ and $\Delta\Sigma(R;
M, z)$ based on equations~\eqref{eq:Sigma_M} and \eqref{eq:dSigma_M},
respectively.  We tabulated all the measurements of
$\mathrm{d}n(M,z)/\mathrm{d}M$ and $\xi_{\rm hm}(R; M, z)$ from the 24 realizations, and built the emulator
that outputs those quantities for an arbitrary input value of halo mass
and redshift, using the cubic spline interpolation for $M$ and the
linear interpolation for $z$, respectively. 
\revise{We checked that, by employing the best-fit  mass-richness relation parameters in Table~\ref{tab:params_prior}
and shifting the halo mass function and the halo-matter cross-correlation in each bin of halo mass,
redshift and separation length by one standard deviation uncertainty
in positive sides before the interpolation above,  
the emulator predicts the lensing profiles in all the richness
and radial bins at the precision better than 2\% in the amplitude (the largest 
uncertainty is from the largest radial bin and better than 0.5-1.0\% for $R<10\ h^{-1}\rm{Mpc}$). We will introduce a nuisance parameter, with 
$\pm 5\%$ prior in the fractional amplitude of the lensing profile, to study how 
the uncertainty in the lensing amplitude affects the results for the mass-richness relation
constraints in Section~\ref{sec:possiblesyserr}.  
Similarly the emulator predicts 
the abundance at the precision better than 2\% (the largest uncertainty is from 
the largest richness bin). This uncertainty is smaller than the covariance amplitude (${\sim}9\%$ in the standard deviation). 
Hence we conclude that the uncertainty in the emulator prediction does not largely affect 
the results we will show below.}

We use this emulator to
compute the model predictions for the cluster abundance and the cluster
lensing profiles, based on equations~\eqref{eq:richnessfunc_model} and
\eqref{eq:lensing_model}. 
Our model of the lensing profiles properly \revise{includes} all the relevant
contributions: the 1-halo term arising from matter 
inside halo boundary, the 2-halo term arising from matter in the
surrounding large-scale structure, and the transition regime between the
1- and 2-halo terms. The 1-halo term, i.e. the average mass
profile, includes the halo mass dependence of halo concentration as well
as scatters in the halo concentrations for different halos of a given mass.

This emulator allows a quick computation of the cluster observables,
abundance and lensing profiles in each richness bin
for a given model of the mass-richness relation in equations~\eqref{eq:p_lambda},
\eqref{eq:mean_relation}, and \eqref{eq:scatter_M}. In turn the emulator
enables us to perform an inference of model parameters given the
measurements using a Markov chain Monte Carlo (MCMC) analysis method, as
we will show below. 

\subsection{Covariance estimation based on the SDSS mock 
catalogs}
\label{sec:m_mock_covariance}

\begin{figure*}
	\centering
 \includegraphics[width=0.995 \textwidth ]{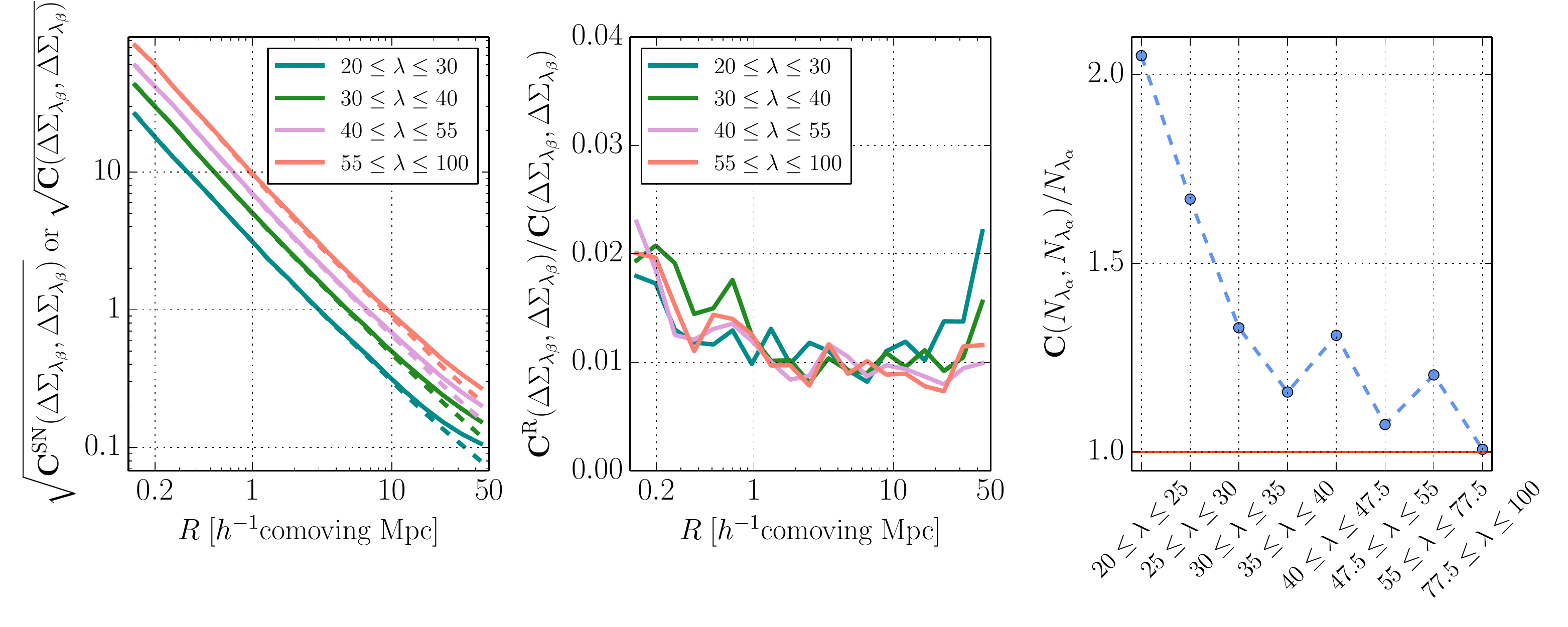}
\caption{The diagonal components of the covariance matrix for
measurements of the stacked cluster lensing profiles and the abundance for the
redMaPPer clusters. 
We used 108 mock catalogs of the redMaPPer clusters and source
galaxies to estimate the sample variance (see text for details). 
{\it Left}: The 
solid curves show the full covariance for the lensing
profile in each richness bin, while the dashed curves denote the
 shape noise contribution alone, ${\bf C}^{\rm SN}$, for comparison.
{\it Middle}: The ratio of the random correction contribution used
in the stacked lensing measurements, ${\bf C}^{\rm R}$, to the full covariance
matrix ${\bf C}$.
{\it Right}: The ratio of the full covariance for the number
counts of clusters (abundance) in each richness bin, relative to the 
Poisson contribution. 
If the ratio is greater than unity, the sample variance gives a dominant contribution to the total power. 
 The curve appears to have an
up-and-down feature, but this is due to our richness binning scheme in Table~\ref{tab:binning} (the
number counts has the similar feature).
}
\label{fig:mock_sn_rand_ratio}
\end{figure*}
\begin{figure}[h]
	\centering
 \includegraphics[width=0.495 \textwidth ]{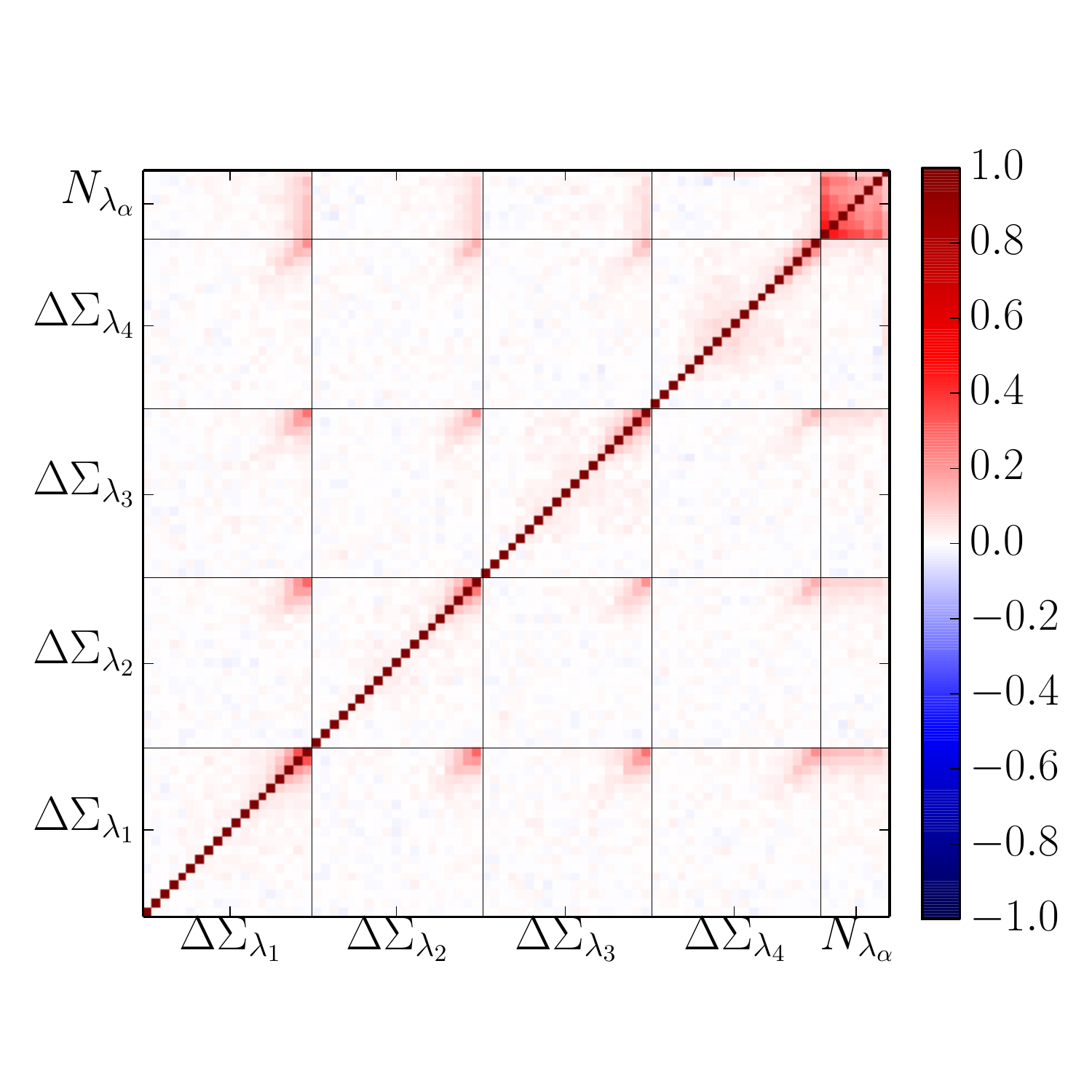}
\caption{The correlation coefficient matrix $r_{ij}$ in
equation~\eqref{eq:rrr} including the cross-covariances between
different richness bins as well as those between the lensing profiles
and the abundances. The blocks enclosed by the solid lines denote the
covariance matrix for $\Delta \Sigma_{\lambda_\beta}$ in the $\beta$-th richness
bin or the matrix for the abundance (upper-right corner). Each pixel for
the elements of $\Delta \Sigma_{\lambda_\beta}$ denotes the radial bin, where we used 19
bins in the range $R=[0.2, 50]~h^{-1}{\rm Mpc}$, while each pixel for the
elements of the abundance ($N_{\lambda_\alpha}$) denote each of the eight richness
bins in Table~\ref{tab:binning}.
}
\label{fig:mock_r}
\end{figure}
We also need to model the error covariance matrices for both the
abundance and the stacked lensing profiles as well as their
cross-covariance matrix.  We use 108 realizations of the SDSS mock
catalogs for source galaxies and clusters, generated based on the method
in \citet{Shirasaki:2016},
in order to estimate the covariance
matrices. We \revise{describe the details below} (also see
Appendix~\ref{app:cov_est}) and summarize the main results.

To make each realization of the mock catalog, we use a full-sky,
light-cone cosmological simulation that is constructed from sets of
$N$-body simulations \citep[see][for details]{Takahashietal:17}.
Each light-cone simulation consists of 27
spherical-shell source planes that are spaced by every 150~$h^{-1}$Mpc
in radial direction from an observer (the center of sphere) up to
redshift $z\simeq 2.4$. The projected matter fields in each plane are
generated by projecting $N$-body particles into the spherical shell of
$N$-body simulation output at redshift corresponding to the radial
distance from an observer. The lensing effects at a given angular
position in each source plane are computed by performing a ray-tracing
simulation through the foreground matter distribution based on the
multiple lens-plane algorithm \citep{Hamana:2001,Shirasaki:2015}. Each
source plane in the full-sky, light-cone simulation is given in the
\texttt{HEALPix} format \citep{Gorski2005},
with angular resolution of about 0.43
arcmin. By fully utilizing the full-sky, light-cone simulation, we
generate \revise{a} mock catalog of SDSS source galaxies according to the
following steps. (1) \revise{First} we assign the celestial coordinates (RA and
Dec) to the full-sky simulation. (2) We insert each source galaxy of the
real SDSS source catalog into the nearest angular pixel in the nearest
source plane according to its angular position (RA and Dec) and
photometric redshift, where we use the best-fitting photo-$z$
point estimate. We repeat this assignment for all the 39 million
galaxies. (3) We randomly rotate ellipticity of each source galaxy in
order to erase the real lensing signal. (4) Simulate the lensing
distortion effect on each source galaxy due to the foreground matter
distribution by adding the lensing shear and the intrinsic ellipticity.
We again repeat this for all the source galaxies. Thus our mock catalogs
take into account the effects of SDSS survey footprint (the right plot
of Figure~\ref{fig:footprint}) and the properties of source galaxies
(distributions of angular position, redshift and ellipticity).

Furthermore we need to make \revise{a} mock catalog of the redMaPPer clusters.  We
use the catalog of halos in each light-cone simulation realization. To
identify halos from $N$-body simulation output at each redshift, we used
the $\mathtt{Rockstar}$ software similarly to what we did in the
emulator building. We preserve the radial distance to each halo in the
line-cone simulation using the position of the halo in the $N$-body simulation box, 
rather than assigning it to the discrete source planes.
In order to assign hypothetical redMaPPer clusters to halos in the
light-cone simulation, we need the mass-richness relation. This \revise{causes}
a bit \revise{of circularity}, because \revise{the} estimation of the mass-richness
relation requires the covariance to perform the parameter estimation,
but an estimation of the covariance requires to know the mass-richness
relation to build the mock catalogs that are needed for the covariance
estimation. Here we \revise{adopt} the following approach. \revise{First}, we use the
real SDSS catalogs to estimate the covariance matrix taking into account
only the shape noise contribution for the lensing measurements
and the Poisson noise for the abundance measurements.
We ignore the off-diagonal parts of the abundance measurements and the cross-covariance parts between the lensing profile and abundance.  
\revise{Then} we estimate the best-fit parameters of the mass-richness relation,
based on the MCMC analysis, using the crude covariance.
The best-fit parameters for $P(\ln \lambda|M)$ 
are estimated from MCMC analysis for the sample of $20 \le \lambda \le 100$:
\revise{$A=3.225$, $B=1.002$, $\sigma_0=0.440$ and $q=-0.175$} as shown in Table \ref{tab:paramsMCMC_full}, with \revise{$\chi^2_{\rm min}=80.6$}
for 75 degrees of freedom ($84-9$ for the 84 data points and \revise{nine} parameters), 
as we will again describe later.
\revise{Then} using the estimated probability distribution $P(\ln \lambda|
M)$, we randomly assign a hypothetical richness $\lambda$ to each halo
that resides inside the SDSS survey footprint
in Figure~\ref{fig:footprint} as well as in the redshift range of
$0.10\leq z\leq 0.33$, in each light-cone realization. 
As stressed in \citet{Singhetal:16}
\citep[also see][]{Shirasaki:2016}, the use of random catalogs is
important for an accurate estimation of the covariance matrix for the
stacked cluster lensing. We construct the mock random catalogs in a way
that each catalog reproduces the redshift and richness distributions of
the real clusters, but randomly distributed the angular distribution of
random points within the SDSS survey footprint (without masks).
Thus we create 108 realizations of the mock catalogs for both the SDSS
source galaxies and the redMaPPer clusters, including the random points, over the entire survey footprint
covering the area of $10,401~{\rm deg^2}$.

We use these mock catalogs to calculate the error covariance as follows.
We apply the jackknife (hereafter JK) resampling method to
each mock catalog to estimate the covariance matrix of each realization,
where we employed 83 subdivisions of the SDSS footprint. Then we average
the covariance matrices estimated from the 108 mock catalogs as an
estimator of the true covariance matrix.  Thus we effectively use about
9,000 (more exactly $83\times 108=8,964$) quasi-independent realizations
(resamplings) for the covariance estimation. Since we use the full-sky simulations, our
covariance includes the effects of survey geometry, the super-sample
covariance contribution \citep{TakadaHu:13}, and the cross-covariance
between the abundance and the stacked lensing.

\begin{deluxetable*}{llccc}[t]
 \tablewidth{0pt}
  \tablecaption{Model parameters and the parameter estimation from the joint fitting of lensing and abundance measurements
  \label{tab:params_prior}}
 \tablehead{\colhead{Parameter}
 & \colhead{Description}
 & \colhead{Prior}
 & \colhead{Median and error}
& \colhead{Median and error}
\\
 \colhead{} 
 &\colhead{} 
 &\colhead{} 
 & \colhead{ $20 \leq \lambda \leq 100$}
 & \colhead{ $30 \leq \lambda \leq 100$}
}
  \startdata
$A$& The amplitude of 
$ \ln{\lambda}$ at pivot mass scale $M_{\rm pivot}=3 \times 10^{14} h^{-1} M_{\odot}$ &  $(0.5, 5.0)$ & \revise{$3.207^{+0.044}_{-0.046}$ $(3.231)$} &  \revise{$3.258^{+0.064}_{-0.057}$ $(3.279)$} \\ 
 &for the mean mass-richness relation   \\
$B$& Power-law index of halo mass dependence for the mean relation &$(0.0, 2.0)$ & \revise{$0.993^{+0.041}_{-0.055}$ $(1.016)$} & \revise{$0.874^{+0.102}_{-0.109}$ $(0.927)$}\\ 
$\sigma_0$& Scatter in $\ln \lambda$ at the pivot mass scale 
in $P(\ln \lambda|M)$
&  $(0.0, 1.5)$ & \revise{$0.456^{+0.047}_{-0.039}$ $(0.429)$} &  \revise{$0.469^{+0.052}_{-0.056}$ $(0.436)$}\\ 
 $q$ & Coefficient of the halo mass dependence in the scatter.
&  $(-1.5, 1.5)$ & \revise{$-0.169^{+0.035}_{-0.026}$ $(-0.184)$} &  \revise{$-0.096^{+0.084}_{-0.066}$ $(-0.132)$}\\     
$f_{\rm cen, 1}$& A fraction of central galaxy being at the true cluster center 
&   $(0.0, 1.0)$ & \revise{$0.58^{+0.27}_{-0.36}$ $(0.78)$} &   -- \\ 
&in the first richness bin \\
$f_{\rm cen, 2}$& Similar to $f_{\rm cen, 1}$, but for the clusters in the second richness bin
& $(0.0, 1.0)$ & \revise{$0.81^{+0.14}_{-0.34}$ $(0.98)$} &   \revise{$0.84^{+0.10}_{-0.18}$ $(0.92)$} \\ 
$f_{\rm cen, 3}$& 
 Similar to $f_{\rm cen, 1}$, but for the clusters in the third richness bin
&  $(0.0, 1.0)$ & \revise{$0.44^{+0.32}_{-0.30}$ $(0.59)$} &   \revise{$0.61^{+0.20}_{-0.37}$ $(0.77)$} \\ 
$f_{\rm cen, 4}$& 
 Similar to $f_{\rm cen, 1}$, but for the clusters in the fourth richness bin
& $(0.0, 1.0)$ & \revise{$0.57^{+0.27}_{-0.35}$ $(0.73)$} &   \revise{$0.74^{+0.16}_{-0.33}$ $(0.84)$}\\
$\alpha_{\rm off}$& Off-centering radius parameter relative to the aperture radius
&  $(10^{-4}, 1.0)$ & \revise{$0.064^{+0.051}_{-0.031}$ $(0.139)$} &   \revise{$0.134^{+0.231}_{-0.077}$ $(0.340)$}
 \enddata  
\tablecomments{
Model parameters (equations~\ref{eq:mean_relation} and
\ref{eq:scatter_M}, and see descriptions above
equation~\ref{eq:R_lambda_beta}), a short description of each parameter,
the prior, and the median of the MCMC samples for the joint fitting to 
the abundances and lensing profiles.
We parametrize the 
mass-richness relation relative to the pivot mass scale,
$M_{\rm
pivot}=3\times 10^{14}h^{-1}M_\odot$, which is a typical halo mass of
the clusters.
For all the model parameters we employ a flat prior in the range denoted. 
Note that we additionally restrict $\sigma_{\ln \lambda|M}>0$ for the range of halo masses we consider,
$10^{12}\le M/[h^{-1}M_\odot]\le 2\times 10^{15}$.
The column labeled as ``Median and error'' denotes the median, the 16th and 84th percentiles of the posterior distribution. 
The number in the round parenthesis is 
the best-fit parameter value for the model with minimum
$\chi^2$.  The fourth and fifth columns denote the results for the sample of $20\le \lambda\le 100$ and for the sample of $30\le
\lambda\le 100$.  
}
\end{deluxetable*}

Figure~\ref{fig:mock_sn_rand_ratio} shows diagonal components of the
covariance matrix estimated based on the above method.
The left panel compares the full covariance with the shape noise contribution for the
components involving the lensing profiles in four richness bins.
The covariance diagonal amplitude, $\sqrt{ {\bf C}(\Delta\Sigma_{\lambda_\beta}, \Delta\Sigma_{\lambda_\beta}) }$,
decreases with radius as $1/R$
for the logarithmically-spacing bins up to $R\simeq 10~h^{-1}{\rm Mpc}$,
reflecting the fact that the shape noise gives a dominant contribution in this radial scale.
The sample variance starts to be non-negligible at the larger radii.
The error amplitude is greater for the larger richness clusters because of the
fewer clusters, thereby leading to the fewer
lens-source pairs (therefore the larger shape noise contribution
in relative).
The maximum scale of $R=50~h^{-1}{\rm Mpc}$ is set
by a size of the JK subregion.  The middle panel displays \revise{the} contribution
of the random covariance for the covariance estimation of the lensing
profile. \revise{We find that the effect of} random subtraction is at a few percent
level at most in the covariance amplitude and is not significant for the
redMaPPer clusters \citep[also see][]{Shirasaki:2016}. The right panel
shows diagonal components of the covariance involving the abundance
${\bf C}(N_{\lambda_\alpha},N_{\lambda_\alpha})$, compared to the Poisson term.
The figure shows that the sample variance is significant for lower richness bins
\citep{HuKravtsov:03,TakadaBridle:07}, while the Poisson contribution
becomes dominate for a larger richness bin due to the fewer clusters.

Another important aspect of the covariance matrix is its off-diagonal
components. It describes cross-correlations between observables at
different bins. The correlated shape noise arising from a clustering of
clusters causes such correlated errors in the lensing profiles at
different bins \citep[see equation~47 in][]{OguriTakada:2011}.
In addition, the sample variance causes such correlated scatters; for
example, the same large-scale structure causes coherent scatters in the
lensing profiles at different radial and/or richness bins as well as in
the abundance \citep{TakadaBridle:07,TakadaHu:13,TakadaSpergel:14}. In particular, the
cross-covariance between the abundance and the lensing profiles is
caused by the super sample
covariance.

The relative contribution of the off-diagonal components to the diagonal
components can be quantified by the cross-correlation coefficient
matrix, defined as
\begin{equation}
r_{i j}\equiv \frac{ {\bf C}( D_i,  D_j) }{ \sqrt{ {\bf C}( D_i,  D_i)
{\bf C}( D_j,  D_j) } },
\label{eq:rrr}
\end{equation}
where $D_i$ is the $i$-th observables ($N_{\lambda_\alpha}$ or $\Delta\Sigma_{\lambda_\beta}$).
Note $r_{ij}=1$ for $i=j$ by definition, and $r_{ij }\rightarrow 1$
means a strong correlation between data at the $i$- and $j$-th bins
($i\ne j$), while $r_{ij}=0$ \revise{denotes} no correlation.
Figure~\ref{fig:mock_r} shows the correlation coefficient matrix.
There are non-vanishing cross-correlations between different bins of
large radii $R\gtrsim 10~h^{-1}{\rm
Mpc}$  for the lensing profiles and all the  richness bins for the abundance. 
We will \revise{discuss the impact of the sample variance on the parameter estimation below}.
\section{Results}
\label{sec:m_results} 
\begin{figure*}
        \centering
    \includegraphics[width=0.83 \textwidth]{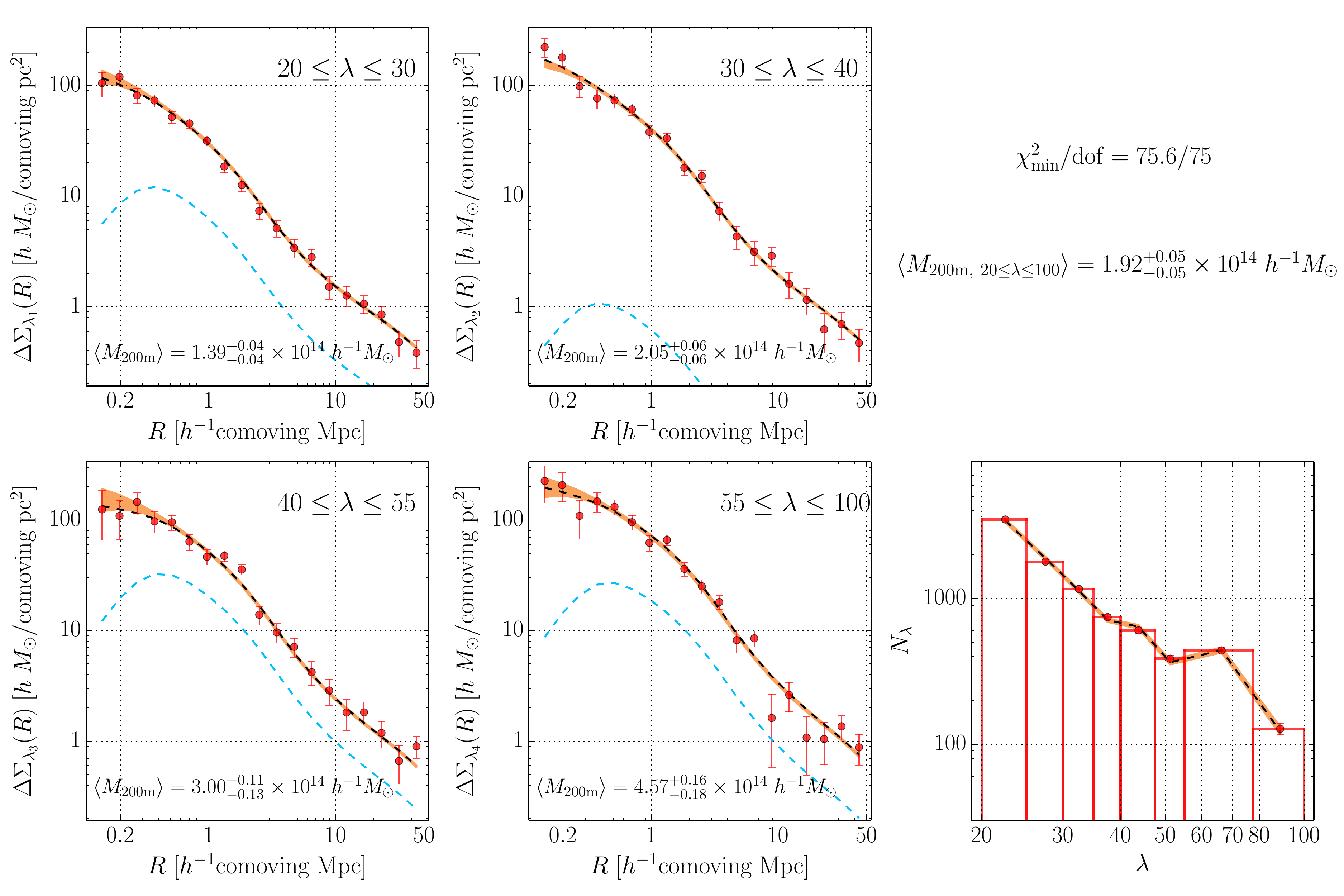}
\caption{Comparison of the best-fit model predictions with the
lensing profile measurements in four richness bins (the left four panels) and
the abundance measurements in eight richness bins (the right-lower panel), for the cluster
sample of $20\leq \lambda \leq 100$.
The orange shaded regions show the 16th and 84th percentiles of the model predictions computed from
the MCMC chains,
while the red points with error bars are the measurements.
The errors are from the diagonal components of the covariance matrix.
The \revise{black dashed} curves in each panel
are from the best-fit model with minimum $\chi^2$.
The light-blue dashed curves in each panel of the
lensing measurements are the best-fit models for the lensing profiles due to off-centered clusters in each richness bin
(see around equation~\ref{eq:R_lambda_beta} for the modeling of the off-centering
effects on the lensing profile).
\revise{Note that the amplitudes of the off-centering lensing profile are not well constrained.}
At the upper right corner, we give
\revise{the minimum value of reduced
chi-square}.
We also give $\langle M_{\rm 200m}\rangle$
in each of the lensing plots and at the upper-right corner; the values are the median of the mean halo mass (weighted
with the cosmological volume and
the halo mass function) and its 16th and 84th percentiles that are computed from the MCMC chains of the mass-richness
relation models.
} \label{fig:fitting_lensing_all}
\end{figure*}
\begin{figure*}
        \centering
    \includegraphics[width=0.83 \textwidth]{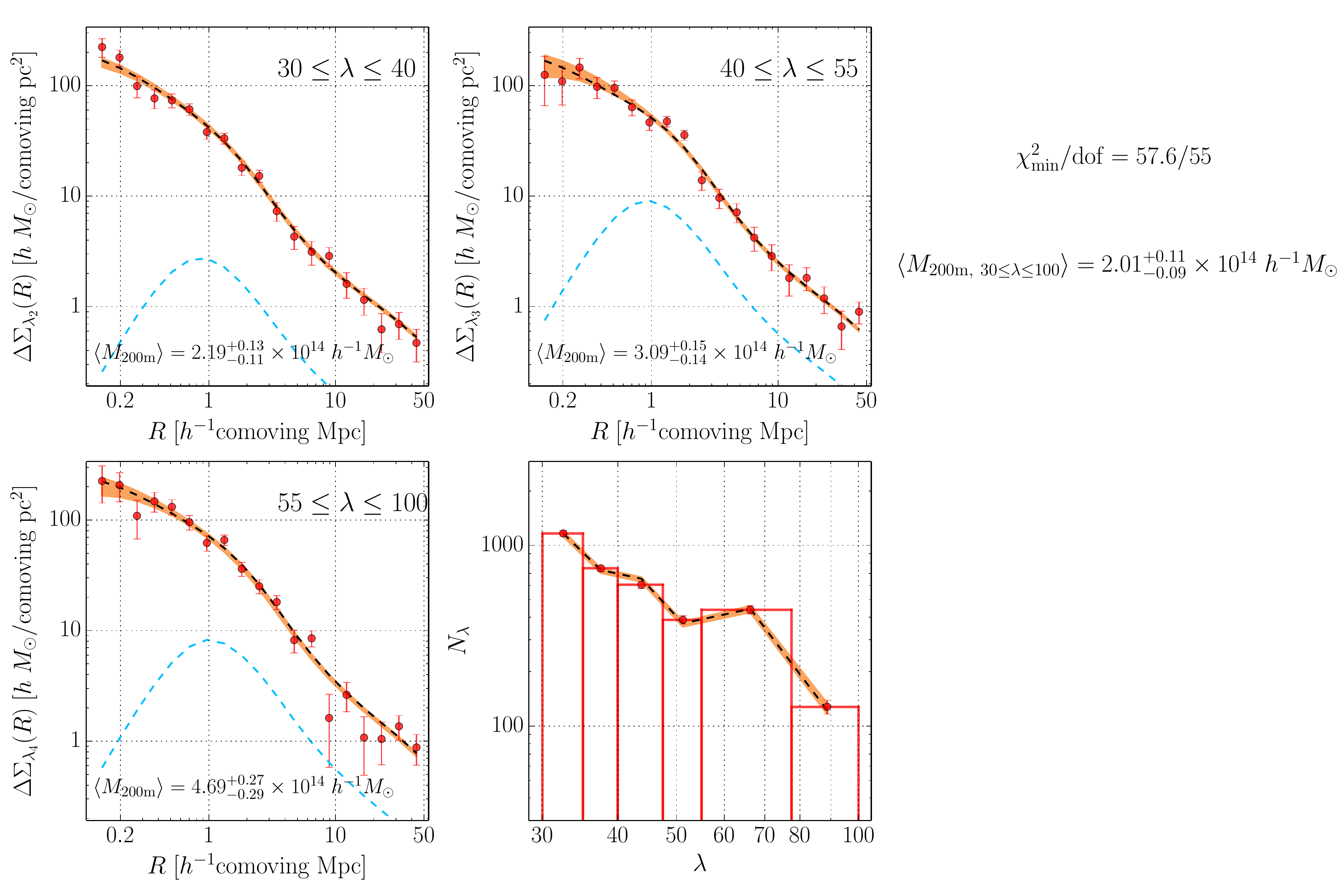}
\caption{Similarly to the previous figure, but for the sample of $30\le \lambda\le 100$.
}
\label{fig:fitting_lensing_sub}
\end{figure*}
In this section, we show the main results of this paper; \revise{constraints on}
the mass-richness relation $P(\ln \lambda|M)$ from a joint
fitting of the model to
the abundance and lensing profiles, based on \revise{our} forward
modeling approach.
In this analysis we do not vary cosmological parameters, and fix those to the \textit{Planck} cosmology.
\begin{figure*}
        \centering
        \includegraphics[width=1.00\textwidth]{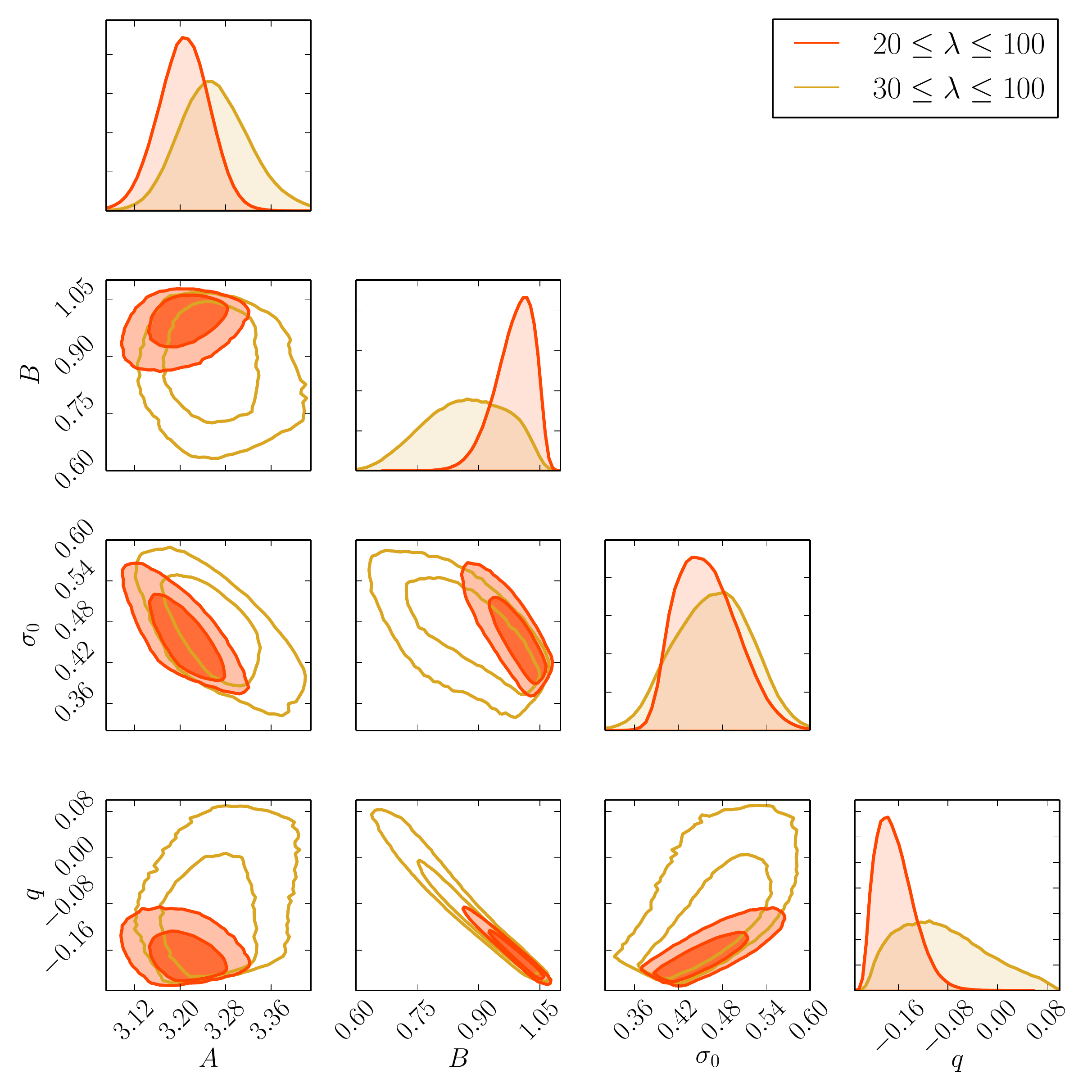}
\caption{
The diagonal panels show the posterior distribution of each parameter of the
mass-richness relation, and the other panels show
the 68\% and 95\%~CL contours of the MCMC chains
in each two-parameter subspace.
The red- and yellow-color results are for
the samples of $20\le \lambda\le 100$ and $30\le \lambda\le 100$, respectively.
The constraints include marginalization
over other parameters including the off-centering parameters.
Each parameter is well constrained compared to the flat priors in Table~\ref{tab:params_prior}.
A sharp bound of the contours in some parameters involving $\sigma_0$ or $q$, e.g.
the lower-left corner of $(\sigma_0, q$) subspace, is
due to the condition $\sigma_{\ln \lambda|M}>0$ over the range of $10^{12}\le M/[h^{-1}M_\odot]\le 2\times 10^{15}$.
}
    \label{fig:fitting_MCMC}
\end{figure*}
\subsection{Parameter estimation}
\label{sec:likelihood}
Once the halo emulator and the error covariance matrix are given as
discussed in Section~\ref{sec:emurator}, we
can constrain model parameters in the mass-richness relation, $P(\ln \lambda|M)$, by comparing the model predictions with the 
measurements of the abundance and the lensing profiles.
We perform \revise{Bayesian} parameter estimation assuming the Gaussian form, $\mathcal{L}\propto \exp(-\chi^2/2)$ \revise{, for the likelihood}: 
\begin{equation}
 \chi^2 =
 \sum_{i, j}~ \Bigl[ {\bf D} - {\bf D}^{\rm model}\Bigr]_i\left({\bf C}^{-1}\right)_{ij}
    \Bigl[{\bf D} - {\bf D}^{\rm model} \Bigr]_j,
\end{equation}
where ${\bf D}$ is the data vector \revise{that consists of the lensing profiles and the abundance in different radial and richness bins}, ${\bf D}^{\rm model}$
is the model predictions, and ${\bf C}^{-1}$ is the inverse of the
covariance matrix.
\revise{Note that we use 19 radial bins in each richness bin, and that
\begin{equation*}
{\bf D}
= \left\{\Delta\Sigma_{\lambda_1}(R_1),
... \Delta\Sigma_{\lambda_1}(R_{19}),
... \Delta\Sigma_{\lambda_4}(R_{19}), N_{\lambda_1} , ... N_{\lambda_8}
\right\},
\end{equation*}
for the sample of $20\le\lambda\leq 100$ and
\begin{equation*}
{\bf D}
= \left\{\Delta\Sigma_{\lambda_2}(R_1),
... \Delta\Sigma_{\lambda_2}(R_{19}),
... \Delta\Sigma_{\lambda_4}(R_{19}), N_{\lambda_3} , ... N_{\lambda_8}
\right\},
\end{equation*}
for the sample of $30\le\lambda\leq 100$.}
The index $i$ \revise{and $j$ run} over the different components of
data, \revise{$i, j=1,2,\dots,84$} for the sample of $20\le\lambda\leq 100$, while
\revise{$i, j=1,2,\dots, 63$} for the sample of $30\leq \lambda\leq 100$.
We include nine model parameters for the former sample
(four for the mass-richness relation and five for
the off-centering effect as described in
Section~\ref{sec:model_lensing}) or eight parameters for the latter sample,
respectively. We use a Markov chain Monte Carlo (hereafter MCMC) method
to perform an interference of the parameters. We use
$\mathit{emcee}$ \citep{Foreman:2013} for our parameter estimation.
Table~\ref{tab:params_prior} summarizes the results of parameter
estimation, which gives a description of each parameter, its prior, and
the median and $68\%$~CL interval after removing the burn-in chains and marginalizing over other parameters.
\revise{We use a flat prior for each parameter that has a sufficiently broad width as given in Table~\ref{tab:params_prior}.}
 The parameters of mass-richness relation
$\{A,B,\sigma_0,q\}$ are well constrained by the cluster observables, for
both the samples of $20\leq \lambda \leq 100 $ and $30\leq \lambda \le
100$.
\revise{The off-centering effects on the lensing profiles need to be included, but some of the off-centering parameters are not well constrained.
This can be compared with the recent work in \cite{vanUitertetal.16}, where the off-centering parameters are well constrained by jointly using 
the lensing profiles and the radial profiles of member galaxies.}

Figure~\ref{fig:fitting_lensing_all} compares the best-fit model
predictions with the measurements for both the lensing profiles in four richness bins and the abundance in eight richness bins, for the sample of $20\le \lambda\le 100$.  The orange-color shaded
regions in each panel denote the 68$\%$~CL intervals of 
the model predictions, 
obtained from the MCMC chains after marginalizing over the model parameters. 
The figure shows that our model remarkably well
reproduces the lensing profiles and the abundance in the different
richness bins simultaneously.
The agreement implies the existence of a
good model of the mass-richness relation that can reproduce the
cluster observables for the \textit{Planck} cosmology. The model mass-richness
relation has sufficient 
flexibility to reproduce the observables
\revise{by} varying the model parameters. 
We would also like to stress that the model lensing profiles, computed from 
the $N$-body simulation based emulator, well reproduce the measurements in the different
richness bins, including the 1-halo and 2-halo terms and the intermediate scales.
The reduced chi-square for the best-fit model is \revise{$\chi^2_{\rm min}/{\rm dof}=75.6/75$} 
for 75 degrees of freedom $(75=84-9)$ 
with nine model parameters, meaning that 
the best-fit model gives a good fit. 
For reference, we show how a shift in each of the mass-richness relation parameters from the best-fit value,
by an amount of the 68\% CL interval in Table~\ref{tab:params_prior}, changes the cluster observables in Appendix~\ref{app:parameters}, 
\revise{while} fixing other parameters to their best-fit values.
Figure~\ref{fig:parameter_dependence} shows that the shift in each parameter 
changes the abundances and the lensing profiles in each richness bin in a complex way. 

Similarly, even when we restrict ourselves to the subsample of \revise{clusters with larger}
richness $30\leq \lambda\leq 100$, our model can reproduce the measurements,
as shown in Figure~\ref{fig:fitting_lensing_sub} and Table~\ref{tab:params_prior}. 
The best-fit model is slightly different
from that for the sample of $20\leq \lambda\leq 100$, but the
two models agree with each other \revise{within their errors}.
\begin{figure*}
	\centering
	{\includegraphics[width=0.495 \textwidth ]{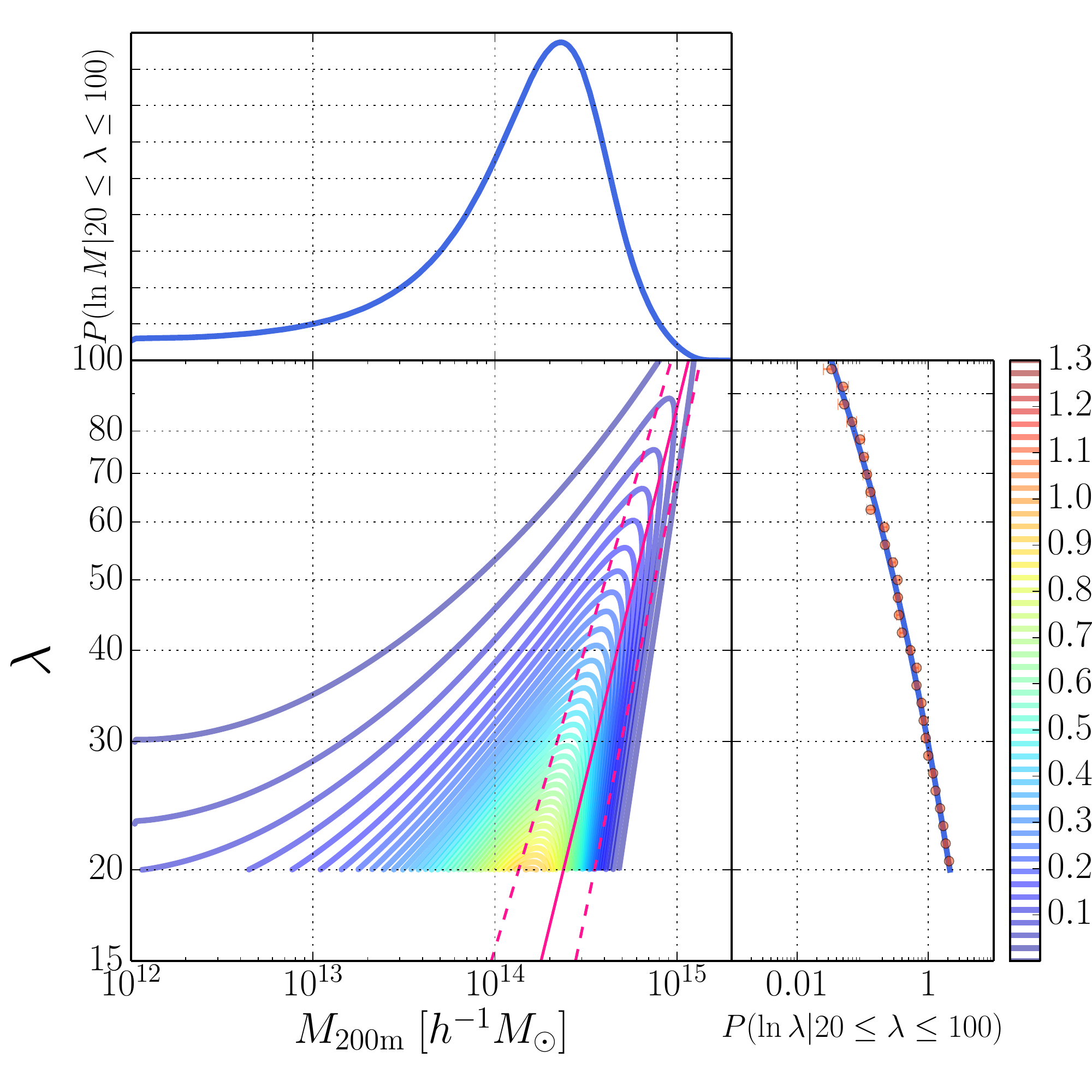}
    	\includegraphics[width=0.495 \textwidth ]{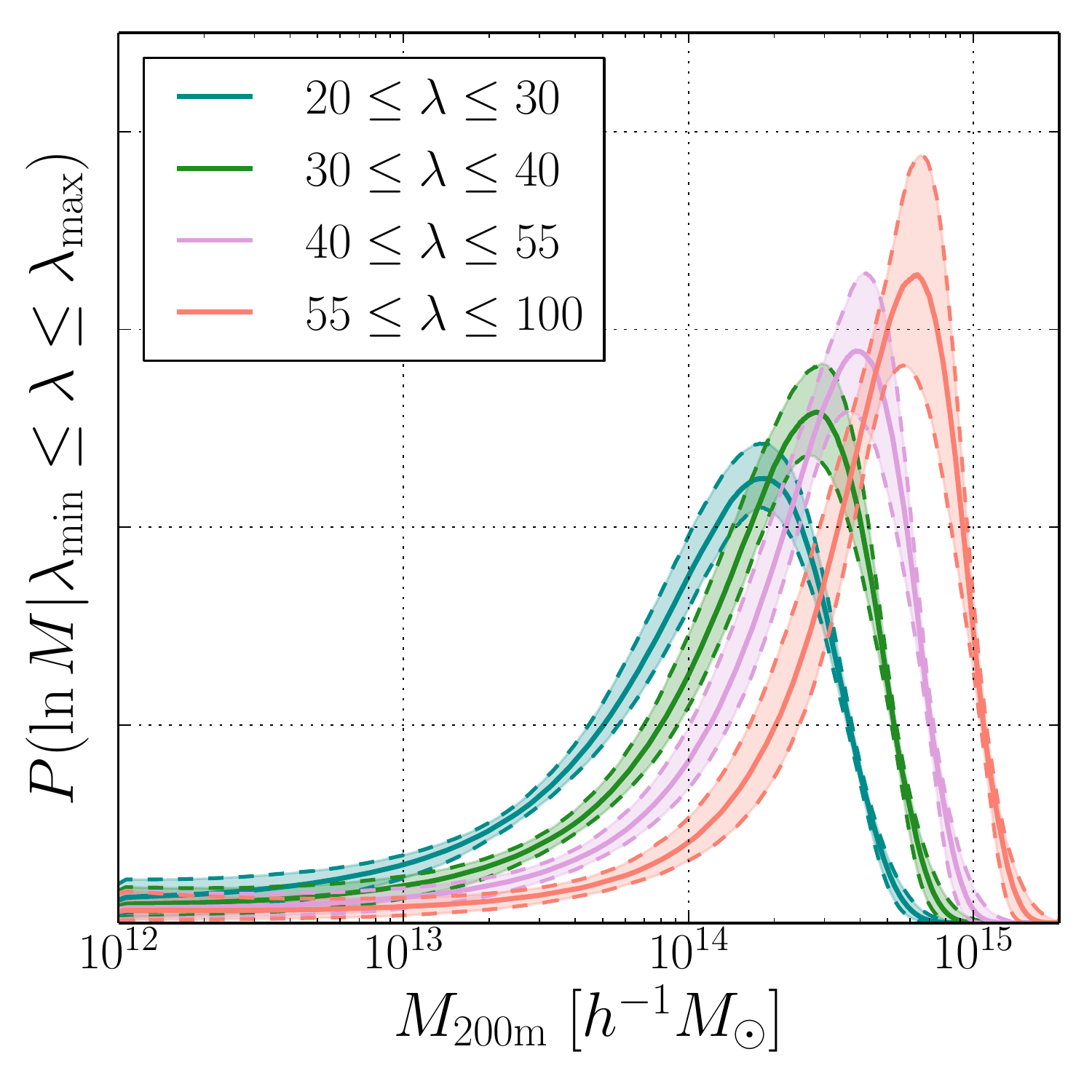}
    }
\caption{The result of the analysis in the sample of $20\le\lambda\le100$. {\it Left}: The {\it joint} probability distribution, $P(\ln M,\ln \lambda)$, from the best-fit model parameters for $P(\ln \lambda|M)$ using equation~\eqref{eq:joint}.
We normalized the joint probability so as to satisfy the normalization condition 
$\int\!\mathrm{d}\ln \lambda\int\!\mathrm{d}\ln M~P(\ln M,\ln\lambda)=1$ for the range of $20\le \lambda \le 100$ and $10^{12}\le M/[h^{-1}M_\odot] \le 2\times 10^{15} $. 
For comparison, the solid red-color line shows the best-fit model for the mean mass-richness relation, $\langle\ln\lambda\rangle(M)$, given 
in equation~\eqref{eq:mean_relation}
 (see Table~\ref{tab:paramsMCMC_full} for the parameter values), 
while the dashed lines show the 16th and 84th percentiles of $\ln \lambda$ distribution at a fixed mass (i.e. the width of $\sigma_{\ln \lambda|M}$ 
in equation~\ref{eq:scatter_M}).
By projecting the joint probability distribution along halo
mass in equation~\eqref{eq:Pr_integrate}, we can compute the probability distribution of richness (i.e. the richness function) as shown in the right panel. It remarkably well
reproduces the measurement denoted by the red points with error bars estimated from the Poisson noise at each of \revise{finer} richness bins.
On the other hand, the upper panel shows the probability distribution of
halo mass for the redMaPPer clusters, computed by projecting the joint
distribution along the richness in equation~\eqref{eq:PMtot}. 
The mean halo mass is found to be about $2\times 10^{14}h^{-1}M_\odot$ as shown in Figure~\ref{fig:fitting_lensing_all}.
{\it Right}: Similarly to the left panel, but the probability
distribution of halo mass for each of the four richness bins used in the
lensing measurements. The solid curve in each color is the median, and
the dashed lines denote the 16th and 84th percentiles 
computed from the MCMC chains.}\label{fig:joint_all}
\end{figure*}
\begin{figure*}
	\centering
	{\includegraphics[width=0.495 \textwidth ]{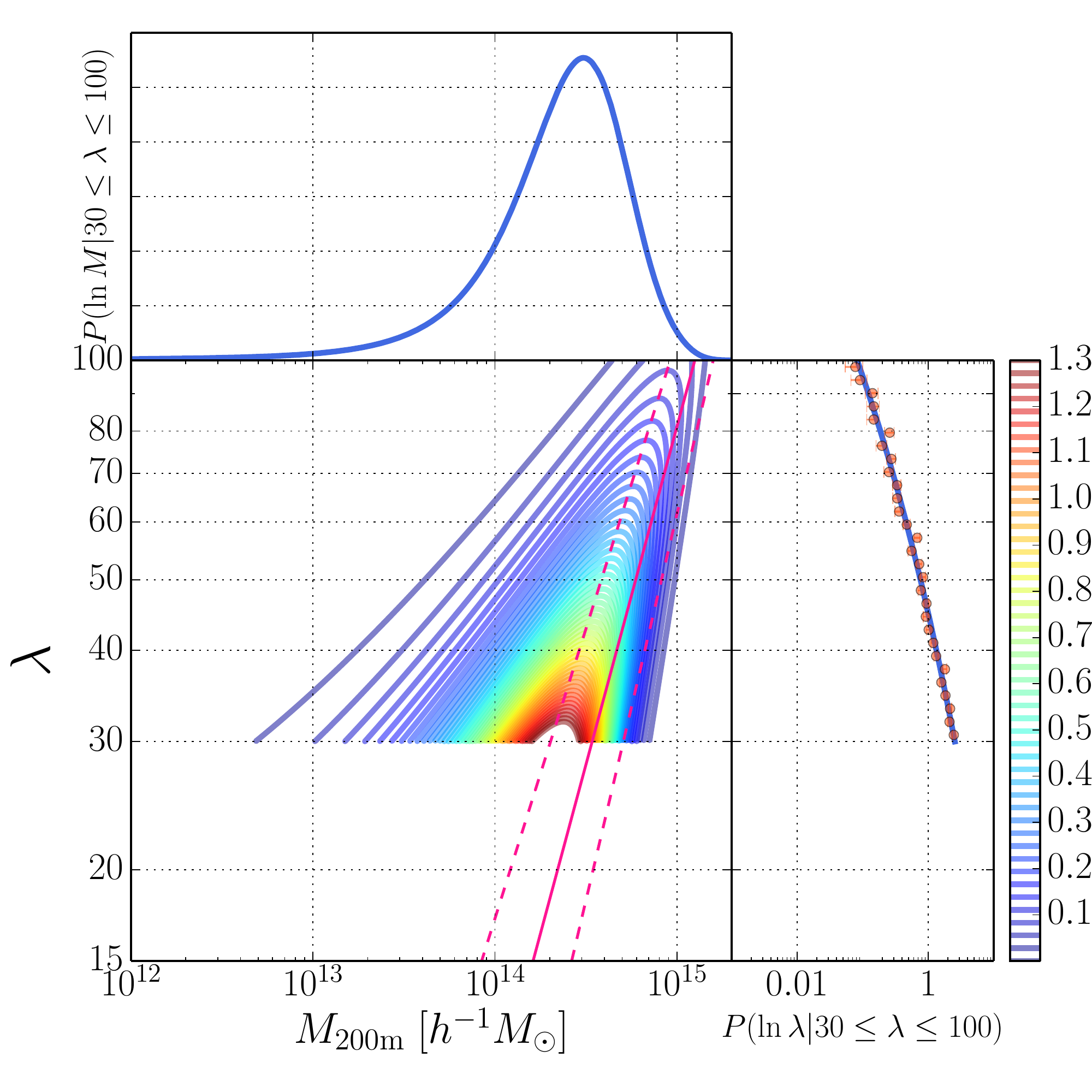}
    	\includegraphics[width=0.495 \textwidth]{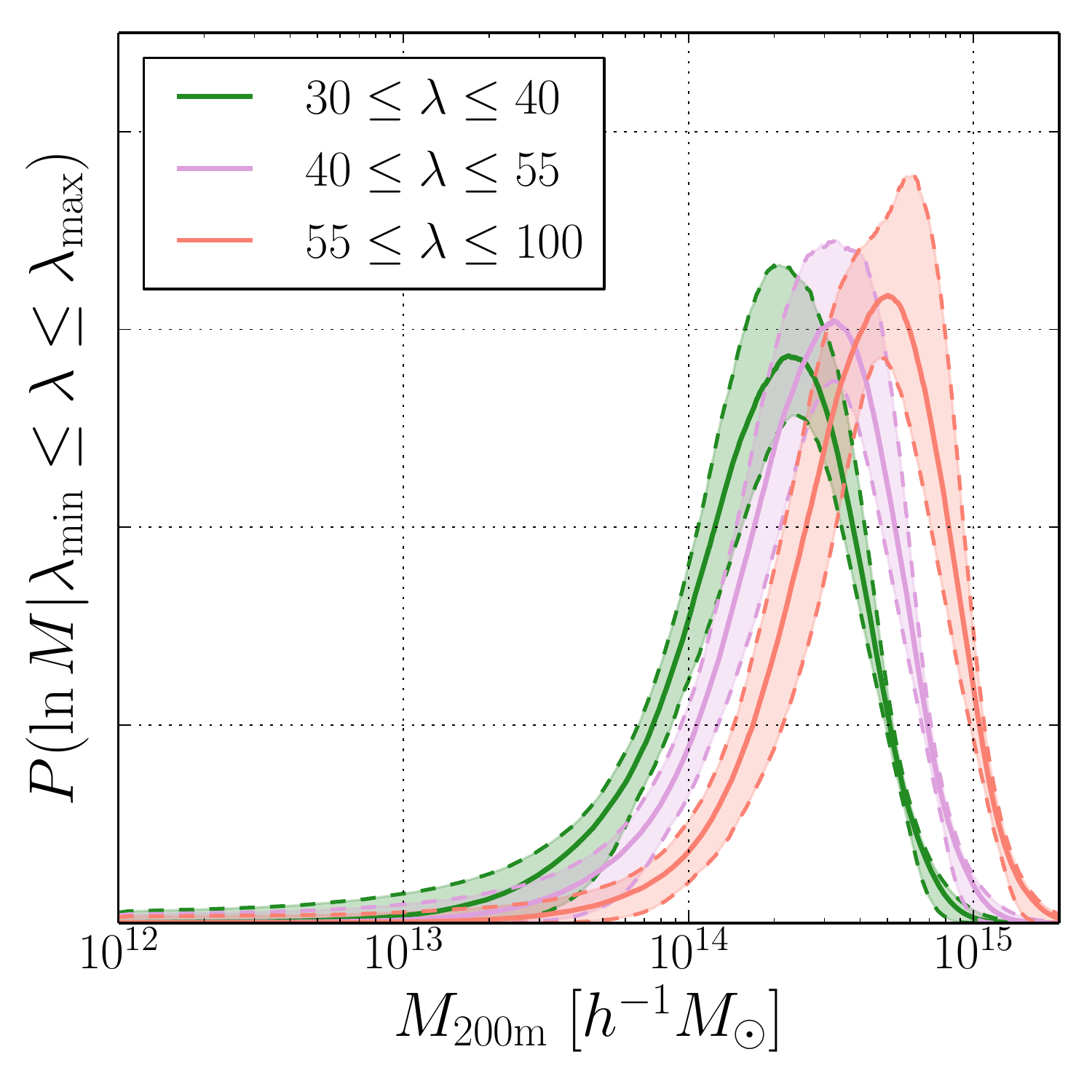}
    }
\caption{Similarly to the previous figure, but for the sample of $30\le \lambda\le 100$. 
The mass distribution at the lower mass is more suppressed compared to the result from the sample of $20\le \lambda\le100$ in Figure~\ref{fig:joint_all}.} 
\label{fig:joint_sub}\textit{}
\end{figure*}

Figure~\ref{fig:fitting_MCMC} shows 68\% and 95\%~CL contours \revise{for each
pair of our four parameters used to model} the mass-richness
relation, after marginalizing over the off-centering parameters, while
the one-dimensional histogram shows the posterior distribution of each parameter.
Even though we employ a flat prior on each parameter, a joint
measurement of the lensing profiles and the abundance allows us to well
constrain each parameter.
For the sample of $20\leq \lambda \leq 100$, a constant
scatter in the mass-richness distribution, i.e. $q=0$, is strongly
disfavored at \revise{about} $5\sigma$. 
 A negative value,
\revise{$q=-0.169^{+0.035}_{-0.026}$} is favored, 
as given in Table~\ref{tab:params_prior},
implying that the scatter \revise{starts to increase} with decreasing halo mass.
On the other hand, for
the subsample of $30\leq\lambda\leq 100$, a constant scatter with
$q=0$ is acceptable within the error bar.  We will \revise{discuss the implication of these results below.}

We can derive the
joint probability distribution of $\lambda$ and $M$: 
\begin{equation}
P( \ln M, \ln \lambda) \propto  P(\ln \lambda| M)P( \ln M),
\label{eq:joint}
\end{equation}
where $P(\ln M)$ is the
probability distribution of halo \revise{mass}.
Note that 
the normalization factor is determined
so as to satisfy the condition $\int\!\mathrm{d}\ln\lambda\int\!\mathrm{d}\ln M~P(\ln M,\ln \lambda)=1$
for the range of $\ln M$ and $\ln \lambda$, and we restrict the domain of $P(\ln M, \ln \lambda)$ to this range.
For $P(\ln M)$, we employ the underlying mass function in the SDSS volume for the \textit{Planck} cosmology:
\begin{eqnarray}
P(\ln M) = 
\frac{\displaystyle \int_{ z_{\rm min} }^{ z_{\rm max} } \mathrm{d}z~ \frac{ \chi^{\revise{2}}(z) }{ H(z) } \frac{ {\rm d}n(M, z) }{ {\rm d}\ln M }  }{ \displaystyle 
\int_{ \ln M_{\rm min} }^{  \ln M_{\rm max} }\mathrm{d}\ln M \int_{ z_{\rm min} }^{ z_{\rm max} } \mathrm{d}z~ \frac{ \chi^{\revise{2}}(z) }{ H(z) } \frac{ {\rm d}n(M, z) }{ {\rm d}\ln M } }.~
\label{eq:prop}
\end{eqnarray}
For $P(\ln \lambda|M)$ in equation~\eqref{eq:joint}, we use the best-fit model in Table~\ref{tab:params_prior}.

The contours in
Figure~\ref{fig:joint_all} show the joint distribution $P(\ln M,
\ln \lambda)$ in equation~\eqref{eq:joint}.
For comparison, the red solid line shows the best-fit model for the mean of the mass-richness relation,
$\langle\ln \lambda\rangle(M)$ (see equation~\ref{eq:mean_relation}),
while the dashed lines show the 68\% percentiles of the distribution of $\ln \lambda$  at a fixed halo mass (i.e. the width of $\sigma_{\ln \lambda|M}$ in equation~\ref{eq:scatter_M}).
\revise{From equation~\eqref{eq:joint}},
the joint probability has a power towards lower halo masses for each $\lambda$ bin due to the contribution of less massive halos via the halo mass function $P(\ln M)\propto \mathrm{d}n/\mathrm{d}\ln M$.  

Integrating the joint probability $P(\ln M,\ln \lambda)$ along either the halo mass or the richness direction 
gives the distribution of $\ln \lambda $ or $\ln M$ for the sample of $20\le \lambda \le 100$: 
\begin{equation}
P(\ln \lambda|\lambda_{\rm min} \leq \lambda \leq \lambda_{\rm max}) = \int_{\ln M_{\rm min}}^{\ln M_{\rm max}} \mathrm{d}\ln M~ P(\ln M, \ln \lambda)
\label{eq:Pr_integrate}
\end{equation}
or
\begin{equation}
P(\ln M|\lambda_{\rm min} \leq \lambda \leq \lambda_{\rm max}) = \displaystyle \int_{\ln \lambda_{\rm min}}^{\ln \lambda_{\rm max}} \mathrm{d}\ln \lambda~ P(\ln M, \ln \lambda).
\label{eq:PMtot}
\end{equation} 
The right panel in the left plot of Figure~\ref{fig:joint_all} gives
the probability $P(\ln \lambda|20\le \lambda\le 100)$, showing that 
 our
model remarkably well reproduces the observed richness function at much
\revise{finer} bins than eight bins in Figure~\ref{fig:fitting_lensing_all}, over
the entire range of richness. Similarly, the upper panel shows our model prediction for 
the probability of halo masses 
$P(\ln M|20\le \lambda \le 100)$  for the redMaPPer clusters. 
The probability 
implies that a typical halo mass of the redMaPPer clusters 
is about $M_{\rm 200 m}=2\times 10^{14} h^{-1}M_\odot $.
However, the distribution displays a long tail towards low mass, even
down to $M=10^{12}h^{-1}M_\odot$ that is much smaller than a cluster
mass scale. This might be an implication of residual systematic errors in our analysis or the redMaPPer catalog. 

In addition, we can compute the expected mass
distribution in each richness bin, $\lambda_{\alpha, \rm min} \le \lambda \le \lambda_{\alpha, \rm max}$, using equation~\eqref{eq:PMtot} by replacing $\{ \lambda_{\rm min}, \lambda_{\rm max}\}$ with $\{\lambda_{\alpha, \rm min}, \lambda_{\alpha, \rm max} \}$. 
The right plot 
of Figures~\ref{fig:joint_all} 
shows the result for the sample of $20 \le \lambda \le 100$. 
The shaded region around each curve shows the 68\%~CL interval computed from the MCMC chains. 
Clearly, the lowest richness bin ($20 \leq \lambda \leq 30$) favors the existence of low mass halos with
$M\lesssim 10^{14} h^{-1}M_\odot$ and it requires about 10\% contribution
from even group- or galaxy-scale halos with $M\lesssim10^{13}
h^{-1}M_{\odot}$.

Figure~\ref{fig:joint_sub} shows similar plots, but for the sample of $30\le \lambda \le 100$. Compared to Figure~\ref{fig:joint_all},
this sample has a suppressed contribution of low mass halos with $M\lesssim 10^{13}h^{-1}M_\odot$. 

\subsection{Comparison of $P(\ln M|\lambda)$ with Simet et~al.}

\begin{figure}
	\centering
	\includegraphics[width=0.495 \textwidth ]{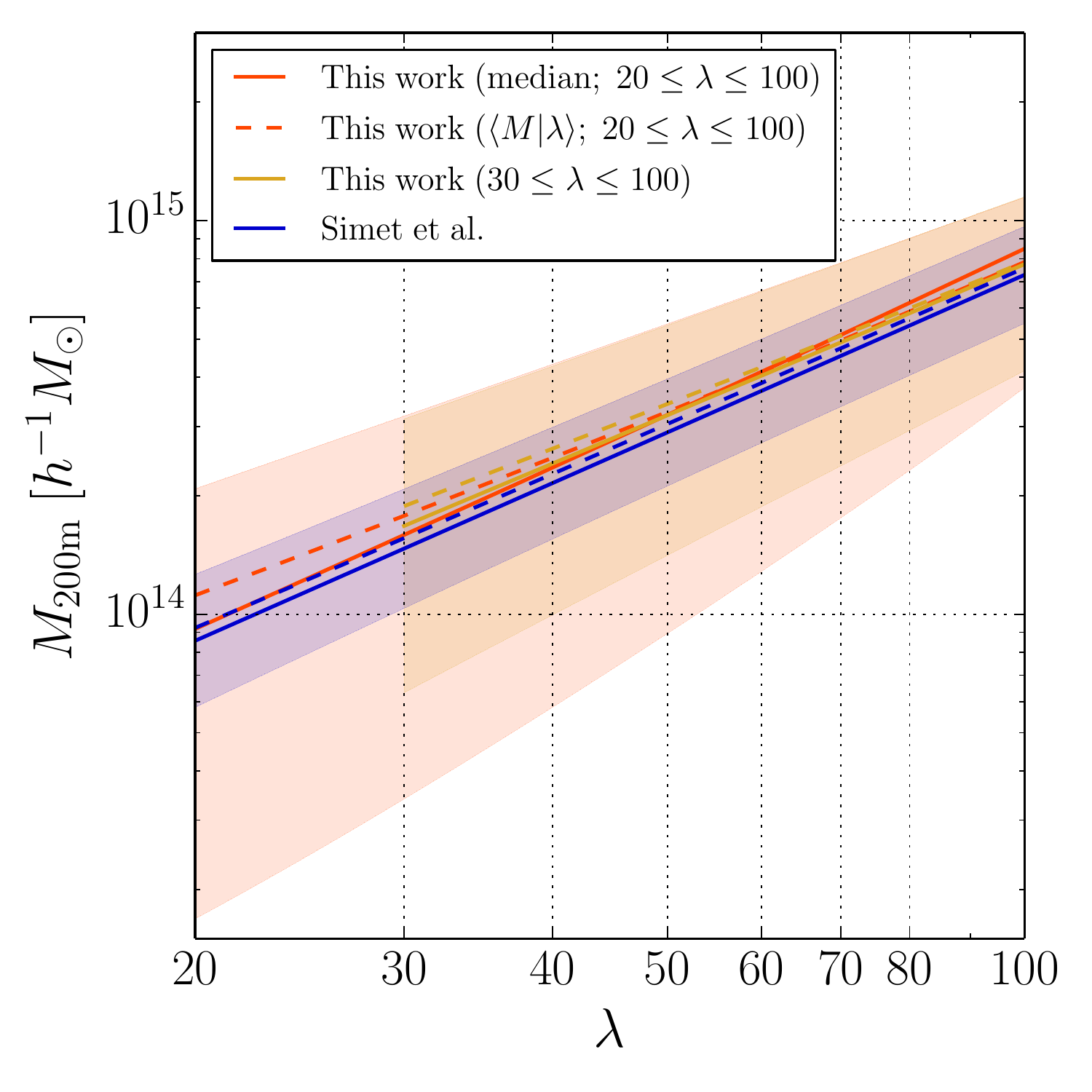}
\caption{The conditional probability distribution $P(\ln M|\lambda)$ 
computed from the best-fit model parameters according to equation~\eqref{eq:bayes}.
The \revise{solid} red or orange line denotes the
median of the mass distribution at a fixed richness for the sample of
$20\le \lambda\le 100$ or $30\le \lambda\le 100$, respectively. 
\revise{The dashed lines denote $\langle M|\lambda \rangle$ in the same colors.}
The respective shaded regions denote the range of the 16th and 84th percentiles of mass distribution at a fixed richness.
Our results for the median \revise{and the mean} relation agree \revise{especially at the high richness}
with the \revise{best-fit} result
in \citet{Simet:2016} \revise{calculated from equation~\eqref{eq:simet_mean},}
denoted by the blue line, 
where $P(\ln M|\lambda)$ was estimated from the lensing information alone based on a backward modeling approach.
However, our result shows the larger scatter than found in \citet{Simet:2016} (see text for the discussion). 
} \label{fig:compSimet}
\end{figure}
\begin{figure*}
        \centering
        {\includegraphics[width=0.495 \textwidth]{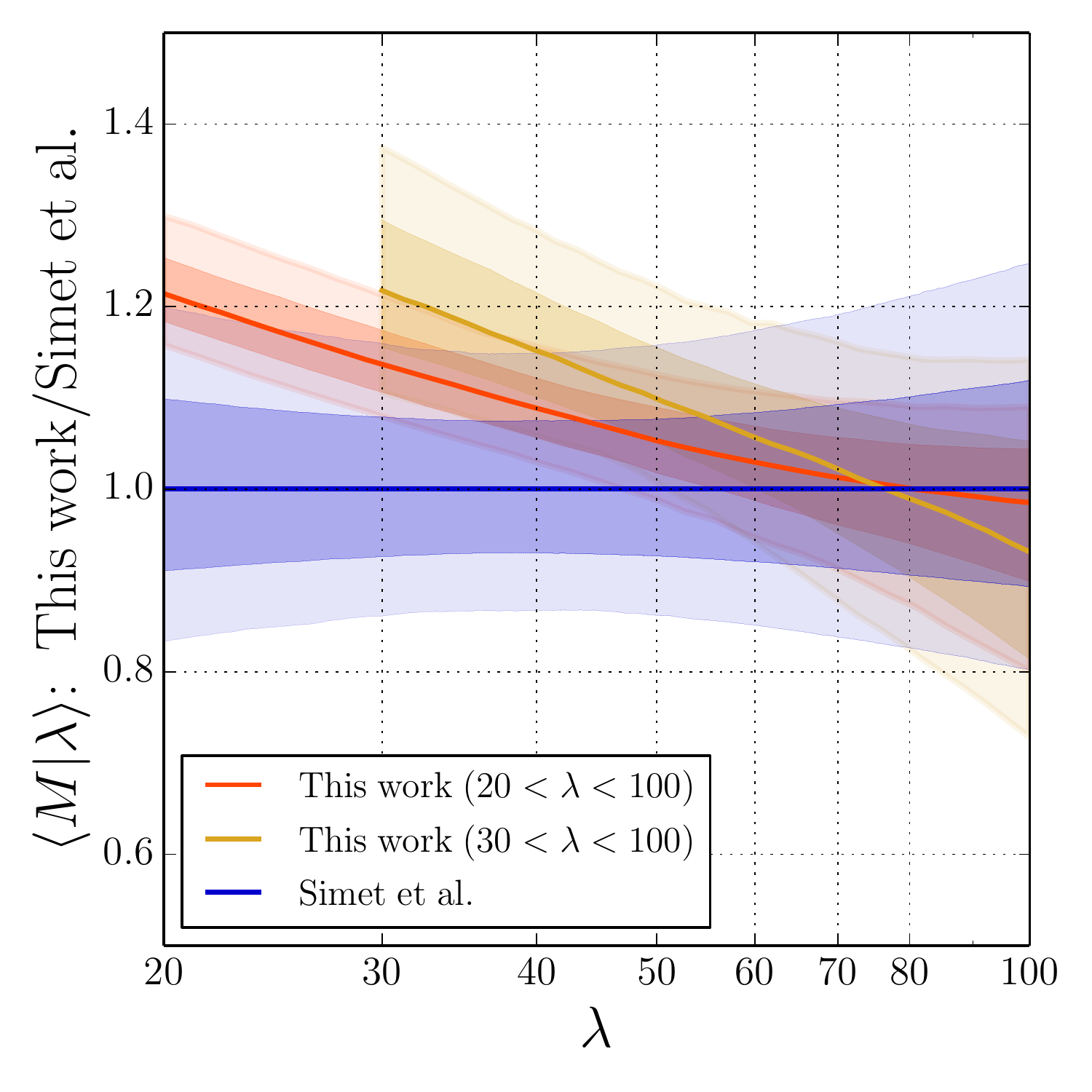}
        \includegraphics[width=0.495 \textwidth]{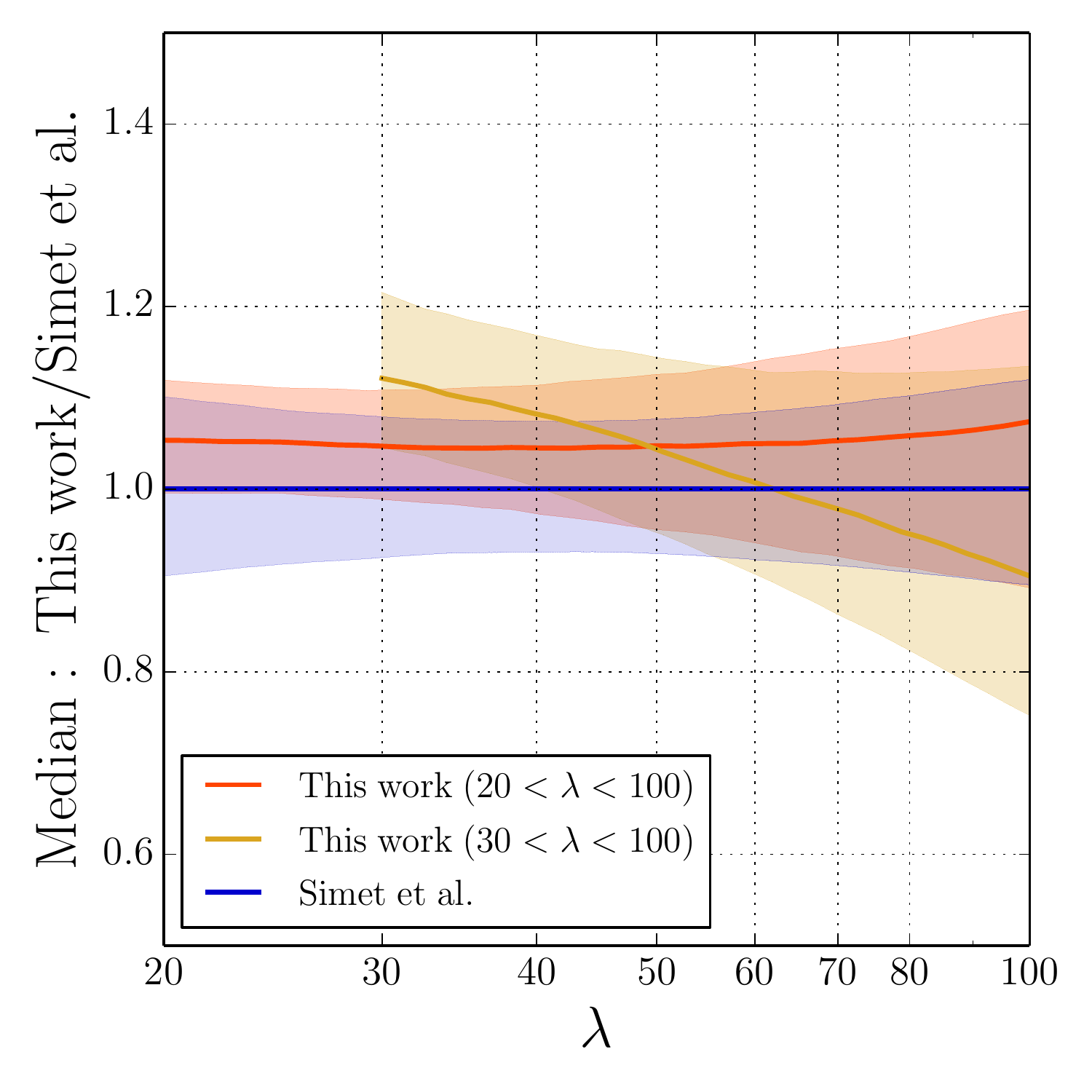} }
        \caption{\revise{Similarly to Figure~\ref{fig:compSimet}, but a more quantitative comparison
          of our result with \citet{Simet:2016} for the mass-richness relation, $P(\ln M|\lambda)$,
         for each of the samples of $20 \le \lambda \le 100$ or $30 \le \lambda \le 100$. 
         {\it Left}: The red- and orange-color inner shaded regions show the 16th and 84th percentiles of the distribution of the mean halo mass $\langle M|\lambda \rangle$, 
          computed from the MCMC chains for each cluster sample.
          The outer shaded region shows the 2nd and 98th percentiles and 
          the middle line is the median of distribution. 
          These results are shown relative to the S17 result,
          which is calculated from equations~\eqref{eq:simet_mean} and \eqref{eq:simet_scatter}
          as $\langle M|\lambda\rangle=\exp(\langle \ln M|\lambda\rangle+\sigma_{\ln M|\lambda}^2/2)$,
          with the best-fit parameters after equation~\eqref{eq:simet_scatter}.
          The blue shaded region around unity shows the same regions for the mean relation from the S17 result,
          inferred from the parameter constraints in S17 described after equation~\eqref{eq:simet_scatter}.
          {\it Right}: 
          Similarly to the left panel, but this shows the results for the median relation. This shows only the 16th and 84th percentiles of the distribution.
          These results are also shown relative to the S17 result, which is calculated as $\exp(\langle \ln M|\lambda\rangle)$, with the best-fit parameters after equation~\eqref{eq:simet_scatter}.
} }
        \label{fig:simet_chains_mean_median}
\end{figure*}
\begin{figure}
        \centering
        \includegraphics[width=0.46 \textwidth]{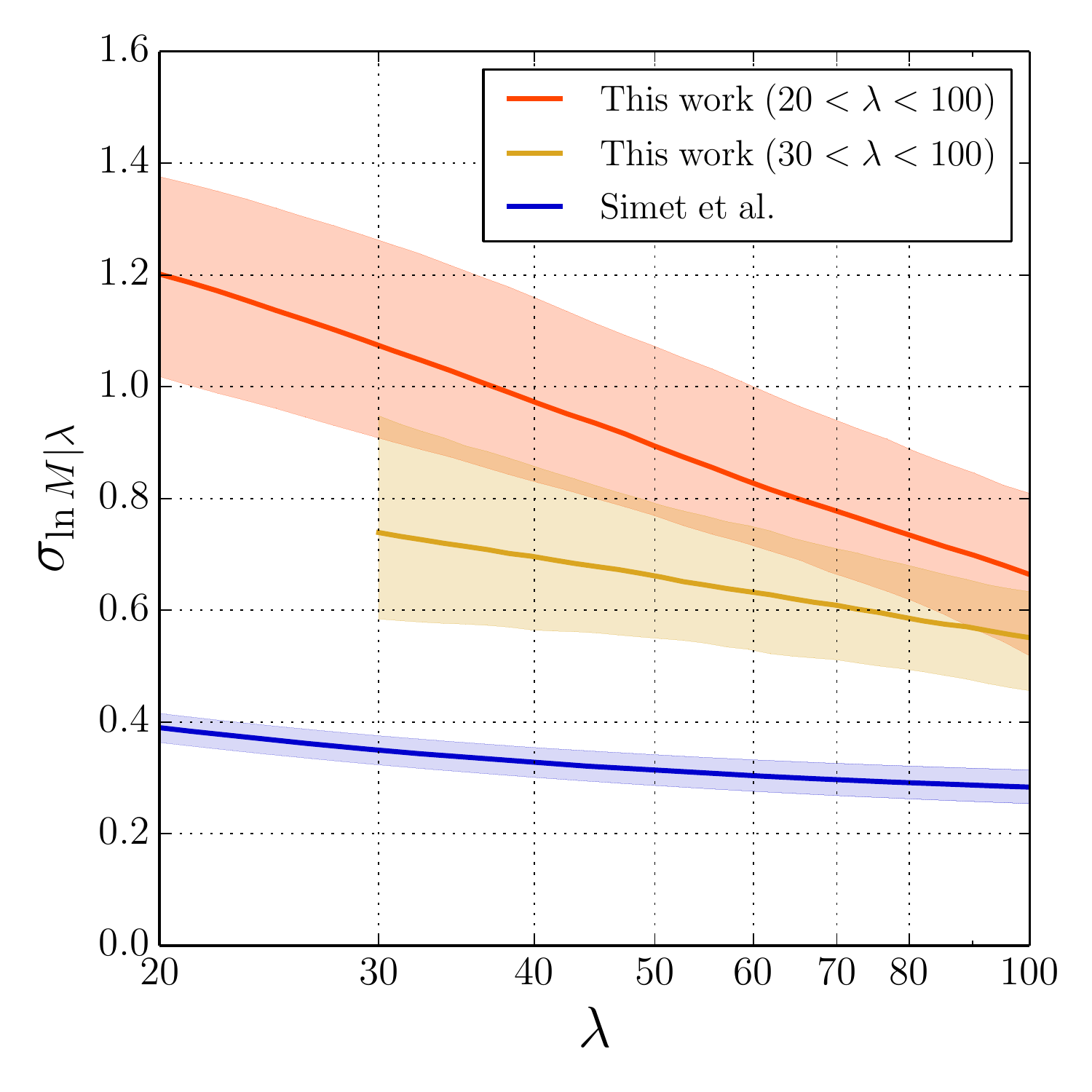}
        \caption{\revise{Similarly to Figure~\ref{fig:compSimet}, but a more quantitative comparison
          of our result with \citet{Simet:2016} for the mass-richness relation, $P(\ln M|\lambda)$,
         for each of the samples of $20 \le \lambda \le 100$ or $30 \le \lambda \le 100$. 
         The red- and orange-color shaded regions 
         show the 16th and 84th percentiles of the distribution of the scatters, $\sigma_{\ln M|\lambda}$,
         that characterize the width of halo mass distribution 
     	 for a fixed richness (the half width of 68\% CL region), computed from the MCMC chains for each sample (see text for details). The blue 
     	 shaded region shows the S17 result from equation~\eqref{eq:simet_scatter},
         where the error on $\sigma_{\ln M|\lambda}$ is estimated by propagating the uncertainties
         in the parameters in S17, which are described after equation~\eqref{eq:simet_scatter}.
         \reviseplus{Note that the constraint of S17 on the scatter is mainly from the prior.}
         The solid lines denote the median for the distributions.}
}
        \label{fig:simet_chains_scatter}
\end{figure}
We now compare our results with \revise{the recent results on redMaPPer clusters presented by} \citet{Simet:2016} (hereafter S17).
S17 constrained the
mass-richness relation  
of the SDSS redMaPPer clusters, $P(\ln M| \lambda)$,
from the weak lensing measurements, \revise{using the backward modeling approach,}
and the abundance information was not included to constrain the
mass-richness relation.
In addition, \revise{S17} used an analytical model for
the lensing profile, the Navarro-White-Frenk (NFW) model \citep{Navarro:1996},
to compare with the measurements, although they instead used a narrow
range of radii ($0.3\lesssim R\lesssim 3~h^{-1}{\rm Mpc}$),
outside which the NFW profile 
ceases to be
accurate, e.g. due to 
an imperfect treatment of 
the mass profile at the transition scales 
between the 1- and 2-halo terms \citep[e.g.,][]{Diemer:2014}.
We should note
that S17 \revise{adopted} the flat $\Lambda {\rm CDM}$ model with $\Omega_{\rm m0}=0.30$
while we
used the model with $\Omega_{\rm m0}=0.3156$.
\revise{In the following comparison, we use a scaling in S17 (see their Section~6) to match their result to the model with $\Omega_{\rm m0}=0.3156$ for the definition of $M_{\rm 200m}$.
This scaling decreases their normalization by 2.5\%.
We also note that the number of the clusters after similar richness and redshift cuts in S17 (5,570)
is smaller than ours (8,312),
mainly because S17 used a conservative shape catalog cut that
removed all clusters in the Southern Galactic Cap.
Here we briefly summarize the constraint in S17.
In their model, the mean in $\log_{10}M$ 
for a fixed $\lambda$\footnote{\revise{\citet{Simet:2016} 
mention in the paper that they constrain $\langle M|\lambda\rangle$ and parameterize it as equation~(12) in their paper. However, 
inadvertently, they used that parameterization to describe $\exp[\langle \ln M|\lambda\rangle ]$ instead and
 hence their results need to be interpreted as such (M. Simet, private communication).}}
is related to the mean in $\ln M$ as
\begin{eqnarray}
\langle \log_{10} M|\lambda\rangle &\equiv& \int_{-\infty}^{+\infty} {\rm d}\ln M~ P(\ln M|\lambda) \log_{10}M \nonumber\\
                             &=& \log_{10} M_0 + \alpha \log_{10}{\left(\frac{\lambda}{40}\right)} \nonumber\\
                             &=& \frac{\langle \ln M|\lambda \rangle}{\ln{10}}.
\label{eq:simet_mean}
\end{eqnarray}
Similarly, the scatter for a fixed $\lambda$
is expressed in terms of their parameters as
\begin{eqnarray}
\sigma_{\ln M|\lambda}=\sqrt{\frac{\alpha^2}{\lambda}+\sigma_{\rm int}^2 }.
\label{eq:simet_scatter}
\end{eqnarray}
In the following we used their best-fit parameters, 
$\{\log_{10} M_{0}, \alpha, \sigma_{\rm int}\}$\footnote{\revise{Note that $\sigma_{\rm int}$ is 
quoted 
 as $\sigma_{\ln M|\lambda}$ in 
\citet{Simet:2016}.}}
are $\{14.344-0.706(\Omega_{\rm m0}-0.3), 1.33, 0.25\}$ for 
$\Omega_{\rm m0}=0.3156$.
The standard deviations
for $\log_{10} M_{0}$ and $\alpha$ are 
approximately $0.031$ (including the systematics after the quadrature sum) 
and $0.095$, 
respectively,
and we checked that 
the distributions are 
approximated by
Gaussian distributions with negligible correlation between $\log_{10} M_{0}$ and $\alpha$
based on the MCMC contour in S17.
The constraint on $\sigma_{\rm int}$ was weak and largely determined by
the flat prior of $[0.2, 0.3]$.
We will use these constraints (the Gaussian distributions 
for $\log_{10}M_0$ and $\alpha$ with the standard deviations,
and the flat prior for $\sigma_{\rm int}$) 
in Figures~\ref{fig:compSimet}, \ref{fig:simet_chains_mean_median} and \ref{fig:simet_chains_scatter}.
Note that the median is calculated for the log-normal distribution 
as $\exp(\langle \ln M|\lambda \rangle)$ and the mean for $M$ at a fixed richness is
$\langle M|\lambda\rangle=\exp(\langle \ln M|\lambda \rangle+\sigma_{\ln M|\lambda}^2/2)$.}

Based on our forward modeling results, we can compute the probability distribution $P(\ln M|\lambda)$  from $P(\ln\lambda|M)$ as
\begin{equation}
P(\ln M|\lambda) = \frac{P(\ln \lambda|M)P(\ln M)}{ \displaystyle \int_{\ln M_{\rm min} }^{\ln M_{\rm max}} {\rm d}\ln M~ P(\ln \lambda|M)P(\ln M) }.
\label{eq:bayes}
\end{equation}
Figure~\ref{fig:compSimet} compares our result for $P(\ln M|\lambda)$
with that in S17 using the best-fit parameters.
Note that $P(\ln M|\lambda)$ inferred from our model is not symmetric in the $\ln M$ space \revise{at a fixed richness}, while \revise{$P(\ln M|\lambda)$} from S17 are symmetric because S17 assumed the log-normal probability for $P(\ln M|\lambda)$ instead of $P(\ln \lambda|M)$ as in our model.
\revise{We also note that the mean $\langle M|\lambda \rangle$ is not the same as the median in both results.}
Here we show the median \revise{and the mean} of the mass-richness relation for the comparison.
Encouragingly \revise{the median and the mean relations}, denoted by the solid or dashed lines, show
a nice agreement with S17. This agreement probably reflects the fact that the average mass for a given richness bin is constrained by the measured amplitude of stacked cluster lensing profiles.
The shaded region around each line denotes the 16th and 84th percentiles of 
\revise{$P(\ln M|\lambda)$ at a fixed richness}.
Our result indicates a much larger scatter than that in S17. 
As we will show below, the scatter is constrained by the joint abundance and lensing 
information. In other words, the scatter is very difficult to constrain with either of the two observables alone. The scatter we constrain with our model should include a total contribution of the richness measurement error, intrinsic scatter, \revise{orientation effects \citep{Dietrich:2014},} 
and also possible projection effects (Eduardo Rozo for private communication).
\citet{Rozo:2014:redMaPPer2}
\citep[also see][]{Rozo:2015:redMaPPer3} studied the scatters
for
overlapping clusters between the redMaPPer clusters and the X-ray or Sunyaev-Zel'dovich clusters with high richness and low redshift, finding a smaller scatter
\revise{$\sigma_{\ln M|\lambda}=0.25\pm0.05$}. 
However, the overlapping X-ray or SZ clusters
are all massive (such as $10^{15}h^{-1}M_\odot$), and not necessarily
representative of the SDSS redMaPPer clusters.
S17 used this \revise{range for the flat prior of $\sigma_{\rm int}$ in equation~\eqref{eq:simet_scatter}, but} the constraint on the scatter \revise{was largely determined} from this prior \revise{rather than the data}. In other words, the scatter is very difficult to constrain with the lensing information alone.

The orange-color shaded region in Figure~\ref{fig:compSimet}
denotes the scatter obtained from the sample of $30\le \lambda\le
100$. 
The scatter for this sample is found to be somewhat smaller than that for the sample with $20\le \lambda\le 100$, implying 
that some contribution of the large scatter for the full sample is from 
low richness halos with $20\le \lambda\le30$. 
We will discuss a possible origin of the apparently large scatters at lower richness bin in our results.

\revise{Figures~\ref{fig:simet_chains_mean_median} and \ref{fig:simet_chains_scatter} give a more quantitative comparison of our result with S17, for 
each of the samples of $20\le \lambda \le 100$ or $30\le \lambda\le 100$, respectively.
The inner and outer shaded regions in the left panel of Figure~\ref{fig:simet_chains_mean_median}
show the 2nd, 16th, 84th and 98th percentiles
of the distribution of the mean for $\langle M|\lambda \rangle$ relation (not $\exp[\langle \ln M|\lambda\rangle ]$), which are
computed from the MCMC chains as a function of richness,
and the middle solid curve is the median of the distribution.
Similarly, the right panel of Figure~\ref{fig:simet_chains_mean_median} shows the 16th and 84th percentiles of the distribution of the median relation.
Although the left panel of Figure~\ref{fig:simet_chains_mean_median} shows that our result is 
consistent with S17 
within the confidence intervals, 
there is a mild disagreement in the mean relation,  
$\langle M|\lambda \rangle$, at low richness.
Since the mean or median of the mass-richness relation is sensitive to the ``shape'' of $\ln M$
distribution in $P(\ln M|\lambda)$ at a fixed richness, 
the mild disagreement might be ascribed to the asymmetric distribution in our model, 
while the shape in S17 is symmetric as shown in Figure~\ref{fig:compSimet}.
Alternatively, this difference might be due to the assumed cosmological models in the parameter estimation
where we employed the {\it Planck} cosmology in order to model the abundance and stacked lensing profiles.
We also found that there is some disagreement in the lensing profiles around $R \sim 3\ h^{-1}{\rm Mpc}$ between the NFW profile and our simulation-calibrated emulator at a fixed halo mass.
A further study would be needed to address the origin of the difference in more detail. 
Similarly, Figure~\ref{fig:simet_chains_scatter} compares the scatter of the mass-richness relation, $\sigma_{\ln M|\lambda}$. Since our model generally predicts a skewed distribution of halo mass for a fixed richness value in $\ln M$ space,
we compute the scatter as follows to compare with S17. First we compute the 16th and 84th percentiles 
of the mass distribution for a fixed richness as we did for the shaded region in Figure~\ref{fig:compSimet} from the MCMC chains. 
Then we assign the width as the half width of $68\%$ CL region in $\ln M$ space
to the scatter $\sigma_{\ln M|\lambda}$; $\sigma_{\ln M|\lambda}\equiv(\ln M_{84}-\ln M_{16})/2$, where $M_{84}$ and $M_{16}$ 
are masses corresponding to the 84th and 16th percentiles, respectively. Then we compute the median and 68\% interval of $\sigma_{\ln M|\lambda}$ 
from the MCMC chains, which are shown by the solid curve and the shaded region in Figure~\ref{fig:simet_chains_scatter}. 
The figure shows that our result implies a larger scatter than that implied in S17.} 

\revise{Finally} we note that $P(\ln \lambda|M)$ is generally difficult to obtain if one uses the backward modeling method. In this case, $P(\ln \lambda|M) \propto P(\ln M|\lambda)P(\ln \lambda)$ needs to be computed and it
requires a knowledge on the underlying distribution of richness
parameters from the catalog for the halos with the smaller richness (i.e. including $\lambda\le20$ for the redMaPPer catalog).
However, $P(\ln \lambda)$ is not generally available, or at least very noisy, for richness below a
threshold richness ($\lambda=20$) in the current redMaPPer cluster
catalog, because such low-richness clusters are by definition difficult
to identify due to fewer member galaxies and masking effect, and would more suffer from 
systematic effects such as the projection effect.

\section{Discussion}
\label{sec:m_discussion}
Our analysis we have so far shown involves some assumptions and
uncertainties, and in this section we discuss the possible impacts on
the results.
\subsection{Information content and complementarity 
of abundance and stacked cluster lensing}
\label{sec:info_lens_abundance}
\begin{figure}
	\centering
	\includegraphics[width=0.495\textwidth]{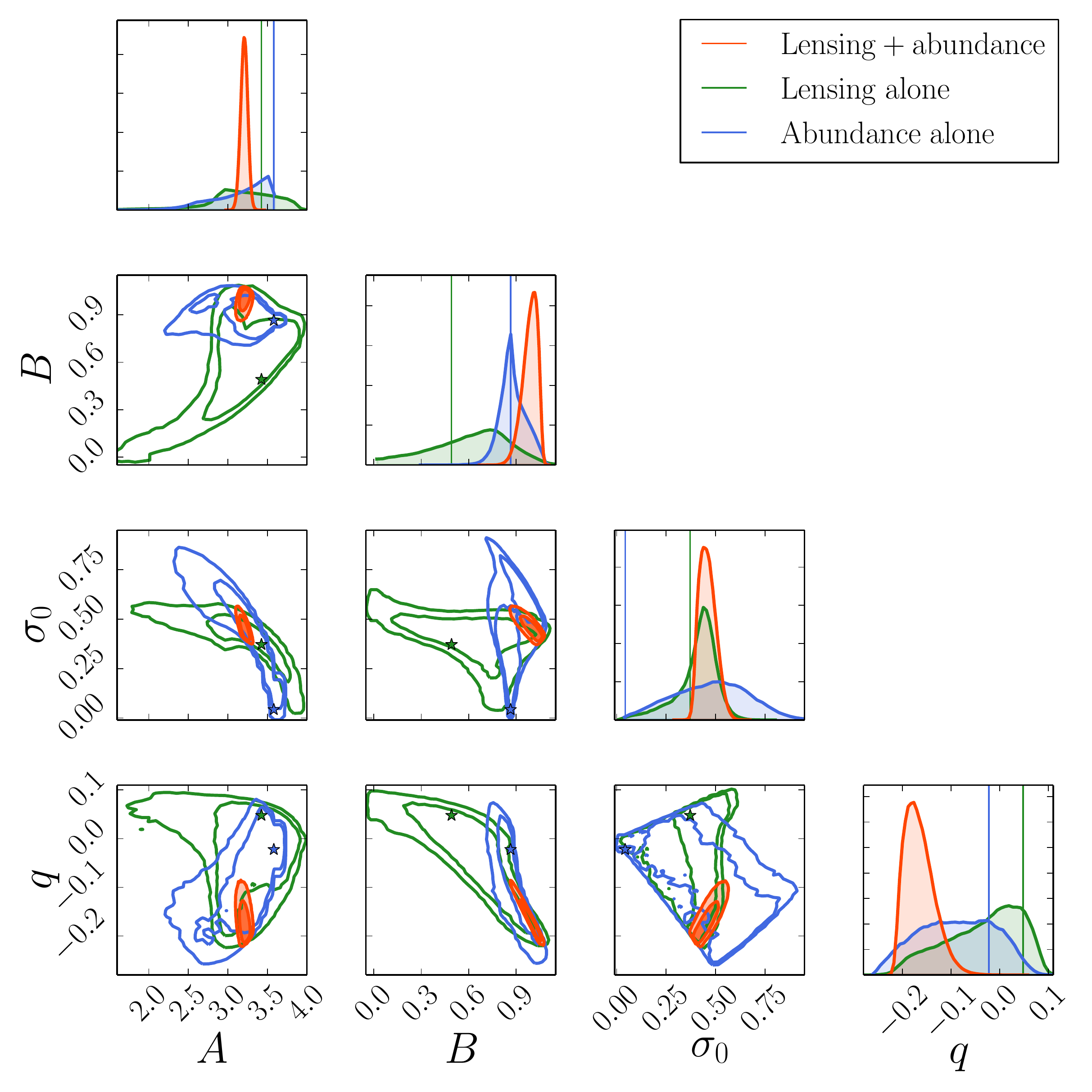}
\caption{
The posterior distribution of each parameter, and the 68\%~and 95\%~CL intervals in each two-parameter subspace for the MCMC chains
when the mass-richness relation is constrained by the model fitting to the lensing (green) or abundance (blue)
information alone. For comparison, the red contours show the results for the joint fitting in Figure~\ref{fig:fitting_MCMC}. Here we used the sample of $20\le \lambda \le 100$. 
The posterior distribution becomes much wider due to
strong degeneracies between the parameters. The star symbols in each color contours or the lines in the one-dimensional 
posterior distribution denote \revise{the} best-fit model parameters. The best-fit model for the abundance 
alone is \revise{around} the corner of the posterior distribution near the sharp bounds from the prior of $\sigma_{\ln \lambda|M}>0$ for the range of the halo mass in consideration.
 }
    \label{fig:fitting_MCMC_alone}
\end{figure}
\begin{figure*}
\centering
\includegraphics[width=0.83 \textwidth]{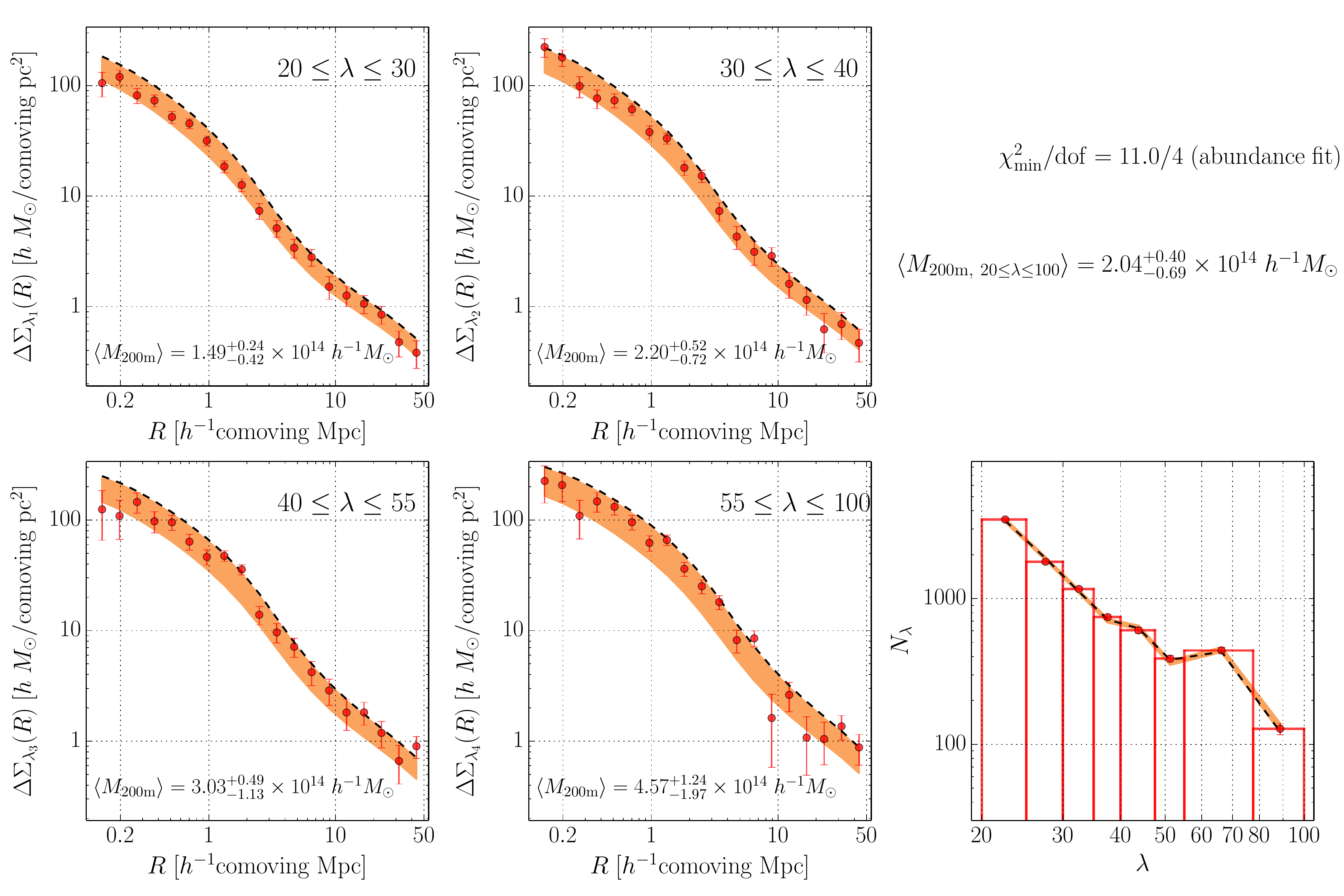}
\caption{
The left-side four panels show the model {\it predictions} for the lensing profiles computed from the MCMC chains, 
when the mass-richness relation is constrained by the abundance information alone, as shown by the blue contours 
in Figure~\ref{fig:fitting_MCMC_alone}.
The right panel shows the {\it fitting} result for the abundance, and therefore
shows that the model well reproduces the measurement. 
The orange-color shaded regions show the 16th and 84th percentiles computed from the MCMC chains. 
The \revise{black dashed} curves in the left-side four panels from the best-fit model for abundance alone appear to be \revise{around} edge of the intervals in the lensing profiles due to the skewed distribution \revise{(see Figure~\ref{fig:lenshist_abundancealone} for the detail)} of the lensing predictions that are computed from the wide posterior distributions of the parameters in Figure~\ref{fig:fitting_MCMC_alone}.
The lensing predictions from the best-fit model for abundance alone show systematically larger amplitudes than the measurement in each richness bin, 
especially for the lowest richness bin of $20\le \lambda\le 30$.
\label{fig:abundance_alone}}
\end{figure*}
\begin{figure*}
\centering
\includegraphics[width=0.83 \textwidth]{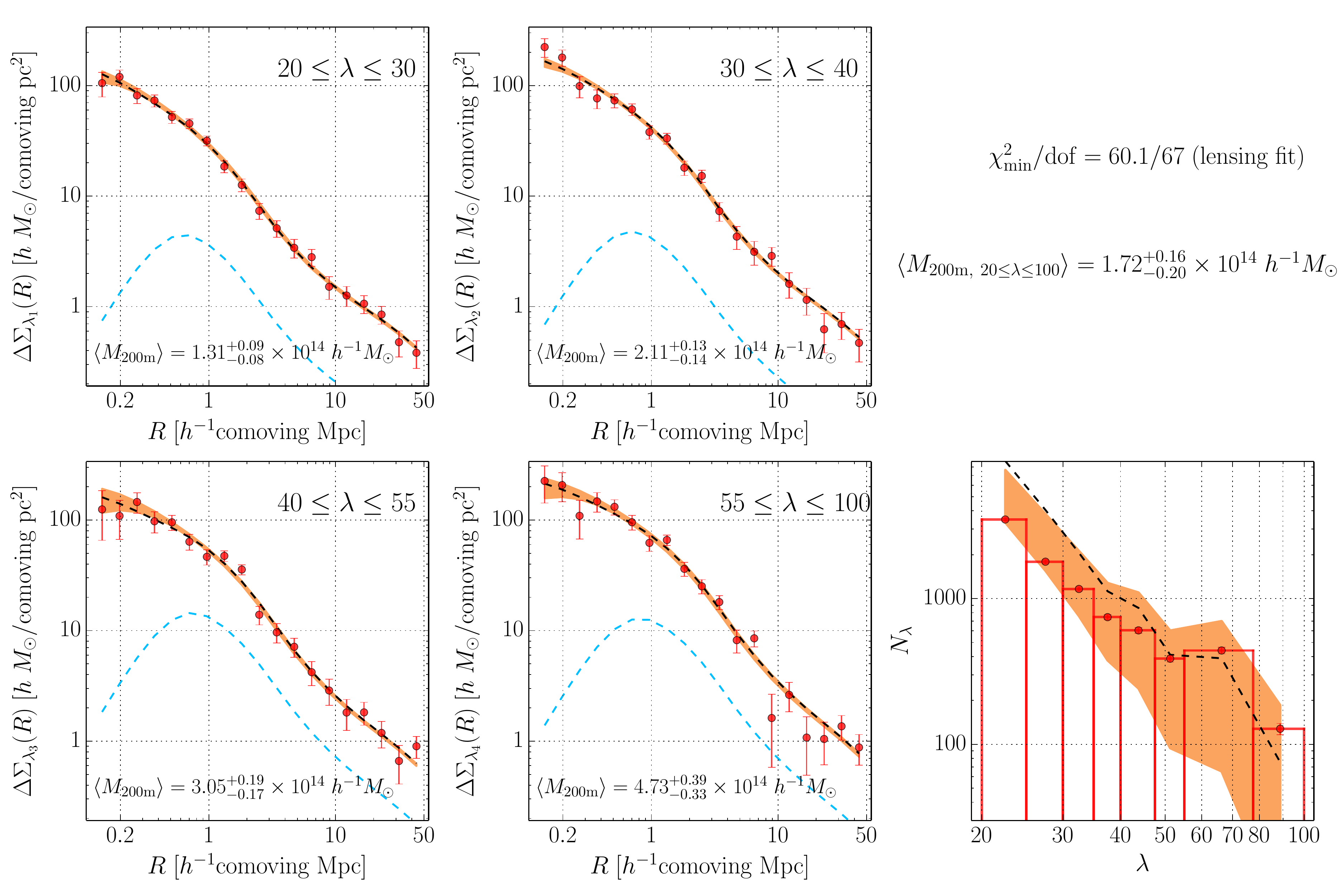}
\caption{As in Figure~\ref{fig:abundance_alone}, but the right panel shows the model predictions of the abundances in each richness bin from the MCMC chains, when the mass-richness relation is constrained by the lensing information alone.
The model inferred from the lensing measurements favors an inclusion of low richness clusters \revise{(see Figure~\ref{fig:abundancehist_lensalone} for the detail) at $20\le\lambda\le30$, although the significance is not very high.}
\label{fig:lensing_alone}
}
\end{figure*}
As we have shown, we can constrain the mass-richness relation from
a {\it joint} fitting of the model predictions to the abundance and the
lensing profiles. Firstly we study how the results are changed if using
either alone of the two observables, or in other words how 
the joint fitting helps lift the parameter degeneracies. 
Figure~\ref{fig:fitting_MCMC_alone} shows the 68\%~and 95\%~CL contours in the 
two parameter subspaces, as in Figure~\ref{fig:fitting_MCMC}.
The figure nicely shows that the two observables are complementary
to each other, and the parameter degeneracies are efficiently broken when combining
the two measurements.
Either of the abundance or lensing information alone gives a wider distribution in each 
two-parameter subspace. 
It is also interesting to find that the parameter $B$ (the halo mass dependence of 
the mean mass-richness relation) or $\sigma_0$ (the normalization of the mass-richness scatter) is relatively
better constrained by either one of the abundance or lensing information than another, respectively. 
The different sensitivities of the two observables to each parameter come from the different 
dependences on model parameters as shown in Figure~\ref{fig:parameter_dependence}.

We further study the complementarity of the two observables 
in Figures~\ref{fig:abundance_alone} and \ref{fig:lensing_alone}.
Figure~\ref{fig:abundance_alone} shows 
the results if the mass-richness relation is constrained 
by the model fitting to the abundance alone. 
Our model well reproduces the abundance measurements
in different richness bins, as shown by the lower right panel. 
The left-side four panels show the model {\it predictions} for the lensing 
profiles in each richness bin. 
The curve and the shaded regions in each panel for the lensing profile are not the fitting results, but rather 
the model predictions computed from  the blue-color 
MCMC chains in Figure~\ref{fig:fitting_MCMC_alone}. 
Note that we did not include
the off-centering parameters in the lensing model predictions, since there is no
information in the abundance that can constrain the parameters,
and we set $f_{\rm cen, \beta}=1$ to compute the model lensing profiles.
The best-fit model for the abundance, denoted by the 
\revise{black dashed}
curve, predicts a systematically larger amplitude of the lensing 
profiles than the measurements
over a range of the radial bins for each richness bin. In particular, 
the systematic offset is the largest
for the  lowest richness bin of 
$20\le \lambda\le 30$. This means that, 
 if we naively reproduce the
number of clusters at each richness bin by our model, the \revise{best-fit} model favors a
higher halo mass on average than the lensing measurements. 
As we discussed above, however, the abundance information alone suffers from severe degeneracies between 
the parameters.
The shaded orange-color regions in each panel show the 
 68\%~CL intervals, which fairly well
include the lensing measurements in each richness bin.  
This apparent agreement comes from the skewed distribution of the lensing profiles from the MCMC chains because the best-fit model is located at a corner in the broad posterior distributions of the parameters
as shown in Figure~\ref{fig:fitting_MCMC_alone}. 
Hence, the best-fit model (blue line) appears to be located around the edge of the 68\% CL interval of the 
lensing profiles in each richness bin. 
\revise{Figure~\ref{fig:lenshist_abundancealone} in Appendix~\ref{app:cluster-abundance_alone} shows 
the posterior distribution of the lensing profile at some representative radial bins for each richness bin, which clearly shows a skewed 
distribution.}

Similarly, Figure~\ref{fig:lensing_alone} shows the results if the mass-richness
relation is constrained by the lensing information alone.
In this case, we also included the off-centering parameters in the model
fitting.
Although our model perfectly reproduces the lensing profile measurements, 
the best-fit model predicts too many low-richness clusters at $\lambda\lesssim 30$. 
Even if we include the 68\% CL intervals of model predictions 
inferred from the lensing data, shown by the orange color region, 
the models tend to predict \revise{somewhat} larger abundances of low-richness clusters at \revise{$20 \le\lambda\lesssim 30$}.
This means that, 
since lower-richness clusters correspond to less massive halos, 
the measured lensing amplitudes favor to include a larger number of less
 massive halos in the low richness bins than the measured abundance. 
This result is consistent with Figure~\ref{fig:abundance_alone}. 
\revise{Figure~\ref{fig:abundancehist_lensalone} in Appendix~\ref{app:cluster-abundance_alone} shows the 
posterior distribution of the abundance in each richness bin. }

Thus, \revise{summarizing the results in Figures~\ref{fig:abundance_alone} and \ref{fig:lensing_alone},  }
the best-fit models from the abundance and lensing measurements indicate a possible tension between the two observables, for the \textit{Planck} cosmology.
\revise{However, this does not appear to be very significant given the errors.}

\subsection{Halo mass information content in the 1- and 2-halo terms of lensing profiles}
\label{sec:lensing_1h2h}

\begin{figure}
        \centering
        \includegraphics[width=0.495\textwidth]{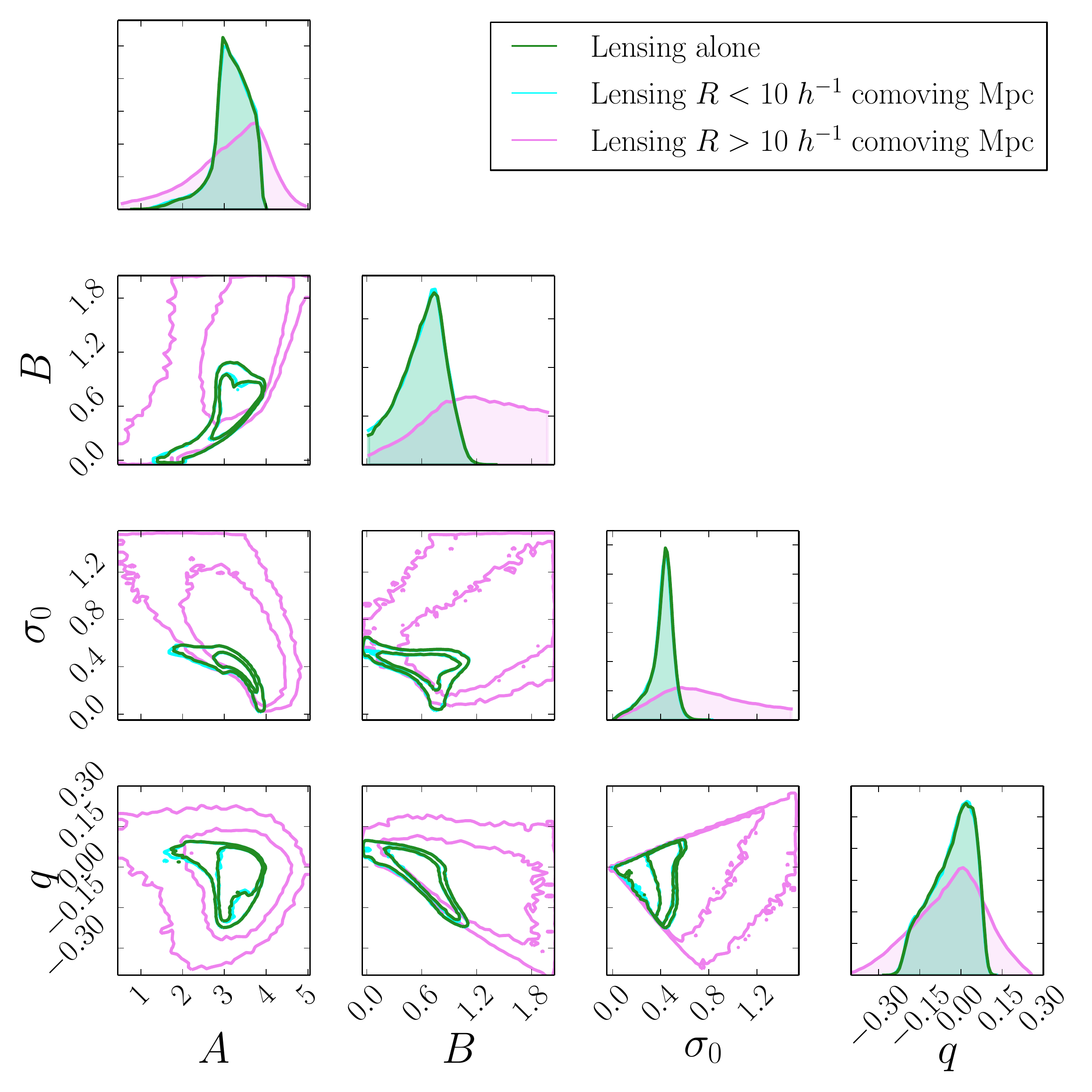}
\caption{
The posterior distribution of each parameter as well as 
the 68\% and 95\%~CL intervals in each two-parameter sub-space when
using the lensing information at either of $R< 10~h^{-1}{\rm Mpc}$ or $R>10~h^{-1}{\rm Mpc}$ 
alone, respectively, for the sample of $20\le \lambda\le 100$. 
Note that we did not use the abundance in these constraints, and 
the prior range is the same as in Table~\ref{tab:params_prior}.
For comparison, the green-color contours are the results  from Figure~\ref{fig:fitting_MCMC_alone} for the lensing alone using all the radial bins.
The constraint from $R>10~h^{-1} {\rm Mpc}$ is so weak 
that the constraint mainly comes from the prior.
The upper-left and lower-left boundaries of the contour in the $\sigma_0$-$q$ subspace 
 are from the prior with the condition $\sigma_{\ln\lambda|M}>0$ 
for the range of halo masses we consider.
 }
    \label{fig:fitting_MCMC_Rcomoving}
\end{figure}
\revise{The} weak gravitational lensing is a unique means of constraining the halo mass
in the sample. 
The halo mass information is from the two regimes
in the lensing profiles: the 1-halo term and the 2-halo term
\citep{OguriTakada:2011}, where 
the 2-halo term amplitude depends on halo
mass via halo bias, because we have fixed the background cosmological
model. 
Since the 1- and 2-halo terms have different dependences on halo mass,
combining the two information might help improve constraints on the halo mass and therefore
break the parameter degeneracies in the mass-richness relation.
These halo mass dependences are properly included in the 
halo emulator we used for the model fitting.
To study which of the 1- or 2-halo term information gives a dominant contribution to constrain the halo mass, we perform
the model fitting only by using the lensing profile information at $R<
10~h^{-1}{\rm Mpc}$ or $R>10~h^{-1}{\rm Mpc}$, 
which roughly corresponds to the transition
between the 1- and 2-halo terms for the redMaPPer cluster-scale halos.
Figure~\ref{fig:fitting_MCMC_Rcomoving} shows the marginalized constraints on the 
parameters in the mass-richness relation, when using either alone of the lensing measurements.
The constraint for the mass-richness relation comes dominantly from the lensing profiles at $R<10~h^{-1} {\rm Mpc}$. The constraint from $R>10~h^{-1} {\rm Mpc}$ comes mainly from the prior we used, 
and the mass information from the lensing at $R> 10~h^{-1}{\rm Mpc}$ is very weak for the scales up to $50~h^{-1}{\rm Mpc}$. 
Future wide and deeper surveys might allow us to use more information in the 2-halo term, and helps improve the parameter 
constraints.

\subsection{The impacts of possible systematic effects in the measurements}
\label{sec:possiblesyserr}
There might be residual systematic effects in
 the lensing and abundance measurements.
 In this subsection, we discuss the impact on the parameter estimation of the mass-richness
 relation.

An imperfect measurement of galaxy shapes or a systematic bias in
photometric redshift estimates affects the lensing amplitudes we can
measure via galaxy shapes. Following S17, we introduce an additional
parameter, $m_{\rm lens}$, 
to model
the possible residual systematic bias, and then study how this affects our estimation of the mass-richness relation parameters:
\begin{equation}
\widetilde{ { \Delta\Sigma }_{\lambda_\beta} }(R; m_{\rm lens}) = (1+m_{ \rm lens}) { \Delta\Sigma_{\lambda_\beta}}(R).
\label{eq:lensingbias}
\end{equation}
We assume a constant parameter $m_{\rm lens}$ for the lensing bias amplitudes in all the richness bins, 
and then allow $m_{\rm
lens}$ to vary in the model fitting. 
We employ a flat prior of $-0.05\le m_{\rm lens}\le 0.05$. 
Table~\ref{tab:paramsMCMC_bias} shows that the parameters in the mass-richness relation are little changed, implying 
that the joint fitting is powerful to extract the parameters or allows for a self-calibration of the systematic errors
\citep{OguriTakada:2011}.

For the abundance measurement, an estimation of the effective survey area
in equation~\eqref{eq:Omega_eff} is somewhat tricky, as it requires to
properly take into account the effects of 
survey masks and depth variations on 
cluster detection efficiency as a function of spatial positions 
in the survey footprint including the boundary regions. 
As can be found from Table~\ref{tab:binning}, the correction for the effective survey area 
is largest for the clusters in the lowest richness bin, $20\le \lambda\le 25$, about 11\% compared to the raw counts.
We study the impact of a possible uncertainty in the effective area correction on the parameter estimation by enlarging the diagonal error bars of the abundance in each richness bin by an amount of the corrections in the counts in Table~\ref{tab:binning}:
\begin{equation}
{\bf C}(N_{\lambda_\alpha},N_{\lambda_\alpha})\rightarrow 
{\bf C}(N_{\lambda_\alpha},N_{\lambda_\alpha})+
\left(N_{\lambda_\alpha}^{\rm corr}-N_{\lambda_\alpha}^{\rm raw}\right)^2.
\label{eq:abundancebias}
\end{equation}
As stated above, the second term amounts to about 11\% fractional error 
in the counts for the lowest richness bin ($20\le\lambda\le25$), 
compared to $1.7\%$ for the Poisson error (from $1/\sqrt{3488.4}$). 
Table~\ref{tab:paramsMCMC_bias} shows that each parameter is consistent with the result for the fiducial analysis in Table~\ref{tab:paramsMCMC_full}, 
to wihin the 68\% CL interval. 

In Table~\ref{tab:paramsMCMC_bias}, we also study how the parameters are changed by including both the possible systematic 
effects in the lensing and abundance measurements. Again the parameters are not largely changed. Thus, the joint fitting seems 
robust against possible systematic effects because the two observables depend on the parameters in different ways. 

\begin{deluxetable*}{cccccccc}
\tablewidth{0 pt}
  \tablecaption{
  Variations in the parameters due to different analyses for the sample of
   $20 \le \lambda \le 100$ 
  \label{tab:paramsMCMC_full}}
\tablehead{\colhead{Parameter}
& \colhead{Crude covariance}
& \colhead{ Lensing alone }
& \colhead{ Abundance alone }
& \colhead{ 1-halo lensing alone }
& \colhead{ 2-halo lensing alone }
\\
\colhead{}
  & 
& \colhead{  }
  & \colhead{ }
& \colhead{ $R < 10\ h^{-1}{\rm Mpc}$ }
  & \colhead{$R > 10\ h^{-1}{\rm Mpc}$ }
}
  \startdata
$A$  &   \revise{$3.215^{+0.036}_{-0.037}$ $(3.225)$} & \revise{$3.191^{+0.415}_{-0.364}\ (3.425)$} & \revise{$3.241^{+0.252}_{-0.482}\ (3.581)$} & \revise{$3.194^{+0.420}_{-0.379}\ (3.410)$} & \revise{$3.275^{+0.699}_{-1.077}\ (3.946)$} \\
$B$ & \revise{$0.994^{+0.037}_{-0.050}$ $(1.002)$} & \revise{$0.636^{+0.208}_{-0.301}\ (0.491)$} & \revise{$0.870^{+0.096}_{-0.066}\ (0.865)$} & \revise{$0.638^{+0.208}_{-0.309}\ (0.446)$} &\revise{$1.181^{+0.535}_{-0.543}\ (0.840)$}\\
$\sigma_0$  & \revise{$0.450^{+0.041}_{-0.035}$ $(0.440)$} & \revise{$0.429^{+0.065}_{-0.114}\ (0.372)$} & \revise{$0.483^{+0.186}_{-0.214}\ (0.043)$} &\revise{$0.425^{+0.066}_{-0.117}\ (0.365)$}&\revise{$0.723^{+0.434}_{-0.328}\ (0.017)$}\\
$q$  & \revise{$-0.170^{+0.031}_{-0.024}$ $(-0.175)$} & \revise{$-0.020^{+0.066}_{-0.101}\ (0.048)$} &\revise{$-0.087^{+0.083}_{-0.086}\ (-0.022)$} &\revise{$-0.021^{+0.065}_{-0.103}\ (0.062)$} &\revise{$-0.026^{+0.112}_{-0.148}\ (-0.001)$}\\     
$f_{\rm cen, 1}$ & \revise{$0.57^{+0.27}_{-0.36}$ $(0.78)$} & \revise{$0.66^{+0.22}_{-0.38}\ (0.86)$} &--&\revise{$0.66^{+0.22}_{-0.38}\ (0.90)$} &\revise{$0.50^{+0.34}_{-0.34}\ (0.00)$}\\
$f_{\rm cen, 2}$ & \revise{$0.81^{+0.14}_{-0.33}$ $(0.99)$} &\revise{$0.81^{+0.13}_{-0.32}\ (0.89)$}  &--&\revise{$0.81^{+0.13}_{-0.31}\ (0.89)$} &\revise{$0.50^{+0.34}_{-0.34}\ (0.70)$}\\
$f_{\rm cen, 3}$ & \revise{$0.44^{+0.32}_{-0.30}$ $(0.70)$} & \revise{$0.43^{+0.30}_{-0.29}\ (0.72)$} &--&\revise{$0.43^{+0.30}_{-0.29}\ (0.69)$} &\revise{$0.49^{+0.35}_{-0.34}\ (0.06)$}\\
$f_{\rm cen, 4}$  & \revise{$0.57^{+0.27}_{-0.35}$ $(0.70)$} &  \revise{$0.58^{+0.26}_{-0.35}\ (0.80)$}&--&\revise{$0.59^{+0.25}_{-0.35}\ (0.82)$} &\revise{$0.50^{+0.34}_{-0.34}\ (0.35)$}\\
$\alpha_{\rm off}$ & \revise{$0.065^{+0.052}_{-0.030}$ $(0.152)$} & \revise{$0.073^{+0.071}_{-0.034}\ (0.266)$} &--&\revise{$0.074^{+0.076}_{-0.034}\ (0.304)$} &\revise{$0.495^{+0.342}_{-0.334}\ (0.218)$} 
\enddata  \tablecomments{
The first column, labeled by ``Crude covariance'', shows the results for parameter estimation when 
using the covariance that includes only the shape noise contribution for the lensing profiles and the Poisson contribution for the abundances as described in Section~\ref{sec:m_mock_covariance}.
The second and third columns show the results when using either alone of the lensing or abundance information, respectively. Furthermore, 
the fourth and fifth columns show the results when using the lensing information at $R < 10~h^{-1}{\rm Mpc}$ or 
$R>10~h^{-1}{\rm Mpc}$ alone, corresponding to the 1- and 2-halo term information, in the analysis of 
the lensing information alone. The priors for the parameters are the same as in Table~\ref{tab:params_prior}. } 
\end{deluxetable*}

\begin{deluxetable*}{cccccccc}
\tablewidth{0.65\textwidth }
  \tablecaption{ 
  The impacts of possible residual systematic errors on the parameters for the sample of $20\le \lambda\le 100$
  \label{tab:paramsMCMC_bias}}
\tablehead{\colhead{Parameter}
& \colhead{+lensing sys.}
& \colhead{+abundance sys.}
& \colhead{+lensing and abundance sys.}
}
  \startdata
$A$&\revise{$3.212^{+0.054}_{-0.057}\ (3.241)$}&\revise{$3.207^{+0.043}_{-0.044}\ (3.236)$}&\revise{$3.210^{+0.055}_{-0.054}\ (3.201)$}\\
$B$&\revise{$0.991^{+0.042}_{-0.056}\ (0.988)$}&\revise{$1.017^{+0.038}_{-0.054}\ (1.030)$}&\revise{$1.014^{+0.039}_{-0.055}\ (1.044)$}\\
$\sigma_0$&\revise{$0.453^{+0.048}_{-0.039}\ (0.442)$}&\revise{$0.452^{+0.044}_{-0.034}\ (0.437)$}&\revise{$0.450^{+0.045}_{-0.036}\ (0.441)$}\\
$q$&\revise{$-0.168^{+0.036}_{-0.027}\ (-0.165)$}&\revise{$-0.179^{+0.032}_{-0.023}\ (-0.184)$}&\revise{$-0.177^{+0.034}_{-0.024}\ (-0.197)$}\\     
$f_{\rm cen, 1}$&\revise{$0.57^{+0.27}_{-0.36}\ (0.80)$}&\revise{$0.57^{+0.27}_{-0.36}\ (0.78)$}&\revise{$0.57^{+0.27}_{-0.36}\ (0.82)$}\\
$f_{\rm cen, 2}$&\revise{$0.81^{+0.14}_{-0.35}\ (0.97)$}&\revise{$0.81^{+0.14}_{-0.35}\ (1.00)$}&\revise{$0.81^{+0.14}_{-0.35}\ (0.98)$}\\
$f_{\rm cen, 3}$&\revise{$0.43^{+0.32}_{-0.30}\ (0.55)$}&\revise{$0.44^{+0.32}_{-0.30}\ (0.53)$}&\revise{$0.45^{+0.32}_{-0.31}\ (0.61)$}\\
$f_{\rm cen, 4}$&\revise{$0.57^{+0.27}_{-0.34}\ (0.63)$}&\revise{$0.57^{+0.27}_{-0.35}\ (0.62)$}&\revise{$0.57^{+0.28}_{-0.35}\ (0.69)$}\\
$\alpha_{\rm off}$&\revise{$0.064^{+0.049}_{-0.031}\ (0.117)$}&\revise{$0.062^{+0.047}_{-0.031}\ (0.113)$}&\revise{$0.062^{+0.051}_{-0.031}\ (0.151)$}\\
$m_{\rm lens}$&\revise{$-0.005^{+0.036}_{-0.031}\ (-0.013)$}&--&\revise{$-0.002^{+0.035}_{-0.033}\ (0.032)$}
\enddata  
\tablecomments{The second column shows how the parameters are changed by introducing additional 
parameter $m_{\rm lens}$ to model a possible residual, multiplicative error in the galaxy shape measurements. We employ the flat prior for $m_{\rm lens}$ as $(-0.05,0.05)$.  
The priors for other parameters are the same as in Table~\ref{tab:params_prior}. 
The third column shows the results when enlarging the errors of the abundance measurements in each richness bin by an amount of the effective area correction according to equation~(\ref{eq:abundancebias}). The fourth 
column shows the results when including the two effects.}
\end{deluxetable*}

\subsection{The impact of the non-Gaussian sample variance}

An accurate estimation of the non-Gaussian sample variance is one of challenging 
issues for ongoing and upcoming surveys. In this paper we used the 108 SDSS mock catalogs to 
estimate the error covariance matrix as described in Section~\ref{sec:m_mock_covariance}. By using the 
full-sky simulations, the covariance we estimated includes the super-sample covariance contribution that arises from 
large-scale density fluctuations over the SDSS footprint, which is difficult to accurately estimate without the full-sky
simulations. In Table~\ref{tab:params_prior}, we show how the parameter estimation is changed when we use the crude covariance  as described in Section~\ref{sec:m_mock_covariance} that includes
only the shape noise contribution for the weak lensing measurements and the Poisson noise for the abundance. The median 
and best-fit values of the model are almost unchanged compared to the fiducial result in Table~\ref{tab:paramsMCMC_full}, implying that the 
shape noise and Poisson noise give a dominant contribution to the covariance. The 68\% CL interval for each parameter from the fiducial analysis is only slightly 
enlarged, by up to about 15\% for some parameter compared to the analysis using the crude covariance. Thus, we conclude that, for the SDSS data which is a relatively shallow survey, 
the impact of the non-Gaussian sample variance is not significant.

\subsection{Projection effect in the SDSS redMaPPer catalog}
\label{sec:projectioneffect}

\begin{deluxetable}{ccc}
\tablewidth{ 0.47 \textwidth }
  \tablecaption{Parameter estimation for the more flexible model of mass-richness relation (equation~\ref{eq:newmodel}) for the sample of $20\le \lambda\le 100$
  \label{tab:paramsMCMC_newmodel}}
\tablehead{\colhead{Parameter}
& \colhead{Prior}
& \colhead{Median and error}
}
  \startdata
$A$  &   $(0.5, 5.0)$&\revise{$3.236^{+0.071}_{-0.075}\ (3.334)$}\\
$B$ &   $(-2.0, 2.0)$&\revise{$0.996^{+0.047}_{-0.058}\ (1.066)$}\\
$C$  &  $(-1.5, 1.5)$ &\revise{$0.001^{+0.035}_{-0.030}\ (-0.004)$}\\
$\sigma_0$  &  $(0.0, 1.5)$&\revise{$0.441^{+0.065}_{-0.088}\ (0.320)$}\\
$q$  &  $(-2.0, 2.0)$&\revise{$-0.158^{+0.046}_{-0.030}\ (-0.187)$}\\
$p$  &  $(-2.0, 2.0)$&\revise{$0.006^{+0.017}_{-0.023}\ (0.011)$}\\
$f_{\rm cen, 1}$ &  $(0.0, 1.0)$ &\revise{$0.56^{+0.27}_{-0.35}\ (0.81)$}\\
$f_{\rm cen, 2}$ &  $(0.0, 1.0)$ &\revise{$0.80^{+0.14}_{-0.34}\ (0.97)$}\\
$f_{\rm cen, 3}$ &  $(0.0, 1.0)$ &\revise{$0.43^{+0.32}_{-0.30}\ (0.56)$}\\
$f_{\rm cen, 4}$  &  $(0.0, 1.0)$ &\revise{$0.58^{+0.28}_{-0.35}\ (0.70)$}\\
$\alpha_{\rm off}$ &  $(10^{-4}, 1.0)$ &\revise{$0.062^{+0.049}_{-0.031}\ (0.134)$}  
\enddata  
\tablecomments{We increased the range of priors for $B$ and $q$ from Table~\ref{tab:params_prior} to be more flexible with $C$ and $p$. In addition, we impose the condition $\sigma_{\ln \lambda|M}>0$ over the range of halo masses we consider; $10^{12}\le M/[h^{-1}M_\odot]\le 2\times 10^{15}$.}
\end{deluxetable}

Our fitting result in Figure~\ref{fig:joint_all} indicates that 
very low mass halos ($M \lesssim10^{13}\
h^{-1}M_{\odot}$) contribute to the sample in 
the lowest richness bin ($20 \leq
\lambda \leq 30$) by about 10\% fraction.
As one of the possibilities, this result may indicate
the projection effects in the SDSS redMaPPer clusters; i.e.
multiple halos along the line-of-sight direction are misidentified as one cluster after projection \citep{Cohn:2007}.  

As we showed in \revise{Figures}~\ref{fig:compSimet} \revise{and \ref{fig:simet_chains_scatter}}, 
our model favors a large scatter in mass at a fixed richness.
There are signatures on the existence of such low mass halos from both the lensing and abundance information as described in Section~\ref{sec:info_lens_abundance}.  
If we naively reproduce the abundance of low richness halos just by the mass-richness relation in our model, 
the model predicts greater amplitudes of the lensing profiles than the measurements (Figure~\ref{fig:abundance_alone}). 
Similarly, if we naively reproduce the lensing amplitudes in the low richness bin, the model 
predicts \revise{somewhat} large abundance of low-richness clusters that correspond to less massive halos.
To reconcile these discrepancies with our model within the \textit{Planck} cosmology, we need to introduce a relatively large
scatter in the mass-richness relation, which yields an inclusion of such low mass halos into the sample, especially 
for the sample of $20\le\lambda\le 100$.
To arrive at these results, we assume a single population of the underlying clusters and the distribution of richness
parameters at a fixed halo mass obeys a log-normal distribution.
In some sense the scatter amplitude we found could be understood as the total contribution that includes
the intrinsic scatter, the richness estimation errors, and a possible contamination of the projection effect.
To study the impact of the projection effects, we need mock catalogs that are carefully designed and built 
in redshift space, using light-cone 
simulations, in order to estimate how the projection effects affect 
the redMaPPer cluster finder as well as the lensing 
and abundance measurements. This is beyond the scope of this paper, and we will study this elsewhere (see below for further discussion). 

\begin{figure}
        \centering
        \includegraphics[width=0.495\textwidth]{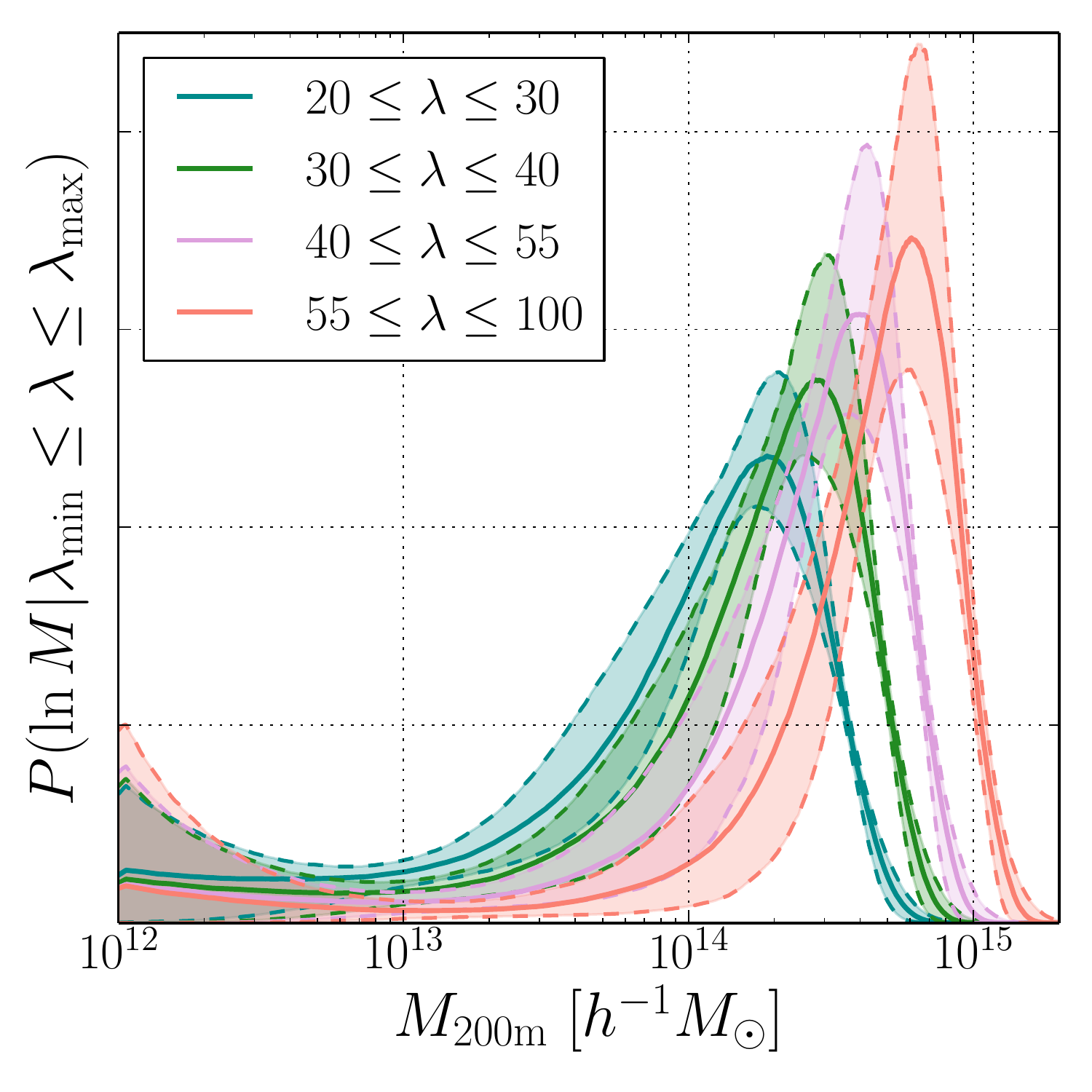}
\caption{
Similarly to the right panel of Figure~\ref{fig:joint_all},  but from the more flexible model.
This more flexible model also favors too low mass contribution ($M_{\rm 200m}\lesssim10^{13}h^{-1}M_{\odot}$) as in the result of the fiducial model in Figure~\ref{fig:joint_all}. 
 }
    \label{fig:massdist_moreparams}
\end{figure}

Instead we here study how a more flexible model of the mass-richness relation changes our results within the log-normal model for $P(\ln \lambda|M)$. 
To study this, we introduce more free parameters to model the mean of the mass-richness relation as well as
the scatters from equations~\eqref{eq:mean_relation} and \eqref{eq:scatter_M}:
\begin{eqnarray}
&&\langle \ln \lambda\rangle (M) = A+B \ln\left(\frac{M}{M_{\rm pivot}}\right)+
C \left[\ln\left(\frac{M}{M_{\rm pivot}}\right)\right]^2,\nonumber\\
&&\sigma_{\ln\lambda|M}=\sigma_0 + q\ln\left(\frac{M}{M_{\rm pivot}}\right)
+p\left[\ln\left(\frac{M}{M_{\rm pivot}}\right)\right]^2,
\label{eq:newmodel}
\end{eqnarray}
The parameters $C$ and $p$ model nonlinear halo mass dependences in the 
$\ln\lambda$-$\ln M$ space for  the mean relation and the scatter, respectively. 
Table~\ref{tab:paramsMCMC_newmodel} summarizes the results of parameter estimation,
showing that each parameter  
is consistent with the result in Table~\ref{tab:params_prior} for the fiducial analysis to within its 68\% CL interval.
The parameters $C$ and $p$ are consistent with zero to within the error bars, 
and thus we do not find a strong evidence on the nonlinear mass dependence in the mean mass-richness relation and its scatter. 
The minimum chi-square value of the more flexible model is \revise{$\chi^2_{\rm min}=75.2$}, which differs from that of the fiducial analysis only by \revise{$\Delta \chi^2_{\rm min}=0.4$} (\revise{$\chi^2_{\rm min}=75.6$} for the fiducial model),
suggesting that our fiducial model has a sufficient flexibility to model the mass-richness relation for the redMaPPer clusters, for the \textit{Planck} cosmology. 
\revise{Figure~\ref{fig:massdist_moreparams}} shows the distribution of halo masses for each richness bin, obtained from the MCMC chains for the more flexible model. Even this model favors an inclusion of low mass halos ($M\lesssim 10^{13}h^{-1}M_{\odot}$) into the sample via a convolution of the 
mass-richness relation with the halo mass function similarly to what we found in Figure~\ref{fig:joint_all}.

\section{Conclusion}
\label{sec:conclusion} 

In this paper, we have developed a method \revise{to constrain} the mass-richness
relation from the joint fitting to the abundance and the lensing profiles 
in different richness bins using the forward \revise{modeling} approach,
where we model the
probability distribution of $\lambda$ at a given mass $M$, $P(\ln\lambda|M)$.
In contrast the backward approach models the probability distribution of mass at a given richness, 
$P(\ln M|\lambda)$, which is
often employed in
the previous works \citep[e.g.,][]{ Baxter:2016, Simet:2016, Melchior:2016}.
The forward method allows for a more direct comparison
of the model prediction with the measurements. 

To accurately model the cluster observables, we have used the halo emulator 
\revise{to interpolate} the halo mass function and the stacked lensing profile as a function of halo mass
\revise{and} redshift \revise{for the {\it Planck} cosmology}. 
We developed \revise{a pipeline that allows a sufficiently fast computation of the cluster observables based on the emulator in order to perform parameter estimation using the MCMC technique.}
We applied this method to
the SDSS redMaPPer clusters, and constrained 
the mass-richness relation 
for the \textit{Planck} cosmology.

We showed that, if we employ the log-normal distribution model for $P(\ln \lambda|M)$, our model 
can well reproduce both the abundance and the lensing profiles in different 
richness bins simultaneously for the \textit{Planck} cosmology 
(see Table~\ref{tab:params_prior} and Figure~\ref{fig:fitting_MCMC}). 
We found that, in these constraints, the lensing and 
abundance information are complementary to each other, and the combination efficiently lifts 
the parameter degeneracies as shown in Figure~\ref{fig:fitting_MCMC_alone}. 
Our method allows us to estimate the probability in the backward method, $P(\ln M|\lambda)$, by transforming
the best-fit model of $P(\ln \lambda|M)$ weighted with 
 the halo mass function, based on the Bayes theorem (equation~\ref{eq:bayes}).
We showed that our \revise{result} is consistent with \revise{that} of the median \revise{and the mean} relation in \citet{Simet:2016} (see \revise{Figures}~\ref{fig:compSimet} \revise{and \ref{fig:simet_chains_mean_median}}), which was estimated 
based on the backward method using the lensing information alone. 

However, the models preferred in our method predict a contribution of less massive halos to the sample. When we use the cluster sample with richness $20\le \lambda\le 100$,
the best-fit model requires a non-negligible contribution ($\simeq 10\%$) of halos with
$M\lesssim10^{13}\ h^{-1}M_{\odot}$ to the sample (see Figure~\ref{fig:joint_all}). 
We also showed that the \revise{best-fit} parameters from either  of the abundance or the lensing profiles alone are slightly inconsistent
with each other (see Figures~\ref{fig:abundance_alone} and \ref{fig:lensing_alone}).
The contributions of too low mass halos are unphysical since such
halos cannot be recognized as massive clusters of galaxies.
The unphysical contribution might be due to the fact that \revise{the} \textit{Planck}
cosmology we have assumed throughout this paper is different from the underlying true cosmology 
of the universe. This would be interesting to further explore. However, 
we think that a more likely origin is due to 
residual systematic effects in the redMaPPer cluster catalog as indicated by the recent several works
\citep{Zu:2016,BuschWhite:17} \revise{(also see Sunayama, More et al. in preparation)}. 
Hence we need to further study the nature of the SDSS redMaPPer clusters. 

Our analysis involves several assumptions. The most critical one is that we assumed that the redMaPPer clusters obey statistical isotropy. More exactly, 
to compute the cluster observables, we first compute the three-dimensional 
mass function and the halo-matter cross-correlation functions from $N$-body 
simulation outputs and then constructed the model predictions of the cluster observables, 
the abundance and the lensing profiles, by projecting the three-dimensional model ingredients
along the line-of-sight direction.
These procedures are violated if the redMaPPer clusters are affected by
the projection effects, such as a mis-identification of different halos (clusters)
along the line-of-sight direction as one halo (cluster). 
The amount of the projection effect would depend on the orientation of a halo shape
and/or the surrounding large-scale structure 
with respect to the line-of-sight direction. 
In addition the projection effects would be more significant
for low-richness clusters, because low mass halos have a higher chance of the projection effect
due to their larger abundances, and have the larger measurement errors in the richness estimation due to 
fewer member galaxies. 
The larger scatters at low mass bins we found in this
work might be a signature of the projection effects.

In order to properly address the projection effects, we need to use the
mock catalogs of the SDSS redMaPPer clusters in the light cone
simulation. For this purpose, the forward method we developed in this
work would be very useful. Firstly, with the initial guess of the 
mass-richness relation $P(\ln \lambda|M)$, we can populate 
hypothetical members galaxies into each halo from their mass $M$ in a
light-cone realization. This richness assignment to halos in the simulation is difficult for the
backward method if using $P(\ln M|\lambda)$. 
Secondly, we project
the mock clusters along the line-of-sight direction to re-define a
``hypothetically-observed'' richness of each cluster on the sky, based on the SDSS
redMaPPer algorithm (circular aperture and redshift width), where multiple
halos can be merged into one detection to have the summed
richness parameter if the halos are aligned along the line-of-sight direction within
a circular aperture ($\simeq 1\ h^{-1}{\rm Mpc}$ radius) in equation~\eqref{eq:aperture} on the sky. Then, we can make the hypothetical
measurements of the lensing profiles and the abundances from the mock
catalogs of the SDSS redMaPPer clusters in the simulations. If the
obtained lensing profiles and the abundances show a deviation from the
actual measurements, we can perturb the input mass-richness relation, and
perform the above procedures again. Such an iterative method enables us to estimate a more accurate model for the mass-richness relation mitigating the projection effects.  We believe that this is doable with the forward method, and this is our future work.

\revise{Our method further assumes} that the cluster observables depend only on halo mass, and the redMaPPer clusters 
are a representative sample of the underlying halos in terms of their respective masses. Properties of 
clusters might also depend on a secondary parameter besides mass, such as the assembly history of each cluster -- the so-called 
assembly bias \citep{Miyatake:2016,More:2016}. If the redMaPPer clusters are affected by the assembly bias, the constraints on the mass-richness
relation would be biased, even if there is no projection effect. This would complicate the method in both theory and observation sides. 
On theory side, we need to properly
take into account such an assembly bias effect in the model prediction. 
On observation side, a secondary parameter besides richness needs to be inferred for each cluster, in order
to track the possible effect in individual cluster basis. For example, the concentration parameter, if estimated from the data itself, 
can be a good proxy of the assembly bias effect. However, such a secondary parameter is difficult to estimate, or at least it causes an additional scatter in relating the observables to the properties of the underlying halos. This also needs to be studied carefully.

The method and results shown in this paper are the first step towards the use of optical cluster for attaining high-precision cluster based cosmological information. Luckily, there are various kinds of cluster observables: X-ray, the SZ effect, the auto-correlation function, the redshift-space distortions \citep{Okumuraetal:17}, and the cross-correlation with other populations of galaxies. 
There is a strong hope that we can combine
these observables to disentangle cosmological information from  systematic/astrophysical effects.
The forward modeling approach used in this paper would be useful for such a study.

\section{Acknowledgment}
We thank Rachel Mandelbaum for making the shape catalog available to us. 
We thank Nick Battaglia, \revise{Eric Baxter}, Matteo Costanzi, Masamune Oguri, Eduardo Rozo, Eli Rykoff, Emmanuel Schaan \revise{and Melanie Simet} for useful discussion.
\revise{We also thank the anonymous referee for helpful comments that improved the quality of this work.}
RM and KO acknowledge supports from Advanced Leading Graduate Course for Photon Science, and RM, MS and KO acknowledge supports from Research Fellowships of the Japan Society for the Promotion of Science for Young Scientists. 
RM also acknowledges a financial support from the University of Tokyo-Princeton strategic partnership grant, and greatly thanks Prof.~David Spergel and members at the cosmology group of the Princeton University for their warm hospitality during his stay, where
this 
work was initiated. 
This work was in
part supported by Grant-in-Aid for Scientific Research from the JSPS
Promotion of Science (No.~23340061, 26610058, 15H03654,  \revise{16H01089,} 17J00658, 16J01512, 17K14273, and 17H01131), MEXT
Grant-in-Aid for Scientific Research on Innovative Areas (No.~15H05887, 15H05893,
15K21733, and 15H05892) and by JSPS Program for Advancing Strategic
International Networks to Accelerate the Circulation of Talented
Researchers. TN  acknowledges financial support from JST CREST Grant Number JPMJCR1414.
HM is supported by the Jet Propulsion Laboratory, California Institute of Technology, under a contract with the National Aeronautics and Space Administration.
Numerical computations presented in this paper were in part carried out on the general-purpose 
PC farm \revise{and Cray XC30} at Center for Computational Astrophysics, CfCA, of National Astronomical Observatory of Japan.

\bibliographystyle{apj}
\bibliography{thesis}

\vspace*{0em}

\appendix

\section{Boost factor}
\label{app:boost}

\begin{figure}[h]
	\centering	
    \includegraphics[width=0.45 \textwidth]{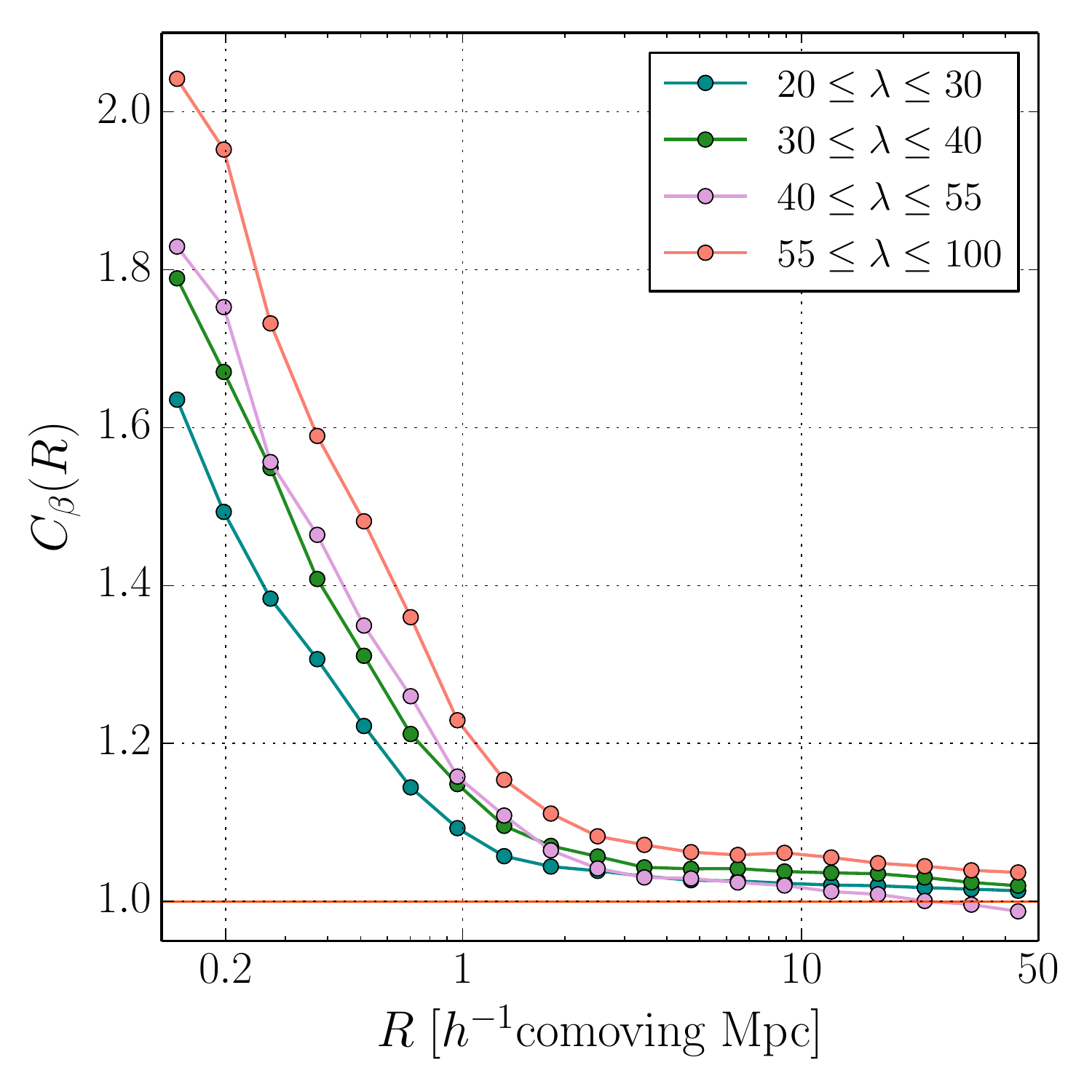} 
\caption{ The measurement of the boost factor according to equation~\eqref{eq:boost}. 
} 
\label{fig:boostfactor}
\end{figure}

Here we show the boost factor that is used to correct for a possible contamination of member galaxies 
in the source catalog for the stacked cluster lensing measurements (equation~\ref{eq:boost}).
Figure~\ref{fig:boostfactor} shows the radial profile of the estimated boost factor for each 
richness bin. The boost factor correction is significant at radii $R\lesssim 1~h^{-1}{\rm Mpc}$ 
for all richness bins, and becomes small at the larger radii.

\section{Covariance estimation}
\label{app:cov_est}

Motivated by the theory for the covariance matrix of cluster abundance and stacked lensing
\citep{HuKravtsov:03,TakadaBridle:07,OguriTakada:2011,TakadaHu:13,Shirasaki:2016,Singhetal:16},
we use the 108 realizations of SDSS mock catalogs to estimate the
covariance matrix for the lensing profiles and the abundance by breaking
down the matrix into different contributions:
\begin{eqnarray}
{\bf C}(D_i,D_j)&\equiv&\langle D_i D_j \rangle-
\langle D_i  \rangle\langle D_j \rangle\nonumber\\
&=&{\bf C}^{\rm SN}+{\bf C}^{\rm SV}-{\bf C}^{\rm R},
\label{eq:full_cov}
\end{eqnarray}
where ${\bf C}$ is the full covariance, ${\bf C}^{\rm SN}$ is the
covariance matrix arising from the shape noise contribution, ${\bf
C}^{\rm SV}$ is the sample variance contribution, and ${\bf C}^{\rm
R}$ is the term corresponding to the random catalog subtraction in \revise{Section~2.3 of} \citet{Singhetal:16}. 
Here, ${\bf D}$ is the data vector that consists of the
lensing profiles and the abundance in different radial and richness bins
(see Table~\ref{tab:binning}):
\begin{equation}
{\bf D}
\equiv \left\{\Delta\Sigma_{\lambda_1}(R_1),
... \Delta\Sigma_{\lambda_1}(R_{19}),
... \Delta\Sigma_{\lambda_4}(R_{19}), N_{\lambda_1} , ... N_{\lambda_8}
\right\},
\end{equation}
and $D_i$ denotes $i$-th component. The dimension of the data vector is
84 for the cluster sample of $20\leq\lambda\leq100$, while 63
for the sample of $30\leq\lambda \leq100$, because we use 19 radial
bins in each of four or three richness bins for the lensing profiles, and eight or six
richness bins for the abundance.
In the following, we describe how we estimate each term in
equation~\eqref{eq:full_cov}.

\subsection{Shape noise contribution:  ${\bf C}^{\rm SN}$}
The intrinsic shape causes statistical uncertainties in the stacked
cluster lensing measurement, due to a finite number of
source-cluster (lens) pairs used in the analysis. For the relatively
shallow SDSS data, where a typical number density of source galaxies is
about $1~$arcmin$^{-2}$, the shape noise gives a dominant source of the
covariance matrix components involving $\Delta\Sigma_{\lambda_\beta}$. To estimate the
shape noise contribution, we use the {\it real} catalog of source
galaxies as well as the redMaPPer clusters, because the real catalog includes
various observational effects such as spatial variations of data quality
and masks. We estimate the shape noise contribution as follows. Firstly,
we randomly rotate the ellipticity orientation of each galaxy, in order
to erase the real lensing signal. Then we measure the
stacked lensing profile for the redMaPPer clusters, in each richness
bin, in the exactly same manner as we did in the actual measurement,
including the corrections of boost factor and photo-$z$ errors
in Section~\ref{sec:lensing_measurement}. Note that we used the \textit{Planck}
cosmology in this estimation and therefore use the exactly same number of
pairs of source galaxies and redMaPPer clusters as in the actual
measurement.
Using \revise{10,000} realizations of the lensing profile measurements after random rotations, 
we estimate the shape noise contribution to the covariance matrix as
\begin{equation}
 {\bf C}^{\rm SN}(D_i, D_j) = \frac{1}{N_{\rm SN}-1} \sum_{a=1}^{N_{\rm
 SN}}[ D^{\rm SN}_{a,i}  - \overline{D}{}^{\rm SN}_{i}][
 {D}^{\rm SN}_{a,j}
 - \overline{D}{}^{\rm SN}_{j}], 
\label{eq:cov_sn}
\end{equation}
where \revise{$N_{\rm SN}=10,000$}, quantities with superscript ``SN'' denote the
measurements from the realization after random rotation of
individual galaxy ellipticities, $D^{\rm SN}_{a,i}$ is the lensing measurement at
the $i$-th lensing bin from the $a$-th realization, i.e. $D^{\rm SN}_{a,i}\equiv
\widehat{\Delta\Sigma}_{(a)\lambda_\beta}(R_j)$, and $\overline{D}{}^{\rm SN}_i$ is the
average, defined as $\overline{D}{}^{\rm SN}_i\equiv (1/N_{\rm
SN})\sum_{a}D^{\rm SN}_{a,i}$. 
\revise{The estimate of the shape noise covariance is clean since the expected noise in its inverse is at the level of 1\% based on \citet{Hartlapetal2007} from $N_{\rm SN}=10,000$ and the total number of richness and radial bins (76 for the sample of $20<\lambda<100$).}
The cross-covariance between the abundance and the shape noise term is
vanishing because of no correlation between the shape noise and the abundance
\citep{TakadaBridle:07,OguriTakada:2011}.  Hence we ignore the cross-covariance between shape noise and abundance.

\subsection{Sample variance: ${\bf C}^{\rm SV}$}

The sample variance arises due to an imperfect sampling of the
fluctuations in large-scale structure from a finite survey volume. This
contribution itself depends on the statistical nature of large-scale
structure, and therefore on cosmology.
To estimate the sample variance effects on the cluster observables, we use
the 108 realizations of the SDSS mock catalogs as follows.  Firstly, we
insert each source galaxy into the light-cone simulation according
to the angular positions and best-fit photo-$z$, and then simulate
lensing distortion effect on the galaxy due to foreground structures.
Here we ignore intrinsic shapes when creating the mock catalogs.
\revise{Secondly}, we use the halos with assigned hypothetical richness to calculate the covariance for abundance and lensing profile in each realization as below.
 
There are various contributions to the sample variance; the Gaussian
contribution arising from products of two-point correlation functions of
lensing fields and cluster distribution, and the non-Gaussian
contribution arising from the four-point functions of matter and cluster
fields at the lens redshift
\citep{TakadaBridle:07,TakadaHu:13,Gruenetal:15,Shirasaki:2016}.  It is
still difficult to study these different contributions separately
(Takahashi, Takada et~al. in prep.). 

In this paper, 
we use the jackknife method (JK) to estimate the sample
variance.
The JK method is one of the statistical techniques \citep{Efron:1982},
which has also been applied to an estimation of the covariance matrix
for the galaxy (cluster)-galaxy weak lensing measurements 
\citep[e.g.][]{Mandelbaum:2013, Coupon:2015, Miyatake:2016,
Clampitt:2016}. As carefully shown in \citet{Shirasaki:2016}, the JK
method gives a fairly accurate estimation of the underlying covariance
matrix. 

We combine the 108 realizations to estimate the sample variance
contribution to the covariance matrix based on the JK method, as
follows. Firstly, we estimate the covariance matrix from the $\alpha$-th
realization; (1) Divide the hypothetical SDSS survey region into
different $N_{\rm sub}$ subregions, where the area of each subregion is
roughly equal. (2) Measure the abundance and the stacked lensing profiles
for mock SDSS redMaPPer clusters in each richness bin, from the survey region
excluding the $\beta$-th JK subregion. (3) Repeat the (2) measurement for
all the $N_{\rm sub}$ subsamples and build the $N_{\rm sub}$ JK
resamples of the measurements. Then we estimate the covariance for the
$\alpha$-th realization as
\begin{eqnarray}
{\bf \bf C}^{\rm JK}_\alpha (D_{i}, D_j)& =& \frac{N_{\rm sub}-1}{N_{\rm sub}}  
\sum_{\beta=1}^{N_{\rm sub}}
\left[D^{\rm SV}_{(\beta),i}-\overline{D}{}^{\rm SV}_{i}\right]
 \left[D^{\rm SV}_{(\beta),j}-\overline{D}{}^{\rm SV}_{j}\right],\nonumber\\
\end{eqnarray}
where quantities with superscript ``SV'' denotes the measurements
from the mock catalog without shape noise for the lensing profiles and the number counts for the abundance, $D^{\rm SV}_{(\beta)i}$ is
the data vector measured from the survey region excluding the $\beta$-th
JK subregion, and $\overline{D}{}^{\rm SV}_i$ is the averaged
measurement for the entire survey region (without excluding any JK
subregion).  In this paper, we use 83 subdivision of the SDSS survey
footprints, i.e. $N_{\rm sub}=83$, following \citet{Miyatake:2016}
\citep[also see the middle panel of Figure~2 in][]{Shirasaki:2016}.
Then we combine the JK covariances from the 108 realizations
to estimate
the sample variance contribution to the covariance matrix:
\begin{equation}
{\bf C}^{\rm SV}(D_i, D_j)= \frac{1}{N_{\rm r}} \sum_{
\alpha=1}^{N_{\rm r}} {\bf \rm C}^{\rm JK}_{\alpha} (D^{\rm
SV}_{i}, D^{\rm SV}_j),
 \label{eq:cov_sv}
\end{equation}
where $N_{\rm r}=108$. With the JK subsamples ($N_{\rm sub}=83$) of each
mock catalog for the 108 realizations, we effectively use about 9000
realizations to estimate the sample-variance covariance matrix for the
data vector of 84 components, and therefore we believe that the matrix
is accurately estimated.

\begin{figure*}
	\centering
   \includegraphics[width=0.90 \textwidth]{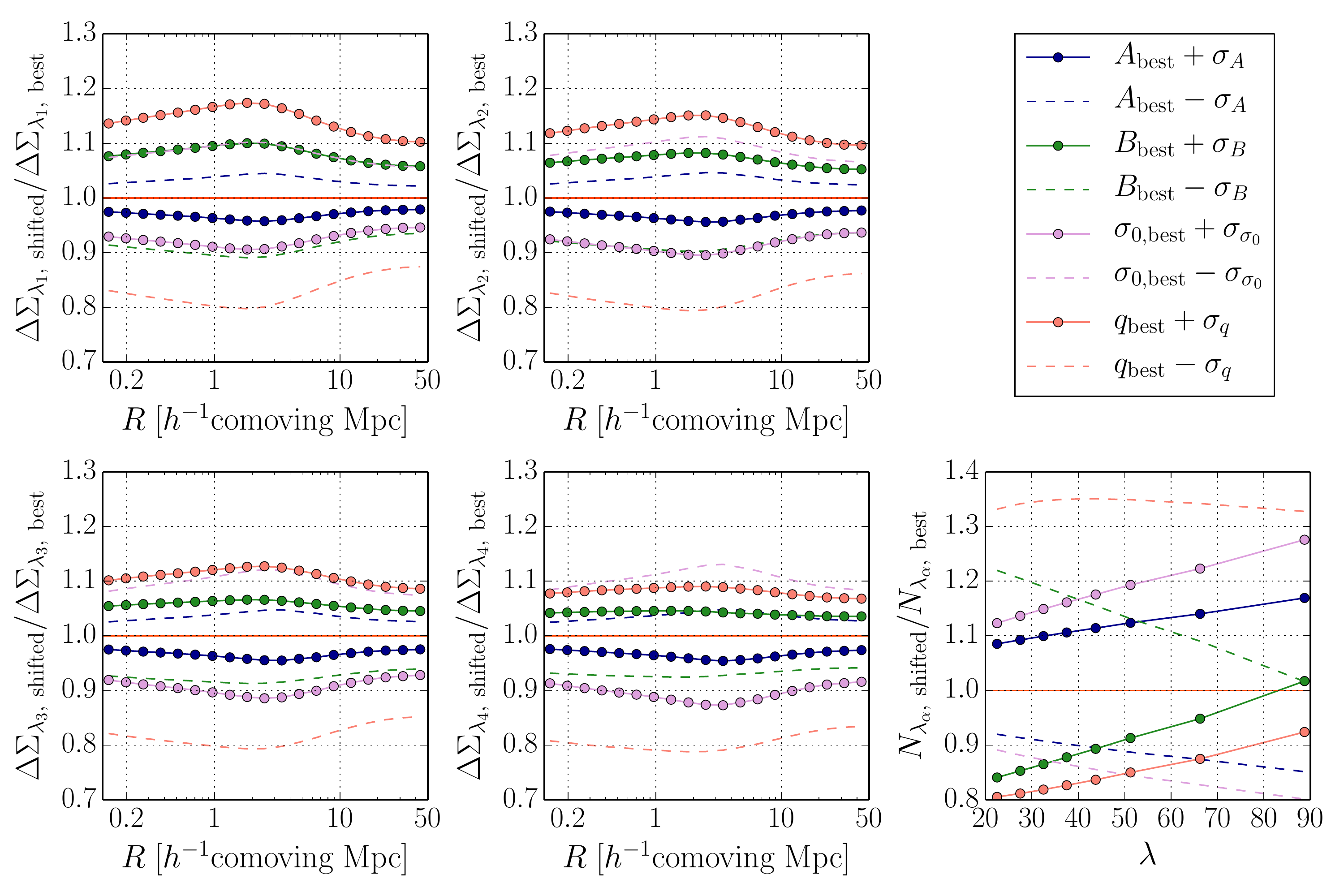}
\caption{
Each panel shows a ratio in each of the lensing profile and the abundance, caused by shifting each model parameter of the mass-richness relation by an amount of $\pm 1\sigma$ error (a half width of 68\% CL region) from the best-fit value in analysis of $20\le\lambda\le100$ (see Table~\ref{tab:params_prior}), with fixing 
other parameters to their best-fit values. Note that the best-fit value of $q$ (see equation~\ref{eq:scatter_M}) is {\it negative}, meaning 
that the scatter increases with decreasing halo mass from the pivot mass scale, $M_{\rm pivot}=3\times 10^{14}h^{-1}M_\odot$. Hence, when we shift the $q$ value by an amount of $+\sigma_q$, it reduces 
the scatter, which yields greater amplitudes in the lensing profiles and reduces the abundances due to a less up-scatter of low mass halos into a given richness bin. Each 
parameter changes the cluster observables in the different richness bins and the different radii in a complex way. 
 \label{fig:obs_params_dep} 
}
\label{fig:parameter_dependence}
\end{figure*}

However, we make a further correction to the sample variance estimation.
The cosmological model assumed in the actual measurement 
and the model fitting is the \textit{Planck} cosmology, which is different from the
\textit{WMAP} cosmology \citep{Hinshaw:2013} used in making the mock catalogs of SDSS data. To correct
for the systematic shift in the covariance amplitudes caused by this
difference of cosmological models, we scale the sub-matrix of the
sample-variance covariance matrix as follows.
In particular this effect is significant for the abundance: the \textit{WMAP}
cosmology has smaller values of $\sigma_8$ and $\Omega_{\rm m0}$ than
those in the \textit{Planck} cosmology, and therefore gives about 7,000 redMaPPer
clusters in each mock catalog, compared to 8,312 in the data.  For the
sub-covariance matrix involving the abundances, ${\bf C}^{\rm
SV}(N_{\lambda_\alpha},N_{\lambda_{\alpha'}})$, we use the formula in \citet{HuKravtsov:03} to correct for
the cosmology difference. We first subtract  $N_{\lambda_\alpha}^{\it WMAP} \delta^K_{\alpha \alpha'}$ as the Poisson noise, where $N_{\lambda_\alpha}^{\it WMAP}$ is the average over the 108 realizations and $\delta^K_{ \alpha \alpha'}$ is Kronecker delta, from the sub-covariance matrix, and then multiply the remaining term by a factor of 
$(N_{\lambda_{\alpha} }^{\textit{Planck} }N_{\lambda_{\alpha'} }^{\textit{Planck} })/(N_{\lambda_{\alpha} }^{\it WMAP}N_{\lambda_{\alpha'}}^{\it WMAP})$
in each richness bin, where $N_{\lambda_{\alpha} }^{\textit{Planck} }$ is the best-fit model value from the crude covariance described in Section~\ref{sec:m_mock_covariance}. 
After this correction, we add the Poisson term for \textit{Planck} cosmology,
$N^{\it Planck}_{\lambda_\alpha}\delta^K_{\alpha \alpha'}$. Similarly, for the cross-covariance matrix, ${\bf
C}^{\rm SV}(N_{\lambda_\alpha}, \Delta\Sigma_{\lambda_\beta})$, we multiply a factor of $N_{\lambda_\alpha}^{\it
Planck}/N_{\lambda_\alpha}^{\it WMAP}$ for the correction of the abundance. 
For the sub-covariance matrix involving the lensing profile $\Delta\Sigma_{\lambda_\beta}$, we multiply each component by a factor of $(
\Delta\Sigma_{\lambda_\beta}^{\it Planck}/\Delta\Sigma_{\lambda_{\beta} }^{\it WMAP})$ for $R>1 h^{-1}{\rm Mpc}$ with enough angular resolution in the mock catalogs. Here $\Delta\Sigma_{\lambda_\beta}^{\it WMAP}$ is the average over the 108 realizations and $\Delta\Sigma_{\lambda_\beta}^{\it Planck}$ is the best-fit model value from the crude covariance.
However, since the shape noise gives a dominant contribution to the covariance involving $\Delta\Sigma_{\lambda_\beta}$, this correction for $\Delta\Sigma_{\lambda_\beta}$is not significant. 

\subsection{Random Covariance:  ${\bf C}^{\rm R}$}
The stacked cluster lensing is a cross-correlation of the number density
fluctuation field of clusters with the matter perturbation. In analogous
to the method in \citet{LandySzalay:93}, the cross-correlation can be
measured by subtracting the stacked lensing profile around random points
from that around the clusters.  The stacked lensing around random
catalogs is generally non-zero due to boundary effects of a general
survey geometry.  
The use of the
random subtraction corrects for these effects.  Thus the use of the
random catalogs in the lensing estimator reduces the \revise{covariance},
as shown in \revise{Section~2.3 of} \citet{Singhetal:16}.

To create random catalogs, we randomly assign angular positions to each
of the ``real'' SDSS redMaPPer clusters within the SDSS survey footprint
(without masks) that is the same footprint of the mock catalog. In this
way we generated 100 times the number of real redMaPPer clusters.
Note that, by construction, the random catalogs have the same distributions
of richness and redshift as those of the real clusters.
We use the same random catalogs for all the 108 mock catalogs. 
We compute the covariance term of random points in
equation~\eqref{eq:full_cov} as
\begin{equation}
{\bf C}^{\rm R}(D_i, D_j)=\frac{1}{N_{\rm r}}\sum_{a=1}^{N_{\rm r}} 
[D^{\rm R}_{a,i}  - \overline{D}{}^{\rm R}_{i}][D^{\rm R}_{a,j}  - \overline{D}{}^{\rm R}_{j}],
\label{eq:cov_r}
\end{equation}
where $D^{\rm R}_{a,i}$ is the lensing measurement from the mock catalogs with shape noise around
random points or the abundance measurement for the total area, in the $a$-th realization, and $\overline{D}{}^{\rm R}_{i}$ is the average over 108 realizations. 

\section{Dependences of cluster observables on the parameters in mass-richness relation}
\label{app:parameters}
To be self-contained within this paper, 
we here study the dependences of the abundance and the lensing profiles on 
model parameters in the mass-richness relation. 
Figure~\ref{fig:parameter_dependence} shows the lensing profiles and the abundance in each richness bin when shifting each model parameter 
by an amount of the 68\% CL interval around the best-fit value  
with fixing other parameters to the best-fit values for the sample of $20 \le \lambda \le 100$ (see Table~\ref{tab:paramsMCMC_full} for the values). 

\revise{
\section{The posterior distributions of cluster observables for the lensing or abundance alone}
\label{app:cluster-abundance_alone}}
\revise{
\begin{figure*}
        \centering
        \includegraphics[width=0.999 \textwidth]{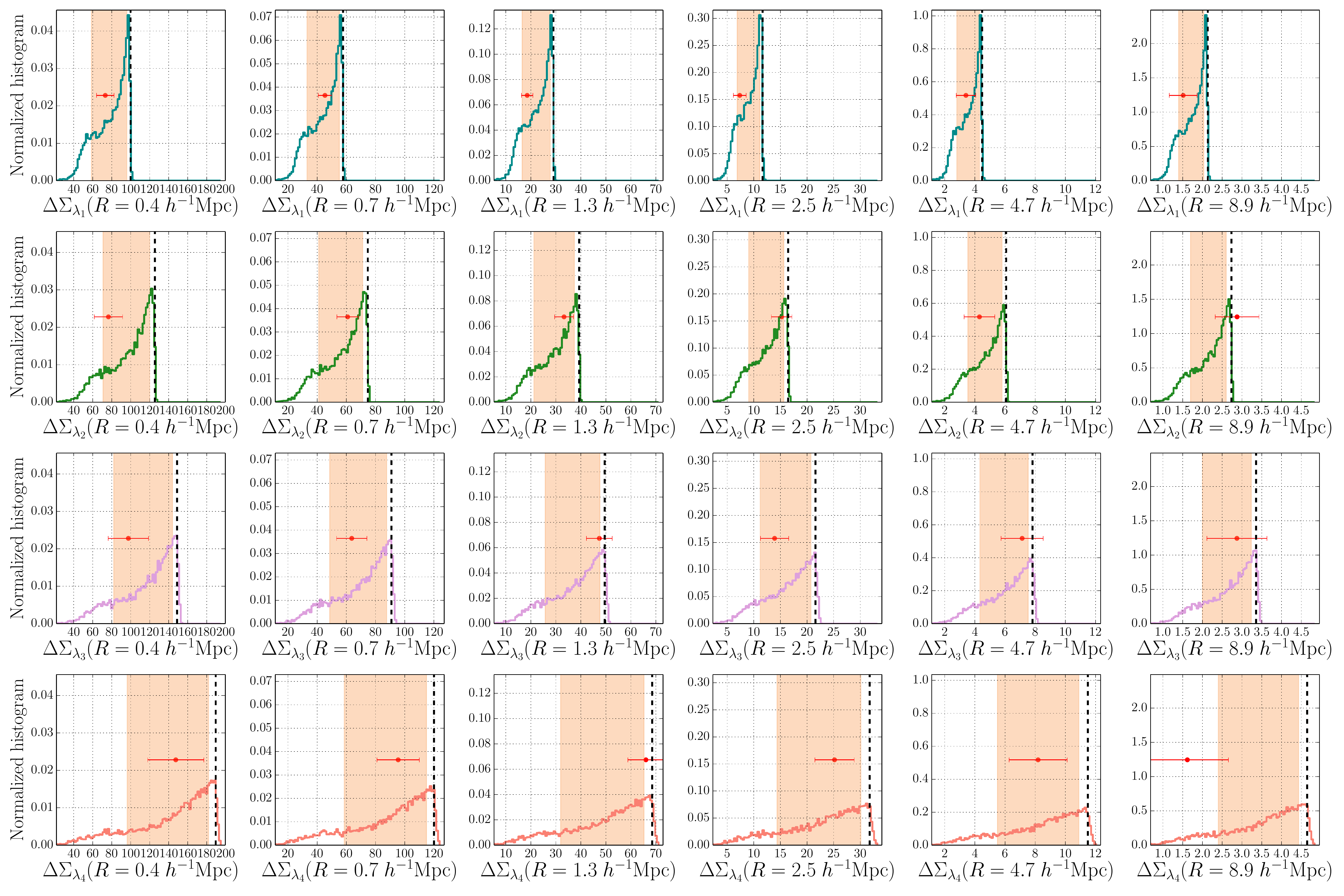}
        \caption{ \revise{The histogram in each panel shows the posterior distribution of the model lensing profile at a given radial bin for each 
        richness cluster sample, which is computed from the MCMC chains when the mass-richness relation is constrained from  
        the model comparison with the abundance information alone, as done in Figure~\ref{fig:abundance_alone}.
        The vertical black dashed line denotes the model
        prediction for the best-fit model that reproduces the abundance. The shaded region shows the 16th and 84th percentiles of the posterior 
        distribution. 
        For comparison the red point with error bar
        denotes the measurement with the estimated error.} }
        \label{fig:lenshist_abundancealone}
\end{figure*}
\begin{figure*}
        \centering
        \includegraphics[width=0.999 \textwidth]{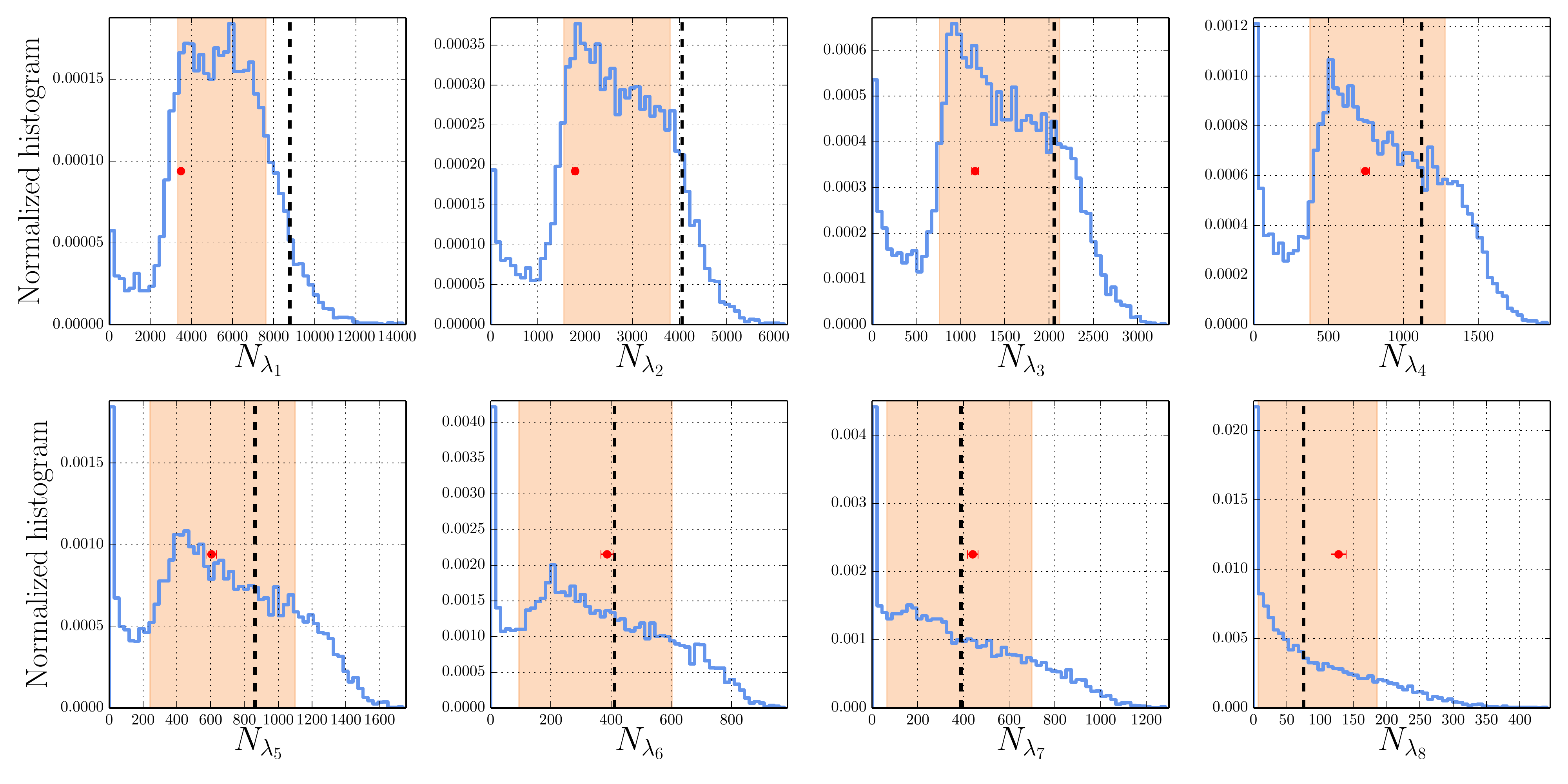}
        \caption{ \revise{Similarly to the previous figure, but the posterior distribution of the abundance in each richness bin, computed 
        from the MCMC chains from the comparison with the lensing information alone, as done in Figure~\ref{fig:lensing_alone}.} }
        \label{fig:abundancehist_lensalone}
\end{figure*}
Figures~\ref{fig:lenshist_abundancealone} and \ref{fig:abundancehist_lensalone} show
the posterior distribution of cluster observables, the lensing profile or the abundance that are computed from the MCMC chains when the mass-richness
relation is constrained from the model comparison with either alone of the abundance or the lensing information. These give a supplementary
information to the results in Figures~\ref{fig:abundance_alone} and \ref{fig:lensing_alone}. 
\reviseplus{
Figure~\ref{fig:lenshist_abundancealone} shows that
the lensing measurements in the lowest richness bin ($20<\lambda<30$)
are systematically smaller than the posterior distributions from the abundance alone fitting result.
Figure~\ref{fig:abundancehist_lensalone} also shows that
the abundance measurements in the first and second richness bins ($20<\lambda<25$ and $25<\lambda<30$) are relatively small values
compared to the posterior distributions from the lensing alone fitting result, although the significance is not very high.
As discussed in Section~\ref{sec:projectioneffect},
these results might be a signature of the residual systematics in the data (e.g. the projection effects at the low richness bins).}}
\end{document}